\RequirePackage{fix-cm}
\documentclass[twocolumn,epjc3]{svjour3}  
\smartqed
\RequirePackage{graphicx}

\pdfoutput = 1

\usepackage[utf8]{inputenc}
\usepackage{color}
\usepackage[dvipsnames]{xcolor}
\usepackage{graphicx}
\usepackage{bm}
\usepackage{caption,float}
\usepackage[normalem]{ulem}
\usepackage{amsmath}
\usepackage{amssymb}
\usepackage{enumerate}

\definecolor{darkblue}{rgb}{0.0,0,0.5}
\definecolor{darkgreen}{rgb}{0.0,0.3,0.0}
\definecolor{redish}{rgb}{0.675,0,0.2}
\definecolor{red}{rgb}{0,0,0}
\definecolor{green}{rgb}{0,0.6,0}
\definecolor{blue}{rgb}{0,0,0.8}

\usepackage[unicode=true, bookmarks=false, linkcolor=darkblue,
  citecolor=redish, breaklinks=false, colorlinks=true,
  hyperfootnotes=true]{hyperref}
\hypersetup{pdftitle={L2 sensitivities},
 pdfauthor={Xiaoxian Jing and others},
 unicode=true, pdfborder={0 0 1}, linkcolor=darkblue, citecolor=redish, breaklinks=false, colorlinks=true}

\journalname{Eur. Phys. J. C}

\begin{document}
\title{Deuterium scattering experiments in CTEQ global QCD analyses: a comparative investigation}

\author{A.~Accardi\thanksref{addr1,addr2}
        \and
        T.~J.~Hobbs\thanksref{addr3,addr2,addr4,addr5,e1}
        \and
        X.~Jing\thanksref{addr3,addr2}
        \and
        P.~M.~Nadolsky\thanksref{addr3}
}

\thankstext{e1}{corresponding author: thobbs@fnal.gov}

\institute{Hampton University, Hampton, VA 23668, U.S.A. \label{addr1}
           \and
           Jefferson Lab, Newport News, VA 23606, U.S.A. \label{addr2}
           \and
           Department of Physics, Southern Methodist University, Dallas, TX 75275, U.S.A. \label{addr3}
           \and
           Fermi National Accelerator Laboratory, Batavia, IL 60510, U.S.A. \label{addr4}
           \and
           Department of Physics, Illinois Institute of Technology, Chicago, IL 60616, U.S.A. \label{addr5}
}

\date{Received: date / Accepted: date}


\maketitle

\begin{small}
\noindent
{\bf Author ORCIDs:}\\
0000-0002-2077-6557 --- A.~Accardi\\
0000-0002-2729-0015 --- T.~J.~Hobbs\\
0000-0003-2679-5262 --- X.~Jing\\
0000-0003-3732-0860 --- P.~M.~Nadolsky\\
\end{small}

\begin{abstract}
    Experimental measurements in deep-inelastic scattering and lepton-pair production on deuterium targets play an important role in the flavor separation of $u$ and $d$ (anti)quarks in global QCD analyses of the parton distribution functions (PDFs) of the nucleon. We investigate the impact of theoretical corrections accounting for the light-nuclear structure of the deuteron upon the fitted $u, d$-quark, gluon, and other PDFs in the CJ15 and CT18 families of next-to-leading order CTEQ global analyses.
    The investigation is done using the $L_2$ sensitivity statistical method, which provides a common metric to quantify the strength of experimental constraints on various PDFs and ratios of PDFs in the two distinct fitting frameworks. Using the $L_2$ sensitivity and other approaches, we examine the compatibility of deuteron data sets with other fitted experiments under varied implementations of the deuteron corrections. We find that freely-fitted deuteron corrections modify the PDF uncertainty at large momentum fractions and will be relevant for future PDFs affecting electroweak precision measurements. 
\end{abstract}

\tableofcontents

\noindent
{\bf Declarations}\\
\begin{small}
\noindent
{\bf Funding:}
This work was funded by the U.S.~Department of Energy under Grant No.~DE-SC0010129 at SMU, and T.~J. Hobbs received financial support from a JLab EIC Center Fellowship.
The work of A.~Accardi was supported by the U.S.~Department of Energy contract DE-AC05-06OR23177,
under which Jefferson Science Associates LLC manages and operates Jefferson Lab. A.~Accardi and and X.~Jing were additionally supported by DOE contract DE-SC0008791.  \\
{\bf Conflicts of interest:}
The authors have no relevant financial or non-financial interests to disclose.\\
{\bf Availability of data and material:}
The data sets generated during and/or analysed during the current study are available from the corresponding author on reasonable request. \\
{\bf Code availability:}
The $L_2$ sensitivity analysis code and supporting files are available at the following location: \url{https://tinyurl.com/CJCT2021}. This or another link will be
continuously maintained.\\
{\bf Authors' contributions:}
We affirm that all authors made important and substantial contributions to the conception, execution, and analysis contained in this study.
\end{small}
%

%
\section{Introduction}
\subsection{Light-parton structure of the nucleon in electroweak precision measurements}
\label{sec:intro}

\begin{figure*}[th]
	\centering
	\includegraphics[width=0.48\textwidth]{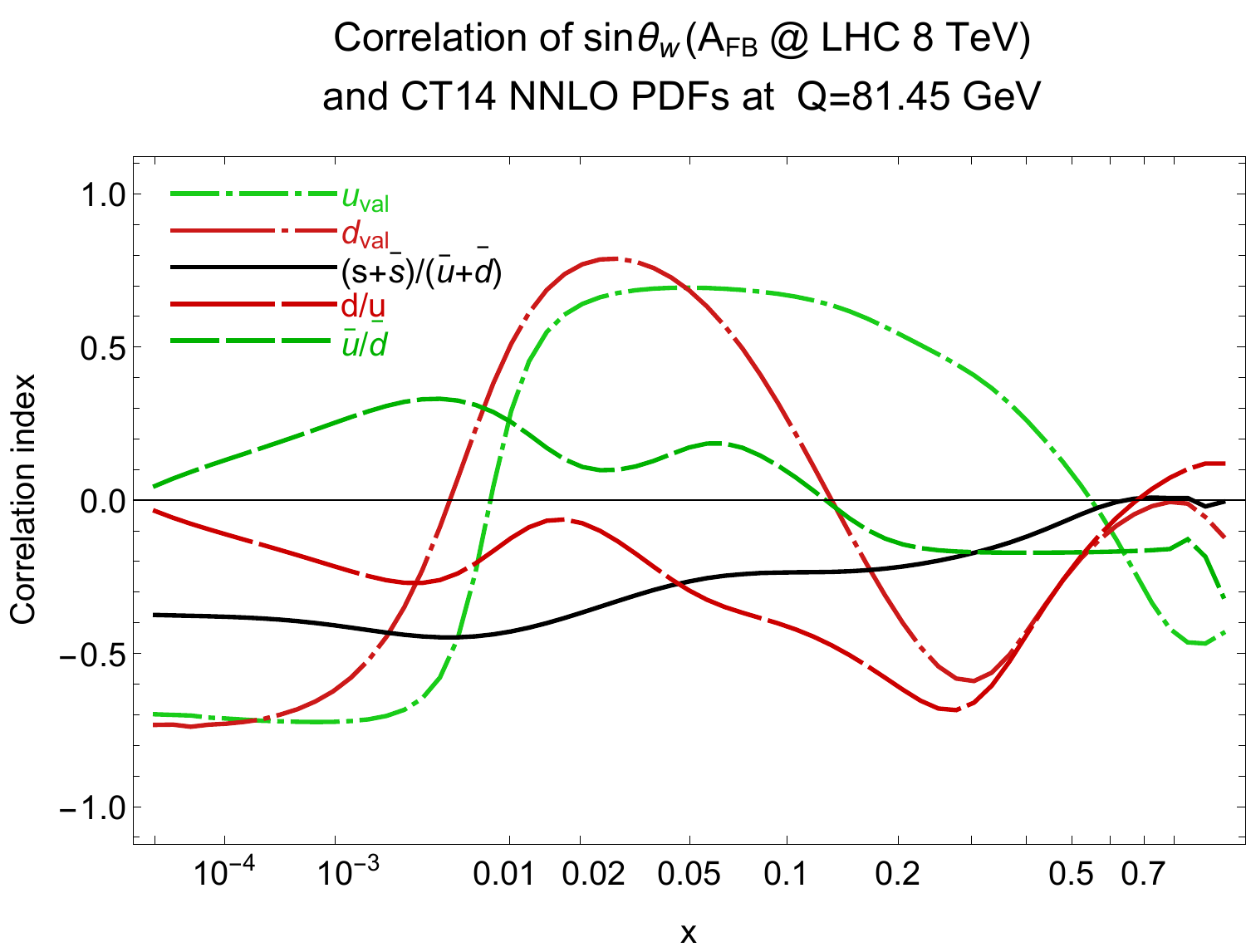} \ \ \ 
	\includegraphics[width=0.48\textwidth]{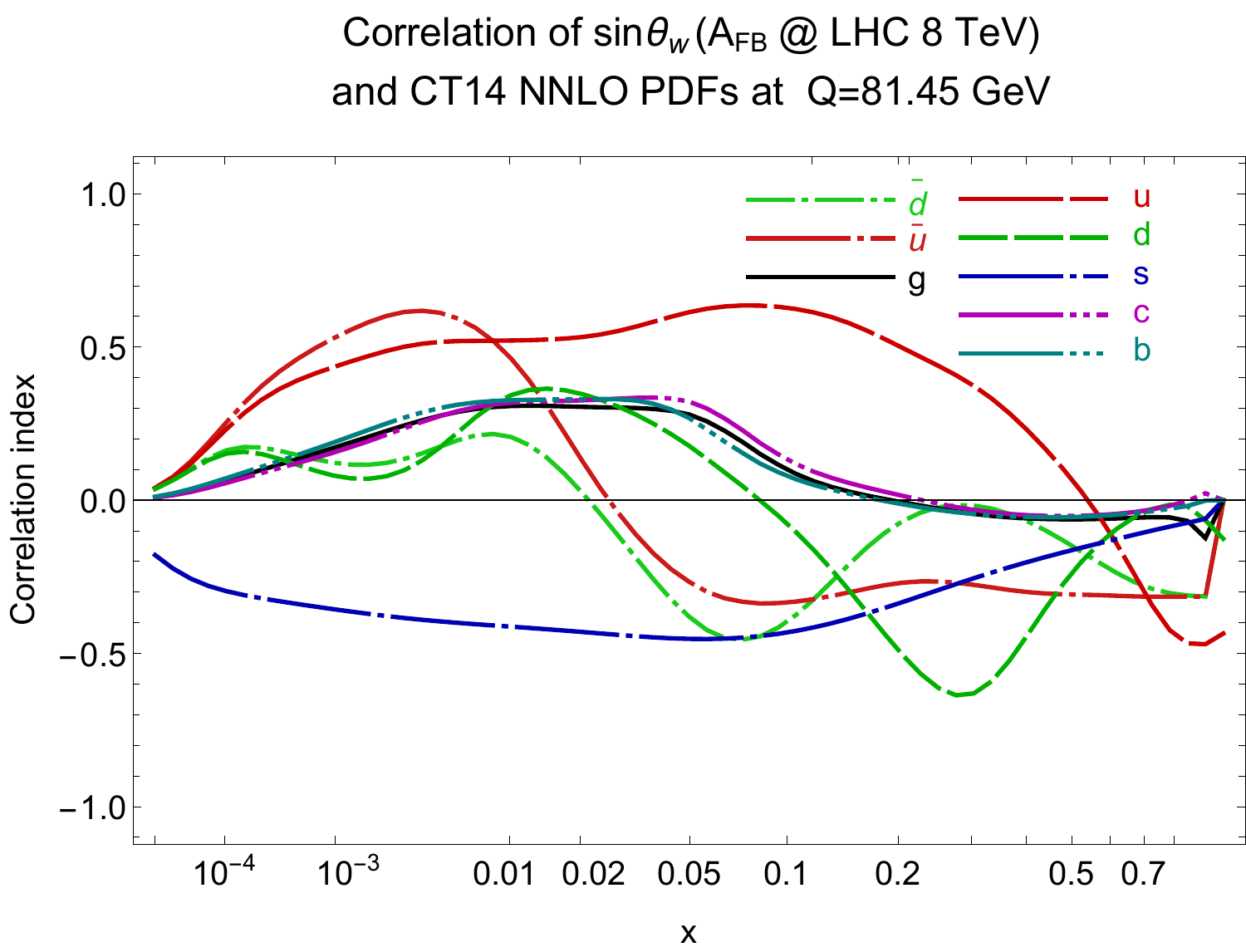} 
    \caption{Hessian correlations \cite{Stump:2001gu,Nadolsky:2001yg,Nadolsky:2008zw} for the values of $\sin\theta_W$ extracted from $Z$ boson production
	at the LHC 8 TeV \cite{Armbruster:2018}. Left: correlations with valence PDFs and PDF ratios at $Q=$81.45 GeV, plotted as a function of $x$ for CT14 NNLO PDFs. Right: the same, for correlations with PDFs of individual flavors.}
\label{fig:s2w_corr}
\end{figure*}

Electroweak precision tests of the Standard Model (SM) at hadron colliders are nontrivially sensitive to the parton flavor composition of initial-state hadrons.
For spin-independent inclusive observables at the CERN Large Hadron Collider (LHC) -- or, indeed, at any high-enough energy facility such as the Relativistic Heavy Ion Collider, the Jefferson Lab CEBAF accelerator, or the future Electron-Ion Collider -- this flavor composition is typically specified by helicity-averaged parton distribution functions (PDFs) of the proton.  The PDFs, $f(x,Q)$, have long been of strong interest from the perspective of both fundamental Quantum Chromo-Dynamics (QCD) as well as particle phenomenology, given
that they quantify the probability 
of resolving a quark or gluon constituent
of flavor $f$ carrying a fraction $x$ of the proton's longitudinal momentum in a scattering process with a squared energy scale $Q^2 \gtrsim 1\mbox{ GeV}^2$.  For this reason, the PDFs play a central role in predicting cross sections for $pp$ collisions at the LHC, and, in particular,  their accuracy influences the ability of LHC measurements or other high-energy data to constrain the SM parameters, including in the electroweak sector.

Due to the challenge of reducing their uncertainties and empirically distinguishing
among their parton flavors, PDFs have historically been determined most robustly through ``global QCD 
fits'' \cite{Jimenez-Delgado:2013sma,Gao:2017yyd,Kovarik:2019xvh,Ethier:2020way}, now increasingly performed at next-to-next-to-leading order (NNLO) accuracy in $\alpha_s$, and drawing upon
large collections of experimental measurements sensitive to QCD and different underlying PDF combinations. In spite of the growing number of 
LHC measurements,  deeply-inelastic scattering (DIS) experiments involving fixed hadronic
or nuclear targets at BCDMS, NMC, SLAC, and JLab continue to provide key information to disentangle the PDFs in recent global QCD analyses such as CJ15 \cite{Accardi:2016qay}, ABMP16 \cite{Alekhin:2017kpj}, CT18 \cite{Hou:2019efy}, NNPDF3.1 \cite{Ball:2017nwa}, and MSHT20 \cite{Bailey:2020ooq}.
The fixed-target experiments complement analogous DIS collisions at the HERA $ep$ collider by extending the momentum fraction coverage to larger $x$ values and adding measurements on deuterium targets. In fact, such experiments provide the leading constraints on the (anti)quark PDFs at low scales $Q$ and large momentum fractions $x\gtrsim 0.05$, as well as on the gluon PDF by observing scaling violations over the same kinematic region \cite{Accardi:2016muk,Wang:2018heo,Hobbs:2019gob}.

\begin{figure*}[t]
	\centering
	\includegraphics[width=0.48\textwidth]{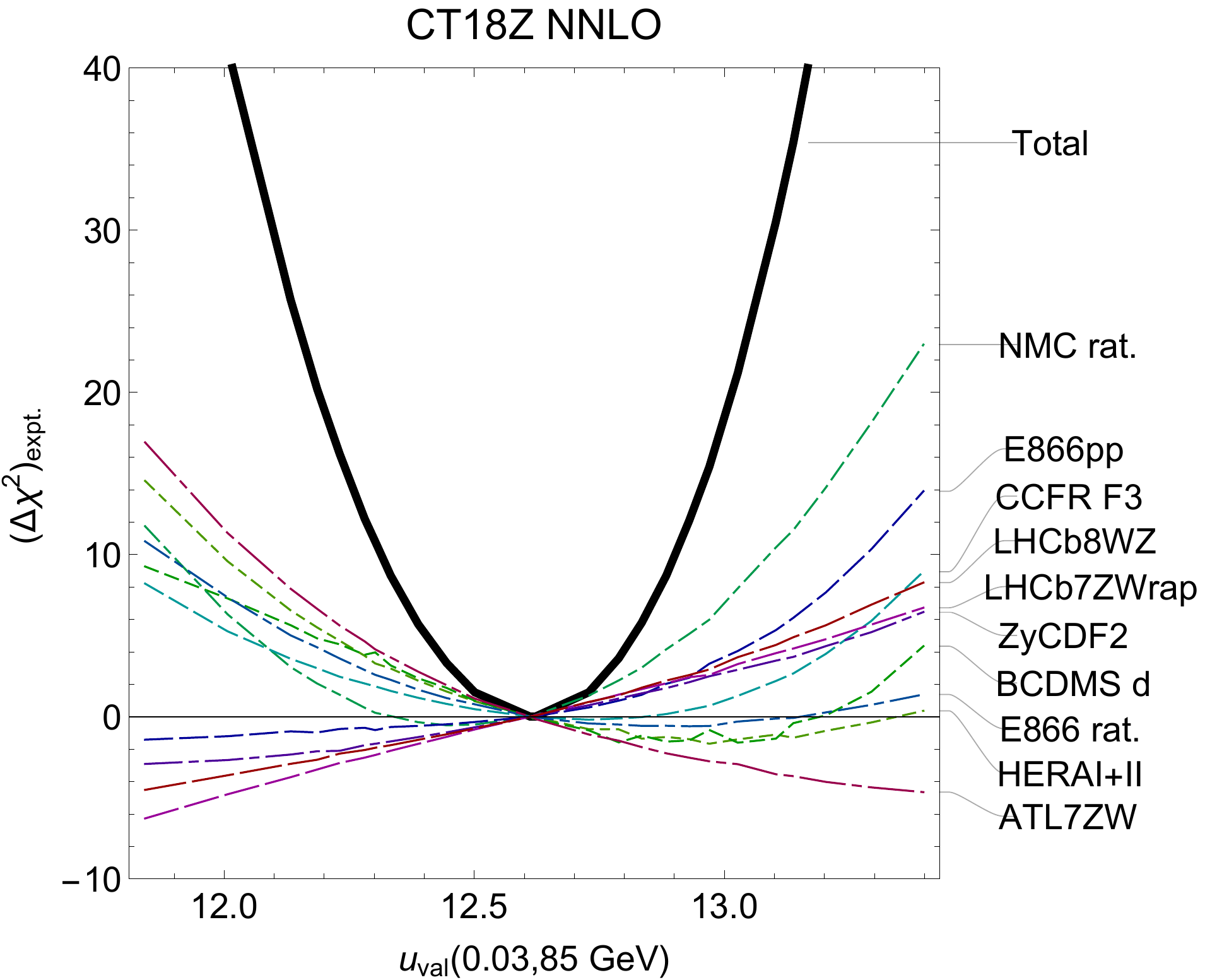} 
	\hspace*{0.01\textwidth}
	\includegraphics[width=0.48\textwidth]{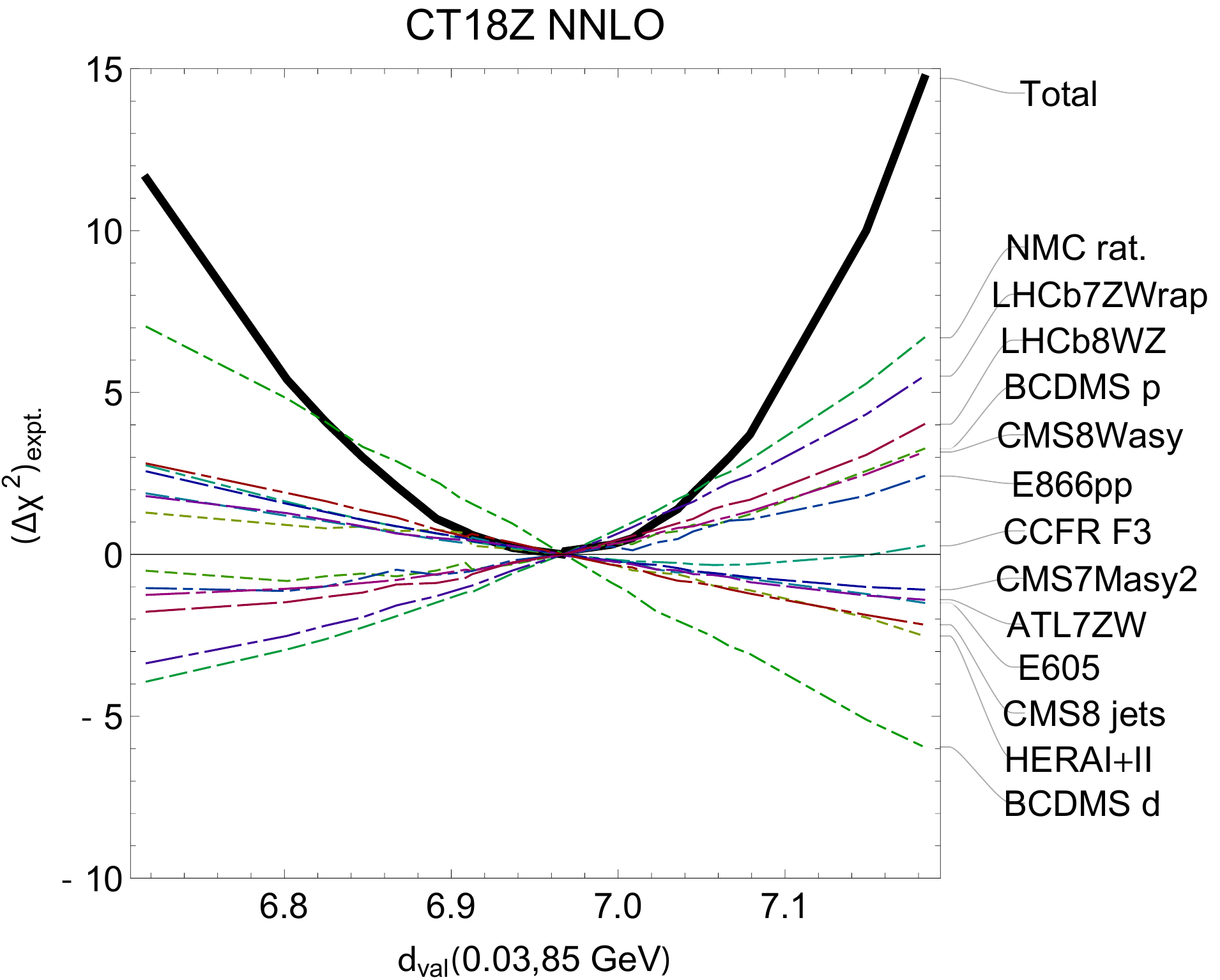} 
\caption{Lagrange Multiplier scans on $d_{val}(x=0.03,Q=85\, \mathrm{GeV})$ (left) and $u_{val}(x=0.03,Q=85\, \mathrm{GeV})$ (right), showing the changes in the $\chi^2$ for all data sets and most sensitive experimental data sets in the CT18Z NNLO global QCD analysis \cite{Hou:2019efy}.
}
\label{fig:qval_LM}
\end{figure*}

In precision tests of the electroweak sector, the substantial PDF dependence of the involved theoretical calculations affects experimental determinations of SM parameters, such as the weak-mixing angle $\theta_W$ extracted from the $A_\mathrm{FB}$ forward-backward asymmetry measured in the production of $Z$ bosons during Runs 1 and 2 of the LHC.  
Fig.~\ref{fig:s2w_corr} illustrates typical Hessian correlations \cite{Stump:2001gu,Nadolsky:2001yg,Nadolsky:2008zw} of PDF combinations (left) and PDFs (right) with the $\sin\theta_W$ values extracted from 8 TeV $A_\mathrm{FB}$ measurements at the LHC. Here, the correlations are computed using the preliminary (unpublished) ATLAS Run-1 data \cite{Armbruster:2018} on $\sin\theta_W$ extracted with individual PDF eigenvector sets of the CT14 NNLO ensemble \cite{Dulat:2015mca}. 

In the left subfigure, we see that the values of the extracted $\sin\theta_W$ are strongly correlated with the valence combinations of light-quark PDFs at $Q=81.45$ GeV, $d_\mathit{val}(x,Q)\equiv d(x,Q)-\bar d(x,Q)$ at $x\approx 0.008-0.05$ and $u_\mathit{val}(x,Q) \equiv u(x,Q)-\bar u(x,Q)$ at $x\approx 0.008-0.1$. In addition, significant correlations with the extracted 
$\sin\theta_W$ exist at higher $x \gtrsim 0.3$ as well, especially for the PDF ratio $d/u$, and again for $d_\mathit{val}$. 
Remarkably, the correlations are weaker with the PDFs of individual parton flavors (shown at right) than with the valence combinations. An anti-correlation with the $d$-quark (green dashed line) at $x\sim 0.3$, affecting $A_\mathrm{FB}$ at smaller $x$ via the valence-quark sum rule, is evident in this case, though not exceptionally strong. 

The sizable correlations between fitted PDFs and $\sin\theta_W$ in
Fig.~\ref{fig:s2w_corr} are consistent with the significant PDF uncertainties
on these and similar BSM-sensitive quantities, including the $W$ boson mass, $M_W$,
and Higgs cross section, $\sigma_H$. For this reason, the realization of
next-generation precision in the determination of these electroweak parameters is
critically dependent on the reduction of their associated PDF uncertainties, including
the high-$x$ uncertainty of the $d$-quark and gluon ($g$) PDFs, as well as of $d_\mathit{val}$
and $d/u$.

We might ask where the experimental constraints on the relevant PDF combinations for LHC electroweak precision tests arise from. While direct measurements at the LHC will supply increasing information on the PDFs affecting $A_\mathrm{FB}$ \cite{Fu:2020mxl} and other observables \cite{Khalek:2018mdn}, recent CTEQ studies \cite{Accardi:2016qay,Hou:2019efy,Wang:2018heo,Hobbs:2019gob,Hou:2019gfw} find that deep-inelastic scattering experiments on nuclear targets will continue to provide strong constraints on the down-quark PDFs in the nucleon in the near future. In a global QCD analysis, experimental measurements made solely on proton targets are at present insufficient for full separation of parton distributions for $d$, $s$, $g$, and anti-quark flavors. Assuming parton-level charge symmetry, $d^p(x,Q)\approx u^n(x,Q)$, between the PDFs in the proton $p$ and the neutron $n$, and correcting for low-energy nuclear effects \cite{Accardi:2016muk}, one can then use scattering processes on the deuteron or heavier nuclei to constrain the down-type PDFs in the proton.

We illustrate the importance of fixed-target data in the determination of the weak mixing angle with the help of Lagrange Multiple (LM) scans \cite{Pumplin:2001ct} in Fig.~\ref{fig:qval_LM}, in which we examine the dependence of the figure-of-merit function $\chi^2$ in the CT18Z NNLO analysis \cite{Hou:2019efy} on the values of the valence $u_{val}$ (left) and $d_{val}$ (right) quarks at $x=0.03$ and $Q=85 $ GeV. We plot the change in $\chi^2$, as compared to the value in the best fit, for all data sets (labeled as ``Total" in the figure) and for individual experiments with the highest sensitivity to this PDF combination. The curves for the experimental data sets are labeled according to the convention in Table~\ref{tab:exps_and_categories}. The LM scans show that a small group of DIS experiments --   NMC ratio of $d$ and $p$ DIS cross sections \cite{Arneodo:1996qe}, inclusive HERA I+II DIS \cite{Abramowicz:2015mha}, BCDMS $p$ and $d$ reduced cross sections \cite{Virchaux:1991jc,virchaux1992measurement,Benvenuti:1989rh,Benvenuti:1989fm}, CCFR $F_3$ structure function \cite{Seligman:1997mc} -- contribute the largest variations in $\Delta \chi^2$ when the valence PDFs are varied, together with several lepton pair production experiments by ATLAS, LHCb, E605, and E866. 
In the case of $d_{val}$ in the right Fig.~\ref{fig:qval_LM}, the BCDMS measurements on $p$ and $d$ show somewhat different preferences, with the deuteron data clearly preferring a higher $d_{val}(0.03,85\mbox{ GeV})$. In these and other cases, the deuteron DIS measurements, with their large numbers of precise data points, have large statistical significance in the global fit. As this LM scan also illustrates, the proton and deuteron data sets in the CT18 fit sometimes prefer somewhat different values of $d_{val}(x,Q)$, which in turn may hamper the efforts for reducing the PDF uncertainty in the relevant EW precision measurements.

In this article, we will employ a statistical technique called the $L_2$ sensitivity \cite{Hobbs:2019gob} that can be viewed as a fast approximation to the LM scans that are usually very computing-intensive. With this technique, we will survey agreement between the constraints on the PDFs from the deuteron and other data sets in a wide range of $x$ and for various treatments of deuteron corrections. While we have investigated sensitivities for various PDF flavors, our presentation will focus on the sensitivities to $u$, $d$, and $g$ PDFs that are most affected. 

Apart from its significance for the LHC and SM phenomenology, the physics of large-$x$ quark PDFs is interesting in its own right. Nonperturbative QCD approaches \cite{Jimenez-Delgado:2013sma,Holt:2010vj,Roberts:2013mja}
and lattice QCD \cite{Lin:2017snn,Lin:2020rut} provide increasingly complete predictions for the flavor composition of unpolarized protons at $x\to 1$, which in turn inform one on the role of color confinement in the binding of the valence quarks. 
These predictions can then be confronted with phenomenological determinations of Mellin moments and PDFs at large $x$.
For example, Ref.~\cite{Courtoy:2020fex} observes that  
the proton (BCDMS $F_2^p$, E866 $pp$ DY, HERAI+II $ep$ DIS) and deuteron (BCDMS $F_2^d$, NMC $p/d$ ratio) experiments in the CT18 NNLO analysis have somewhat different preferences for the effective exponents $\beta$ controlling the $(1-x)^\beta$ falloff of the
valence up and down quark PDFs at $x\to 1$. In turn, these differences impact comparisons of phenomenological PDFs against large-$x$ predictions from quark counting rules \cite{Brodsky:1973kr,Brodsky:1974vy,Farrar:1975yb} and other nonperturbative approaches \cite{Holt:2010vj,Roberts:2013mja}. The analysis methods utilized in this paper can 
shed light on these issues.

\subsection{The role of nuclear-medium effects}

Extracting parton-level information from nuclear data sets involving the deuteron or heavier targets
requires an understanding of the effects of the nuclear environment \cite{Ethier:2020way,Accardi:2016muk}. The trivial dependence on the nuclear atomic number $A$ and charge $Z$ is normally implemented by constructing a nuclear PDF as a linear combination of the PDFs on free protons and neutrons, as reviewed, e.g., in Sec.~5.~A of Ref.~\cite{Kovarik:2019xvh}. On top of this trivial $(A,Z)$ dependence, low-energy interactions in the nuclear medium may modify 
the quark and gluon distributions relatively to those in free nucleons. The nuclear corrections that account for these deviations can be computed with increasing sophistication and connection to the formal theory describing low-energy dynamics. One can, for example, utilize phenomenological, data-driven ratios to convert nuclear-target cross sections to free-nucleon ones~\cite{Hou:2019efy,Ball:2020xqw};
parametrize and fit the nuclear deformation of the PDFs either inside the deuteron in a nucleon PDF fit ~\cite{Bailey:2020ooq,Martin:2009iq} or in a heavy nucleus in a nuclear PDF fit \cite{Eskola:2016oht,Kovarik:2015cma,Deflorian:2011fp,Hirai:2007sx,Khanpour:2016pph,AbdulKhalek:2019mzd};
or, finally, adopt a dynamical model of the low-energy nucleon-nucleon interactions and calculate the hard cross-section as a double convolution of parton distributions inside the nucleons and nucleon wave functions inside the nuclear targets 
\cite{Accardi:2016qay,Alekhin:2017fpf,Kulagin:2004ie}. 
The resulting extractions of the nucleon PDFs from nuclear data then have a dependence
on the assumed nuclear corrections. 

%
%
The specific methods used typically differ when analyzing light nuclear targets ({\it e.g.}, the deuteron) or heavy nuclei ({\it e.g.}, Fe). Here, we concentrate on and summarize the techniques utilized in the CT18 and CJ15 fits which form the starting point for the study presented in this paper, and then briefly mention other approaches:

\begin{enumerate}
    \item {\bf Deuteron corrections}.
A dynamic deuteron correction in the CJ15 next-to-leading order (NLO) PDF fit \cite{Accardi:2016qay} was applied to any process involving interaction with a deuterium target,
including both DIS and Drell-Yan experiments, as detailed at greater length in Sec.~\ref{sec:deut}.
This correction allows the CJ15 NLO fit to include the fixed-target DIS data from SLAC and JLab at the largest $x$ not admitted by other groups. The correction 
can be understood as arising from several dynamical effects,
such as the relativistic Fermi motion of bound nucleons, binding corrections, 
and nucleon off-shellness effects. In practice these mechanisms are taken into account via
convolutions of free-hadron cross sections with nuclear smearing functions calculated starting from bound nucleon wave-functions. Nuclear correction mostly affects the intermediate and large regions of $x$. 

The CJ15 analysis also applies a phenomenological parametrization for the off-shell deformation of the scattered nucleon's structure function (in short, ``off-shell corrections") with parameters fitted to data to increase the model flexibility. Care is taken that the valence quark number inside the nucleon is not modified; since the off-shell function is flavor independent and has no significant dependence on $Q^2$, it must then change the sign in the interval of $[0,1]$, meaning that it is essentially given by a polynomial with one or more roots in this range.

An analysis similar in spirit to the CJ15 fit is available from the AKP group \cite{Alekhin:2017fpf}, and differences in the extracted off-shell functions are currently being investigated by the two groups jointly. Alternatively, the deuteron correction can be fitted using a purely phenomenological parametrization as in the MSHT20 analysis \cite{Bailey:2020ooq}, or the additional uncertainty associated with the deuteron effects can be learned from the global analysis data themselves, as done in the NNPDF study \cite{Ball:2020xqw}. Other groups do not include deuteron corrections by selecting the fitted data in a $\{x,Q\}$ region where the deuteron corrections are small compared to the precision of data, as is done, e.g., in the CT18 analysis \cite{Hou:2019efy}.

\item {\bf Heavy-nucleus
effects}.
Nucleon PDF fits may include DIS experiments performed on heavy nuclear targets, such as CCFR~\cite{osti_879078} and NuTeV~\cite{Goncharov:2001qe}, involving $^{56}$Fe, and CHORUS~\cite{Kayis_Topaksu_2008}, with $^{82}$Pb (which is not presently fitted in either CT or CJ packages studied here).
It has been known empirically for some time that the structure functions of these heavier nuclear targets exhibit $x$- and $A$-dependent deviations from the structure function of the
physical deuteron, owing to a variety of physical processes characterizing the nuclear medium \cite{Frankfurt:1988nt,Arneodo:1992wf,Kopeliovich:2012kw,Malace:2014uea}, including the heavy-nucleus analogue of the EMC and Fermi-motion effects discussed for the deuteron at high $x$, and nuclear (anti)shadowing phenomena at lower $x$.

To address these effects, the CT group corrects DIS cross sections on iron (CCFR~\cite{osti_879078}, CDHSW~\cite{Berge:1989hr}, and NuTeV~\cite{Goncharov:2001qe}) and proton-copper Drell-Yan (E605 \cite{Moreno:1990sf}) to the corresponding cross sections on deuterium using a phenomenological parametrization of the nuclear-to-deuteron cross section ratios based on results in \cite{Arneodo:1992wf} (see also \cite{Rondio:1993mf}, Fig. 2a), which depends on $x$ but not on $Q^2$ in the fitted region. To include the heavy-nuclear data in the MSHT20 \cite{Bailey:2020ooq} and earlier MMHT fits, a nuclear correction factor, $R_f$, \cite{Martin:2009iq} having the form
$f^A(x,Q^2) = R_f(x,Q^2,A)f(x,Q^2)$, where $f^A(x,Q^2)$ is defined to be the PDF of a proton bound in a nucleus of mass number $A$, was determined. This was assessed using the de Florian {\it et al.} nuclear PDF (nPDF) of Ref.~\cite{Deflorian:2011fp}, then weighted by a 3-parameter modification factor as in Ref.~\cite{Martin:2009iq}, which is actively refitted along with the PDF-associated parameters.
NNPDF \cite{Ball:2018twp} determines the uncertainty due to heavy-nuclear effects using a similar statistical procedure as for the deuteron. 
\end{enumerate}
%
%

As can be seen from this summary, global analyses vary in their treatments of the nuclear effects, but with the frequent conclusion that the resulting differences are marginal in comparison to contemporary PDF uncertainties. Indeed, the experiments in question are fitted reasonably well, while the higher-order QCD and parametrization uncertainties on the PDFs are comparable for the most part to the nuclear ones at NLO or even NNLO. In the present study, however, we identify several areas where the deuteron corrections will play a prominent role in the near future, as the field advances toward higher accuracy in the determination of nucleon PDFs. 
We compare, in particular, the effects of deuteron corrections in two independent PDF global fits by the CTEQ-JLab (CJ) and CTEQ-TEA (CT groups which differ in their phenomenological focus, data selection and, crucially, the treatment of scattering process in nuclear targets. We find that {\bf the deuteron effects will have pronounced consequences for both the fitted PDFs in the large-$x$ region and the correlations among the PDFs and quantities derived from them in an extended $x$ range.} 

More specifically, this paper will elucidate constraints on the $d$-quark, gluon, and other PDFs arising in the CT18 and CJ15 global fits. We will accomplish this by analyzing the $L_2$ sensitivity to various PDFs \cite{Hobbs:2019gob}, a simple informative figure of merit that allows us to look inside the CJ and CT fits and understand the constraints from the fitted experiments on various parton flavors in an expansive region of $x$ and $Q$.

\subsection{Paper organization}

After this introduction, the remainder of the article is as follows. In Sec.~\ref{sec:low}, we briefly
present the deuteron-structure corrections (Sec.~\ref{sec:deut}) with which this investigation is primarily concerned, as well as power-suppressed QCD effects, also relevant to fits involving nuclear data and/or at lower
$Q$ and $W$ (in Sec.~\ref{sec:wnuc}). In Sec.~\ref{sec:methods}, we
summarize the relevant features of the CT and CJ fitting frameworks and the special modifications we made in the two analyses to allow their direct comparisons. In Sec.~\ref{sec:L2}, we review the $L_2$ sensitivity method. In Sec.~\ref{sec:results}, we apply this method to analyze the impact of deuteron-structure
corrections on the fit results, and examine the patterns of PDF pulls obtained in the several iterations of CT/CJ under different assumed treatments of the deuteron corrections.
As representative cases, we will concentrate on the $d/u$ ratio in Secs.~\ref{sec:d/u_PDF} and \ref{sec:d/u_pulls}, and on the gluon in Secs.~\ref{sec:gluon_PDF} and \ref{sec:gluon_pulls}.
In Section~\ref{sec:weakmix} we also explore the implications  of our analysis to precision measurements of the weak mixing angle. The conclusions in Sec.~\ref{sec:conc} are followed by a technical appendix describing a numerical procedure to reconcile the error analyses in the CJ and CT approaches.

\section{Low-energy QCD effects}
\label{sec:low}

\subsection{Deuteron-structure effects}
\label{sec:deut}
The critical low-energy effect considered in this study, which arises due to MeV-scale dynamics characterizing nuclear bound states, is the modification of the parton-level substructure of nucleons embedded in the nuclear medium --- in particular, those effects arising in the most weakly bound nuclear system, the two-body deuteron.

\begin{figure*}[htb]
	\centering
	\includegraphics[width=0.48\textwidth]{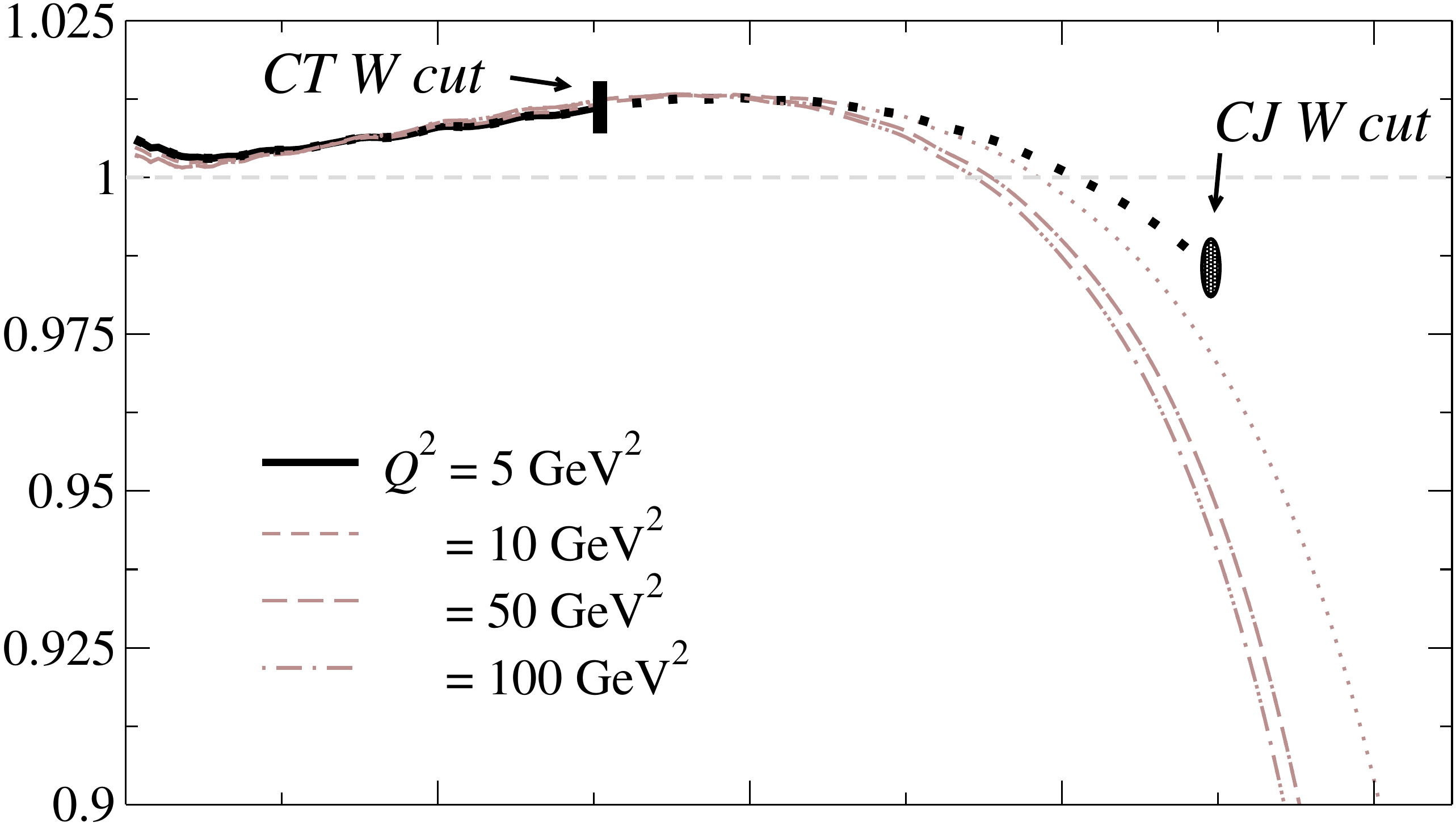}
	\includegraphics[width=0.48\textwidth]{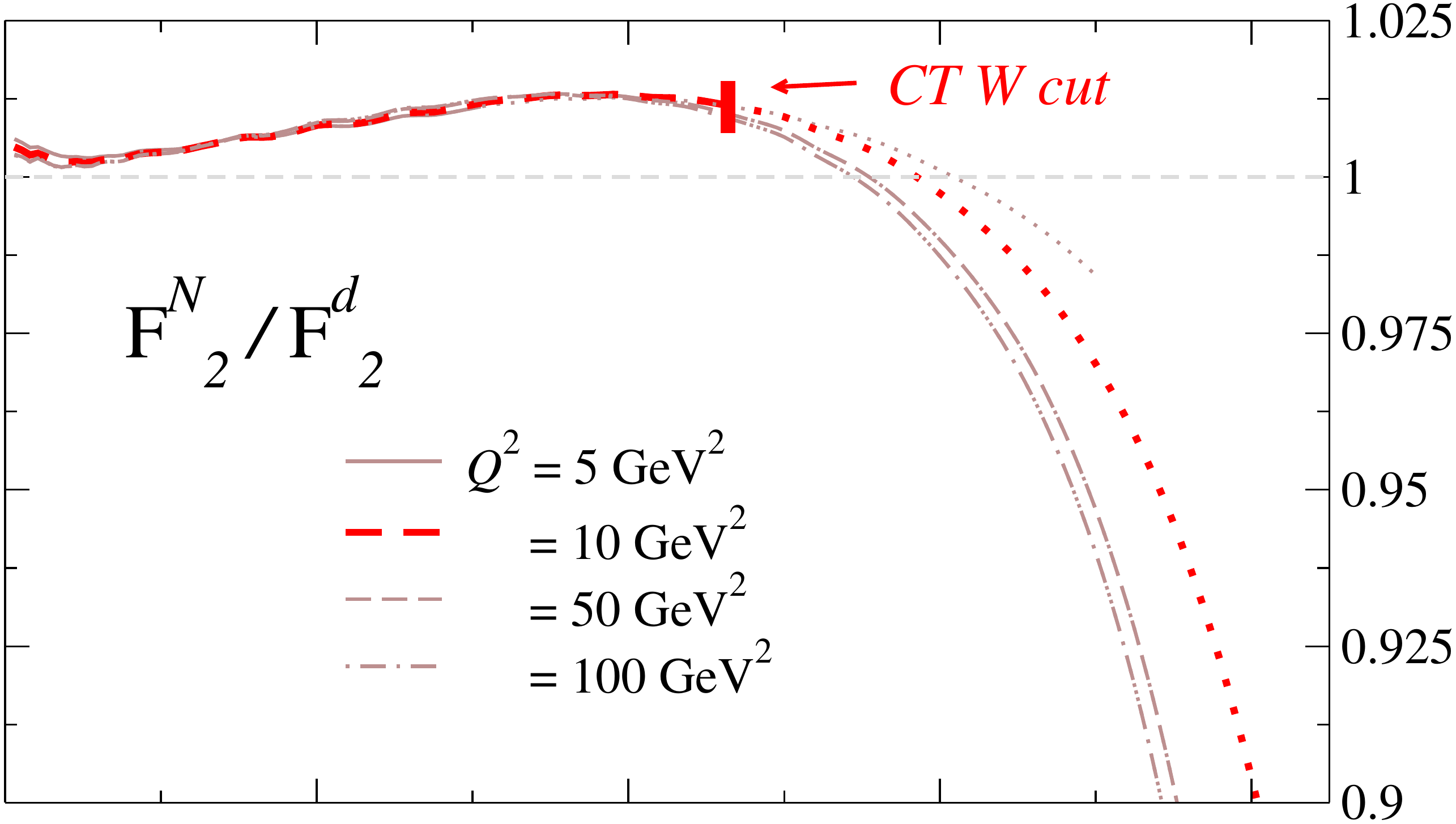} \\
\vspace*{0.1cm}
	\includegraphics[width=0.48\textwidth]{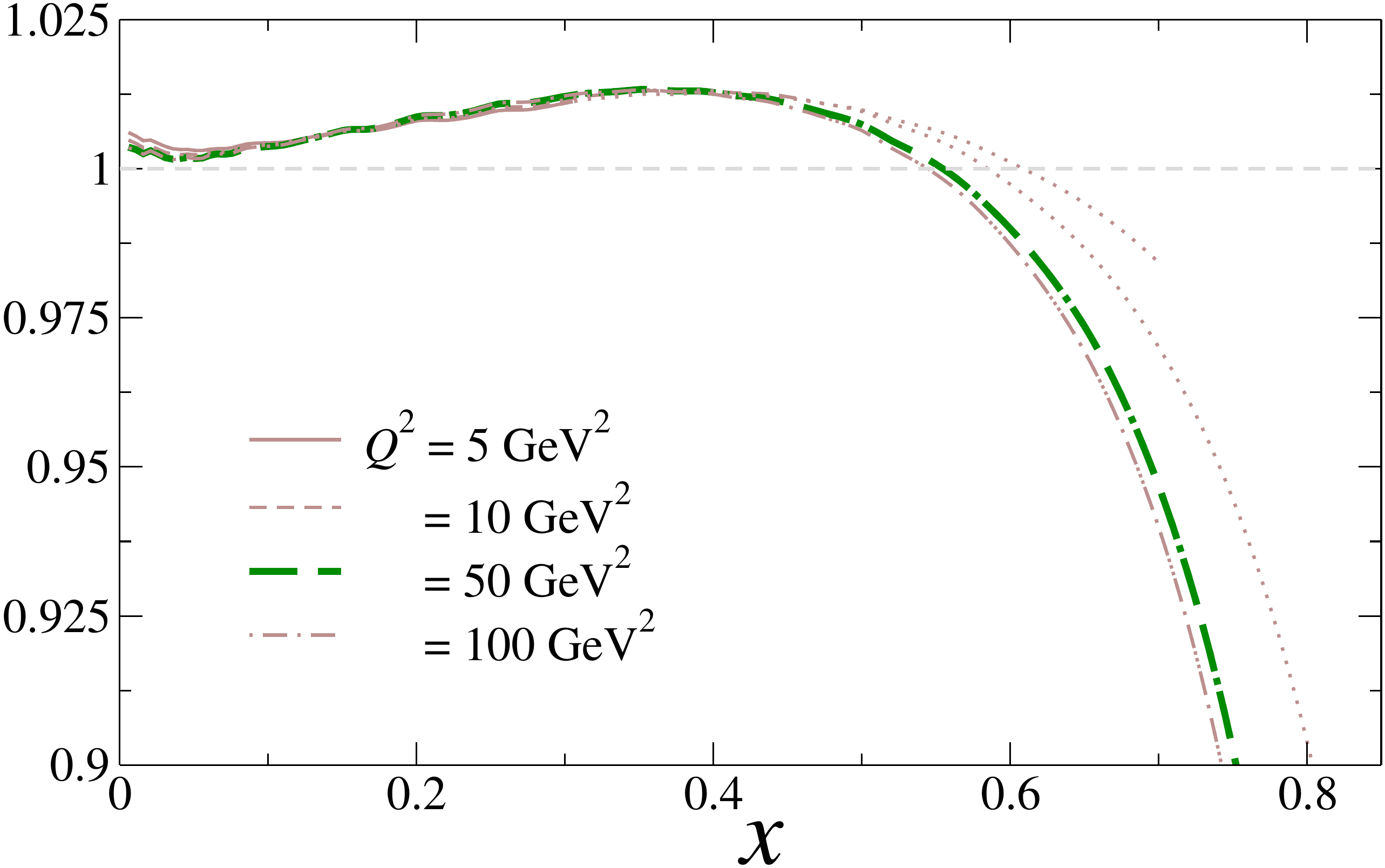}
	\includegraphics[width=0.48\textwidth]{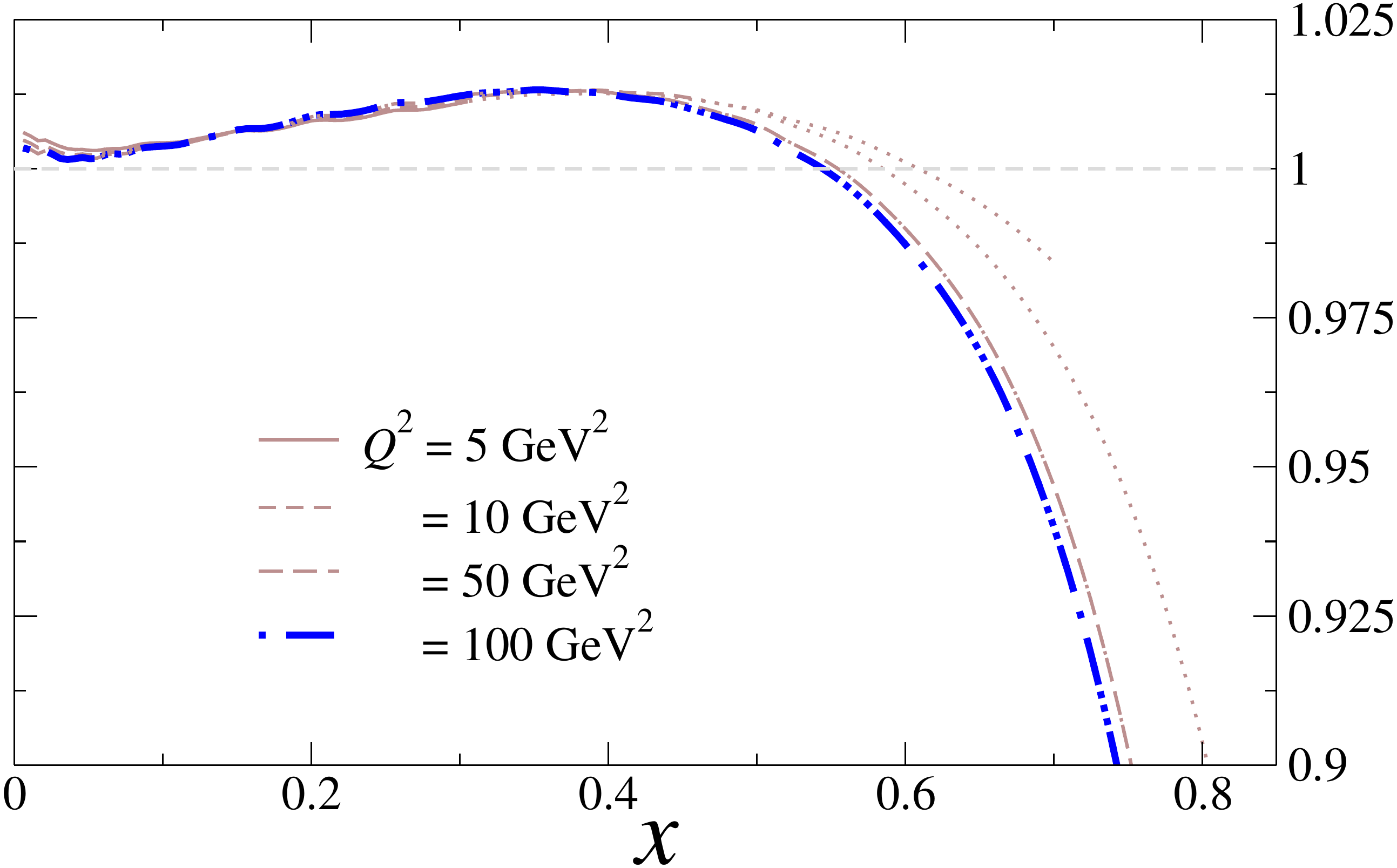}
\caption{
    We plot the nuclear correction ratio, $F^N_2/F^d_2$, calculated using the central CJ15 fit results for several selections of the $Q^2$ scale. Each of the four panels above
    highlights a given value of $Q^2$, while graying out the curves for other scales in order to retain visual information on the scale dependence of the correction factor at large $x$.
    In the upper two panels, which focus on lower scales, $Q^2=5, 10$ GeV$^2$, the dotted lines indicate the range of $x$ that is only accessible to CJ ($W^2\!>\!3$ GeV$^2$) but not CT ($W^2\!>\!12.25$ GeV$^2$), due to the more conservative cut of the latter.
}
\label{fig:deut-SF}
\end{figure*}

In the CJ framework, these corrections are treated as nuclear wave-function effects, and the deuteron parton distributions $f^d$ are calculated as a convolution of the bound nucleon's parton distributions, $\widetilde f^N$, with a suitable nucleonic ``smearing function,'' ${\cal S}^{N/d}$:
\begin{align}
     f^d(x,Q^2) &= \int \frac{dz}{z}\, \int dp^2_N \\
     &\hspace{2cm}{\cal S}^{N/d}(z,p^2_N) \, \widetilde{f}^{N}(x/z,p^2_N,Q^2)\ . \nonumber
\end{align}
Here, $z$ represents the momentum fraction of the (isoscalar) nucleon within the deuteron, defined as $z \equiv (M_d/M_N) (p_N \cdot q / p_d \cdot q)$; $p_{d,N}$ are the deuteron and nucleon four-momenta; and $M_{d,N}$ are their respective on-shell masses.
This representation is founded on the so-called Weak Binding Approximation (WBA) to the calculation of nuclear structure functions~\cite{Kulagin:2004ie,Kulagin:1989mu}, where the ${\cal S}^{N/d}$ smearing function is calculable based on an assumed nuclear potential; as in Ref.~\cite{Accardi:2016qay}, we assume the AV18 potential. 
Since $p_N$ is generically off-shell for a bound nucleon, but typically by only a small amount, one can further expand the bound-nucleon PDF, $\widetilde f^N$, in powers of its off-shellness, $\omega = (p_N^2-M_N^2)/M_N^2$, as
\begin{align}
    \widetilde{f}^{q/N}(y,p^2_N,Q^2)
    &= f^N(y,Q^2) \\
    &+ \frac{p_N^2-M_N^2}{M_N^2} \, \delta f^N(y,Q^2) + O(\omega^2) \ . \nonumber
\end{align}
The first term, corresponding to $p_N^2=M_N^2$, gives the PDF of the free, on-shell nucleon. In the second term, the $\mathcal{O}(\omega)$ coefficient (also known as ``off-shell function'') can be phenomenologically parametrized and determined in a global fit from the interplay of data involving deuterium targets and information involving free-nucleon-based observables like $W$ boson production at the Tevatron, the Relativistic Heavy Ion Collider (RHIC) or the LHC. Like in Ref.~\cite{Accardi:2016qay}, we assume the flavor-independent 3-parameter shape function
\begin{equation}
    \delta f^N (x) = C (x-x_0)(x-x_1)(1+x_0-x)\ ,
\label{eq:offshell}
\end{equation}
with $x_1$ fixed by requiring the off-shell PDFs to satisfy the quark-number sum rule. Further technical details and a discussion of the fit results can be found in Ref.~\cite{Accardi:2016qay}.

\begin{figure*}[htb]
	\centering
	\includegraphics[width=0.565\textwidth]{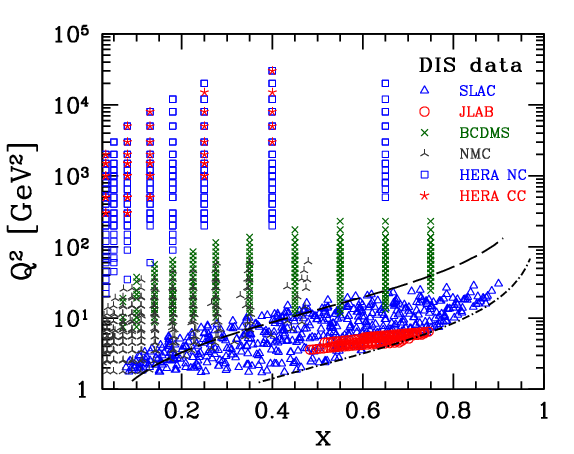} 
\caption{
	Kinematics of the DIS  data included in the fits discussed in this paper. The HERA DIS collider data were taken on proton targets; the fixed-target SLAC, JLab, BCDMS and NMC experiments include both proton and deuterium target data at approximately the same kinematics. The $W^2=12.25$ GeV$^2$ and $W^2=3$ GeV$^2$ cuts adopted, respectively, by the CT and CJ fits are shown by dashed and dot-dashed lines, respectively. The figure is taken from Ref.~\cite{Owens:2012bv}.
}
\label{fig:xQdata}
\end{figure*}

Section~\ref{sec:results} considers three main scenarios for implementing the deuteron corrections (d.c.) discussed above: 
\begin{enumerate}[(i)]
\item
an uncorrected scenario for
which no nuclear effects are included for the deuteron; 
\item
a fixed scenario in which
the nuclear wave-function effects (on- {\it and} off-shell) are frozen to the AV18-informed choice of Ref.~\cite{Accardi:2016qay},
and the off-shellness correction, $\delta f^N(x)$, is set to the CJ15 central fit; and 
\item
a free scenario particular
to CJ, in which the parameters in Eq.~(\ref{eq:offshell}) for the off-shell nucleon are allowed to vary.
\end{enumerate}
The dynamical deuteron corrections are natively implemented in the CJ framework, and the off-shell parameters can be simultaneously fitted with the PDFs. So far, however, the CT code only supports deuteron corrections given in the form of analytic interpolations, such as the one obtained from the correction in \cite{Melnitchouk:1994rv}. To implement the {\it fixed} CJ15 deuteron correction in the CT framework and render it more directly comparable to CJ with respect to its
treatment of deuteron target data, we instead
multiply the experimental DIS deuteron structure function by the $F_2^N/F_2^D$ nucleon-to-deuteron ratio plotted in Fig.~\ref{fig:deut-SF}:
\begin{align}
    F_{2}^N \equiv F_{2,\text{exp}}^d 
    \left( \frac{F_2^N}{F_2^d} \right)_\text{CJ15},
\end{align}
with $F_2^N = F_2^p+F_2^n$. The effective isoscalar combination of proton and neutron structure functions thus defined
can then be directly compared to uncorrected theoretical calculations of the isoscalar deuteron DIS structure function.
On this logic, the CT and CJ fits with a fixed CJ15 correction are placed on similar theoretical footing regarding the
implementation of the deuteron effects, with the main difference being whether the correction is imposed within the theoretical
structure function calculation or in the $F^d_2$ experimental data -- a fact which is immaterial for the sake of evaluating the $\chi^2$-function and allows us to compare the impact of the same fixed correction on the CJ and CT frameworks. 
While a full analysis of the nuclear correction uncertainties is outside the scope of this article, the effect of letting the nuclear off-shell parameters free to vary in the present analysis can be appreciated by comparing the CJ fits in the fixed and free nuclear corrections scenarios.

The size and $x$ dependence of the deuteron corrections, as quantified by the isoscalar nucleon-to-deuteron structure function ratio $F^N_2/F^d_2$, are shown in Fig.~\ref{fig:deut-SF} for several representative choices of $Q^2$. One immediately notices that deuteron corrections depend on the DIS scale and, at large $x$, increase with $Q^2$ toward a fixed point in the $Q^2\! \to\! \infty$ Bjorken limit; as such, deuteron corrections become effectively scale independent for $Q^2 \gtrsim 50$ GeV$^2$. For each plotted value of $Q^2$, the figure also indicates the maximum $x$ values below which data are accepted in the CJ and CT fits according to their $W^2> 3 \mbox{\ and\ } W^2>12.25$ GeV$^2$ kinematic cuts, respectively. For CJ, which extends the analyzed DIS data set to the low-$Q^2$ and large-$x$ values shown in Fig.~\ref{fig:xQdata}, it is imperative to correctly account for the $Q^2$ dependence of the deuteron correction in order to avoid conflicts with the leading-twist logarithmic $Q^2$ evolution  that constrains the fitted gluon distribution in DIS experiments. For CT, with its larger $W^2$ cut, the deuteron corrections are small and nearly scale independent, as seen in Fig.~\ref{fig:deut-SF}, except for the less precise BCDMS deuteron points with $x \gtrsim 0.6$ (see the kinematical map in Fig.~\ref{fig:xQdata}), where some influence from the deuteron correction is expected and indeed quantified in Sec.~\ref{sec:results}.

\subsection{Power-suppressed effects}
\label{sec:wnuc}

Due to their less conservative kinematical restrictions on 
$Q^2$ and $W^2$, the CJ global fits extend into a region for which power-suppressed corrections are non-negligible, as depicted in Fig.~\ref{fig:xQdata}.
On the one hand, dynamical higher-twist corrections of $\mathcal{O}(\Lambda^2/Q^2)$ emerge because of the presence of multi-parton correlations within the soft portion of the factorized DIS
process, for which the first subleading contribution to the twist expansion for unpolarized scattering are matrix elements of twist-$4$ operators \cite{Ellis:1982cd,Qiu:1990xxa}. As in CJ, these are often determined phenomenologically using forms like $F_2(x,Q^2) = F_2^\text{LT}(x,Q^2) \left[ 1 + C(x)\big/Q^2 \right]$, where $F_2^\text{LT}$ represents the leading-twist structure function, and a fitted coefficient, $C(x)=\alpha x^\beta(1+\gamma x)$, parametrizes the power-suppressed twist-4 corrections.
On the other hand, target-mass corrections of $\mathcal{O}(M^2_N/Q^2)$ are due to the non-negligible mass, $M_N$, of the struck nucleon, and are implemented via
the operator product expansion of Georgi and Politzer \cite{Georgi:1976ve,DeRujula:1976baf} or related prescriptions, as extensively reviewed in \cite{Schienbein:2007gr,Brady:2011uy}. 
Both corrections are natively implemented in the CJ framework.

In contrast, CT imposes more restrictive kinematical cuts in $W^2$, such that the standard CT data sets lie beyond the region for which the finite $\sim\! 1/Q^2$ corrections are significant. In the past CT studies it mattered little whether the deuteron correction was included according to a specific model or not included at all (the default choice). While we expect some interplay between the deuteron and power-suppressed effects, we do not systematically isolate the latter and leave their investigation to future studies.

\section{Methods}
\label{sec:methods}

\subsection{Selections of experimental data sets}
\label{sec:data}
The CT18 and CJ15 global data sets consist of both high-energy measurements as well as data
down to the few-GeV region.  Despite their somewhat differing phenomenological emphases  --- with CT generally aimed toward high-energy processes, and CJ toward the low-$Q$ and large-$x$ region probed at
facilities like JLab and SLAC --- there is a substantial overlap with respect to the key experiments they include.  In Table~\ref{tab:exps_and_categories}, we
provide a complete listing of the experiments included in each global analysis. Of particular importance
for interpreting our results in Sec.~\ref{sec:results}, the leftmost column of Table~\ref{tab:exps_and_categories} designates a process-based
category label for each experiment, placing, {\it e.g.}, the BCDMS $F^d_2$ inclusive structure function data in Group 4: DIS Deuterium. We will investigate the agreement between these groups of experiments with the help of $L_2$ sensitivities. In contrast, previous studies \cite{Wang:2018heo,Hobbs:2019gob,Courtoy:2020fex} employing the same technique focused primarily on the individual experiments. 

While the CJ and CT analyses share 10 experiments, Table~\ref{tab:exps_and_categories} shows that the CJ fits include additional DIS data at fixed-target energies from SLAC, HERMES, JLab, and NMC. They also include Tevatron measurements of charge asymmetries reconstructed to the level of $W$ bosons and cross sections for photon plus leading jet production. 
The CT PDF fits more extensively cover collider observables. They include HERA heavy-quark production, an assortment of cross sections and cross section asymmetries in electroweak boson production at the LHC, as well as LHC cross sections for inclusive jet and $t\bar t$ production. The CT fits include CCFR and NuTeV cross sections for 
both inclusive (Group \#{\bf 8}) and semi-inclusive (Group \#{\bf 10}, for opposite-sign dimuon production) charge-current DIS on iron. 
The CT fits,  however, implement only the 
most direct measurements of Tevatron and LHC charge asymmetries in $W \to \ell$ lepton decay, presented as a function of the rapidity and transverse momentum of the charged lepton. They do not include the CDF and D\O~boson-level charge asymmetries fitted by CJ, which directly probe the large-$x$ PDF ratios, while they also involve additional recursive unfolding of the data that utilizes a PDF-dependent calculation to reconstruct the weak boson's rapidity.

\begin{table*}[hp]
\begin{tabular}{c|ll|c|c|c}
Group \#     & Experiment Group     & Experiment Name                                                                                  & CT, $N_\mathit{pt}$ & CJ, $N_\mathit{pt}$   & Ref.                            \\
\hline
{\bf 1}      & $\gamma$+jet         & DØ $\gamma$+jet                                                                                  &                     & 56                    & \cite{Abazov:2008er}            \\
\hline                                                                                                                                                                                
{\bf 2}      & Jets Tevatron        & CDF Run-2 inclusive jet production                                                               & 72                  & 72                    & \cite{Aaltonen:2008eq}          \\
             &                      & DØ Run-2 inclusive jet production                                                                & 110                 & 110                   & \cite{Abazov:2008ae}            \\
\hline                                                                                                                                                                                
{\bf 3}      & DIS Proton           & HERA Run I+II                                                                                    & 1120                & 1185                  & \cite{Abramowicz:2015mha}       \\
             &                      & H1 $F_L$                                                                                         & 9                   &                       & \cite{Collaboration:2010ry}     \\
             &                      & H1 $\sigma_r^b$                                                                                  & 10                  &                       & \cite{Aktas:2004az}             \\
             &                      & Combined HERA $\sigma_r^c$               & 47                  &                       & \cite{Abramowicz:1900rp}        \\
             &                      & JLab proton                                                                                      &                     & 136                   & \cite{Malace:2009kw}            \\
             &                      & SLAC proton                                                                                      &                     & 564                   & \cite{Whitlow:1991uw}           \\
             &                      & HERMES proton                                                                                    &                     & 37                    & \cite{airapetian2011inclusive}  \\
             &                      & NMC $F_2$                                                                                        &                     & 275                   & \cite{Rondio:1997se}                 \\
             &                      & BCDMS $F_2^p$                                                                                    & 337                 & 351                   & \cite{virchaux1992measurement,Benvenuti:1989rh}  \\
\hline                                                                                                                                                                                
{\bf 4}      & DIS Deuterium        & BoNuS $F^n_2/F^d_2$                                                                              &                     & 191                                & \cite{tkachenko2014measurement} \\
             &                      & JLab deuteron                                                                                    &                     & 136                   & \cite{Malace:2009kw}            \\
             &                      & SLAC deuteron                                                                                    &                     & 582                   & \cite{Whitlow:1991uw}           \\
             &                      & HERMES deuteron                                                                                  &                     & 37                    & \cite{airapetian2011inclusive}  \\
             &                      & NMC $F_2^d$/$F_2^p$                                                                              & 123                 & 189                   & \cite{Arneodo:1996qe}             \\
             &                      & BCDMS $F_2^d$                                                                                    & 250                 & 254                   & \cite{Virchaux:1991jc,Benvenuti:1989fm}          \\

\hline                                                                                                                                                                                
{\bf 5}      & WZ Tevatron          & CDF Run-1 lepton $A_{ch}$ , $p_{Tl}>$  25 GeV                                                    & 11                  &                       & \cite{Abe:1996us,Abe:1998rv}               \\
             &                      & CDF Run-2 electron $A_{ch}$ , $p_{Tl}>$  25 GeV                                                  & 11                  & 11                    & \cite{Acosta:2005ud}            \\
             &                      & CDF Run-2 Z rapidity                                                                             & 29                  & 28                    & \cite{Aaltonen:2010zza}         \\
             &                      & DØ Run-2  9.7 $\mathrm{fb}^{-1}$ electron $A_{ch}$ , $p_{Tl}>$  25 GeV                           & 13                  & 13                    & \cite{D0:2014kma}               \\
             &                      & DØ Run-2 muon $A_{ch}$ , $p_{Tl}>$  20 GeV                                                       & 9                   &                       & \cite{Abazov:2007pm}            \\
             &                      & DØ Run-2 $Z$ rapidity                                                                            &    28                 & 28                    & \cite{Abazov:2007jy}    \\
             &                      & CDF $W$ asymmetry                                                                                &                     & 13                    & \cite{Aaltonen:2009ta}          \\
             &                      & DØ $W$ asymmetry                                                                                 &                     & 14                    & \cite{Abazov:2013dsa}           \\
             &                      & DØ muon charge asymmetry                                                                         &                     & 10                    & \cite{Abazov:2013rja}           \\
\hline                                                                                                                                                                                
    {\bf 6}  & WZ LHC               & LHCb 7 TeV 1.0 $\mathrm{fb}^{-1}$ $W/Z$ forward rapidity cross sec.                              & 33                  &                       & \cite{Aaij:2015gna}             \\
             &                      & LHCb 8 TeV 2.0 $\mathrm{fb}^{-1}$ $Z \to e^-e^+$ forward rapidity cross sec.                     & 17                  &                       & \cite{Aaij:2015vua}             \\
             &                      & CMS 8 TeV 18.8 $\mathrm{fb}^{-1}$ muon charge asymmetry $A_{ch}$                                 & 11                  &                       & \cite{Khachatryan:2016pev}      \\
             &                      & LHCb 8 TeV 2.0 $\mathrm{fb}^{-1}$ $W/Z$ cross sec.                                               & 34                  &                       & \cite{Aaij:2015zlq}             \\
             &                      & ATLAS 8 TeV 20.3 $\mathrm{fb}^{-1}$, $Z$ $p_T$ cross sec.                                        & 27                  &                       & \cite{Aad:2015auj}              \\
             &                      & CMS 7 TeV 4.7 $\mathrm{fb}^{-1}$ muon $A_{ch}$ , $p_{Tl}>$  35 GeV                               & 11                  &                       & \cite{Chatrchyan:2013mza}       \\
             &                      & CMS 7 TeV 840 $\mathrm{pb}^{-1}$ electron $A_{ch}$ , $p_{Tl}>$  35 GeV                           & 11                  &                       & \cite{Chatrchyan:2012xt}        \\
             &                      & ATLAS 7 TeV 35 $\mathrm{pb}^{-1}$ $W/Z$ cross sec., $A_{ch}$                                     & 41                  &                       & \cite{Aad:2011dm}           \\
\hline                                                                                                                                                                                
{\bf 7}      & Drell-Yan            & E605   p+Cu                                                                                      & 119                 &                       & \cite{Moreno:1990sf}            \\
             &                      & E866,   $\sigma^{pd}/\sigma^{pp}$                                                                & 15                  &                       & \cite{Towell:2001nh}            \\
             &                      & E866,   $\sigma^{pp}$                                                                            & 184                 & 121                   & \cite{E866lanl,Webb:2003ps}                 \\
             &                      & E866,    $\sigma^{pd}$                                                                           &                     & 129                   & \cite{E866lanl,Webb:2003ps}                 \\
\hline                                                                                                                                                                                
{\bf 8}      & $\nu$-A incl.~DIS    & CDHSW $F_2$                                                                                      & 85                  &                       & \cite{Berge:1989hr}     \\
             &                      & CDHSW $xF_3$                                                                                     & 96                  &                       & \cite{Berge:1989hr}     \\
             &                      & CCFR $F_2$                                                                                       & 69                  &                       & \cite{Yang:2000ju}              \\
             &                      & CCFR $xF_3$                                                                                      & 86                  &                       & \cite{Seligman:1997mc}          \\
\hline                                                                                                                                                                                
{\bf 9}      & $t\bar{t}$ production     & CMS 8 TeV 19.7 $\mathrm{fb}^{-1}$ , $t\bar{t}$ norm.~top $p_T$ and $y$ cross sec.                & 16                  &                       & \cite{Sirunyan:2017azo}         \\
             &                      & ATLAS 8 TeV 20.3 $\mathrm{fb}^{-1}$, $t\bar{t}$ $p^t_T$, $m_{t\bar{t}} $ abs.~spectrum           & 15                  &                       & \cite{Aad:2015mbv}              \\
\hline                                                                                                                                                                                
{\bf 10}     & $\nu$-A dimuon SIDIS & NuTeV $\nu \mu \mu$ SIDIS                                                                        & 38                  &                       & \cite{osti_879078}              \\
             &                      & NuTeV $\nu \bar{\mu} \mu$ SIDIS                                                                  & 33                  &                       & \cite{osti_879078}              \\
             &                      & CCFR  $\nu \mu \mu$ SIDIS                                                                        & 40                  &                       & \cite{Goncharov:2001qe}         \\
             &                      & CCFR $\nu \bar{\mu} \mu$ SIDIS                                                                   & 38                  &                       & \cite{Goncharov:2001qe}         \\
\hline                                                                                                                                                                                
{\bf 11}     & Jets LHC             & CMS 7 TeV 5 $\mathrm{fb}^{-1}$, single incl.~jet cross sec., R = 0.7                             & 158                 &                       & \cite{Chatrchyan:2014gia}          \\
             &                      & ATLAS 7 TeV 4.5 $\mathrm{fb}^{-1}$, single incl.~jet cross sec., R = 0.6                         & 140                 &                       & \cite{Aad:2014vwa}                 \\
             &                      & CMS 8 TeV 19.7 $\mathrm{fb}^{-1}$, single incl.~jet cross sec., R = 0.7                          & 185                 &                       & \cite{Khachatryan:2016mlc}             
\end{tabular}
\caption{A comprehensive listing of experiments included within the CT and CJ frameworks for this study,
grouped according to the experimental process they represent.  For each experiment, we give the number of experimental
data points, $N_\mathit{pt}$, included into the CT and CJ fits presented in this study, with blank entries indicating
that the given experiment is omitted from the respective global fit.
}
\label{tab:exps_and_categories}
\end{table*}

\subsection{Modifications in the fitting methodologies}
\label{sec:mods}
For the study presented in this article, we modified some default settings of the CJ15 and CT18 fits, fully described respectively in Ref.~\cite{Accardi:2016qay} and \cite{Hou:2019efy}, to place the two fitting frameworks on a common footing and  isolate the impact of various assumed treatments of the deuteron structure. 
\begin{enumerate}

    \item We match perturbative orders between the two fits at NLO in $\alpha_s$. In practice,
    this means that in the CT fits, performed by default at NNLO,  we instead compute the hard cross sections, perturbative PDF evolution, and running of $\alpha_s$ at $\mathcal{O}(\alpha_s)$ accuracy to agree with the default NLO settings used in CJ.
    
    \item We perform supplementary fits by excluding some data sets that appear in one fit only. While both CJ and CT fits include Tevatron lepton charge asymmetry measurements presented as a function of the charged lepton's rapidity, the CJ fit also includes the fixed-target low $W^2$ and $Q^2$ DIS data from SLAC \cite{Whitlow:1991uw} and JLab \cite{Malace:2009kw}, as well as the CDF \cite{Aaltonen:2009ta} and DØ \cite{Abazov:2013dsa} $W$ boson charge asymmetry with reconstructed weak boson kinematics. On the other hand, CT makes use of neutrino-initiated DIS data sets on heavy nuclear targets (both inclusive and semi-inclusive DIS [SIDIS] di-muon  production in $\nu$-A scattering). 
    In CT, data on heavy-nuclear targets are fitted at the isoscalar level after being corrected in the fit using a phenomenological parametrization of the $F^A(x,Q^2)/F^d(x,Q^2)$ ratio from Ref.~\cite{Rondio:1993mf}.
    To isolate the impact of these extra experiments, we performed CJ fits without the $W$ asymmetry and SLAC DIS data sets, and CT fits without the inclusive $\nu$-A DIS data.
    
    \item As in the original CJ and CT publications, we estimate the final PDF uncertainties using the Hessian method \cite{Stump:2001gu}, but in this paper we fix the tolerance to be $T^2\! =\! 10$ for both global analyses, in between the nominal $T^2 = 2.71$ in the CJ15 fit and the $T^2=37$ value (at the 68\% probability level) used in the CT18 fits. Furthermore, we do not include the additional ``Tier-2" tolerance contribution \cite{Hou:2019efy,Lai:2010nw} that is applied in the CT18 fits to prevent the error PDFs from running into strong disagreements with individual experiments, but content ourselves with the ``Tier-1" tolerance as defined in \cite{Pumplin:2002vw}. 
    
\end{enumerate}
Regarding the lattermost point, in the CJ analysis it is additionally necessary to implement a numerical prescription at the level of individual eigenvector directions of the diagonalized Hessian
matrix to guarantee $\Delta \chi^2 = T^2$ to the needed accuracy and to ensure the validity of the analysis methods utilized in this article.  These technical details are reviewed in the appendix.

\subsection{The $L_2$ sensitivity technique}
\label{sec:L2}

In the next two sections, we will investigate the impact of the DIS deuteron data on the PDFs with the help of the ``$L_2$ sensitivity'' \cite{Hobbs:2019gob}. The method is easy to use and has already been applied to clarify the sensitivity of the global data sets to the
CT18 NNLO PDFs \cite{Hou:2019efy}, LHC parton luminosities \cite{Amoroso:2020lgh} (Sec.~II.2), and effective exponents of the high-$x$ PDF falloff~\cite{Courtoy:2020fex}. Here we give a quick summary of the $L_2$ technique and refer the interested reader to Ref.~\cite{Hobbs:2019gob} for additional details.

\begin{figure*}[htb]
	\centering
	\includegraphics[width=0.48\textwidth]{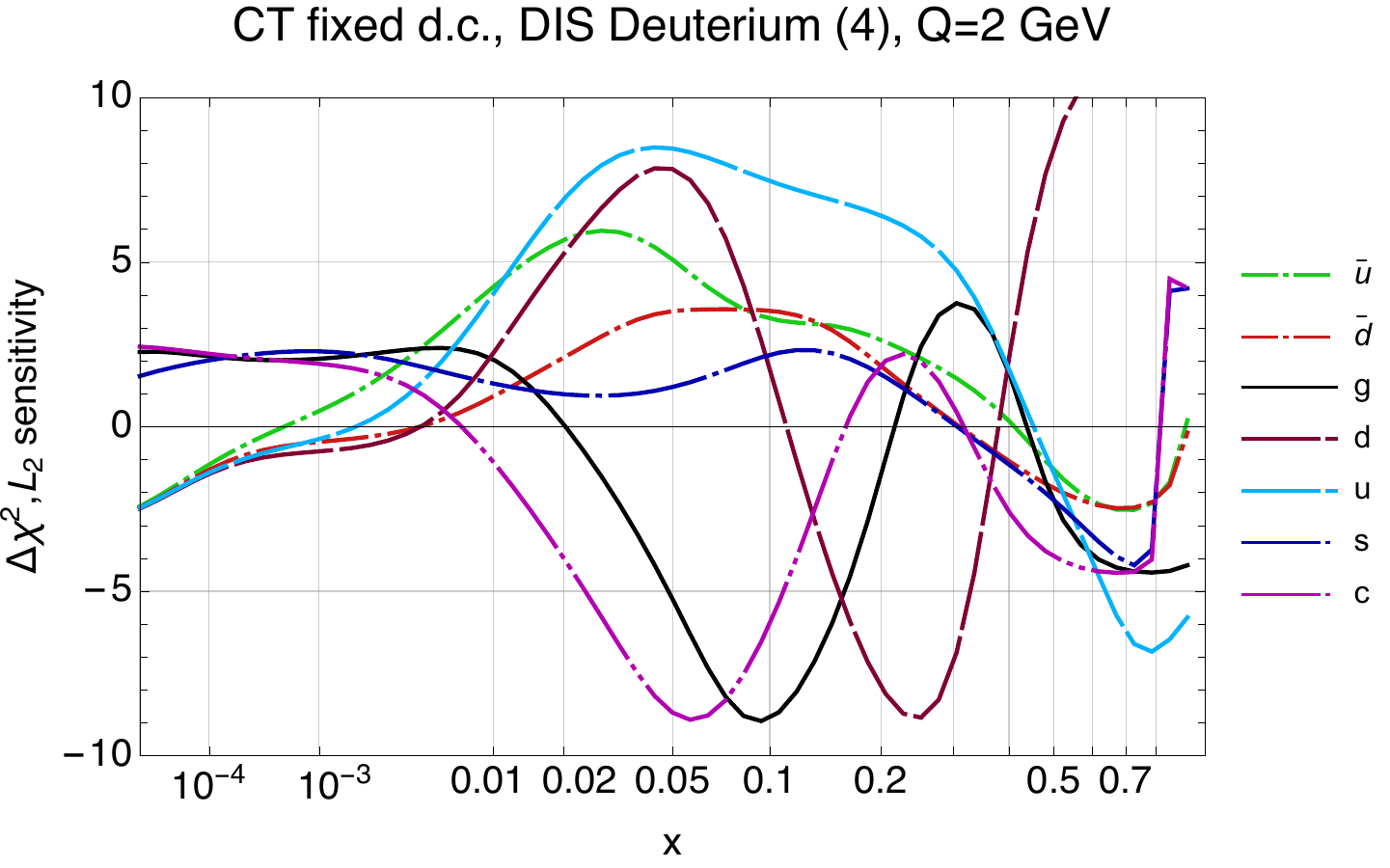} \ \ \
	\includegraphics[width=0.48\textwidth]{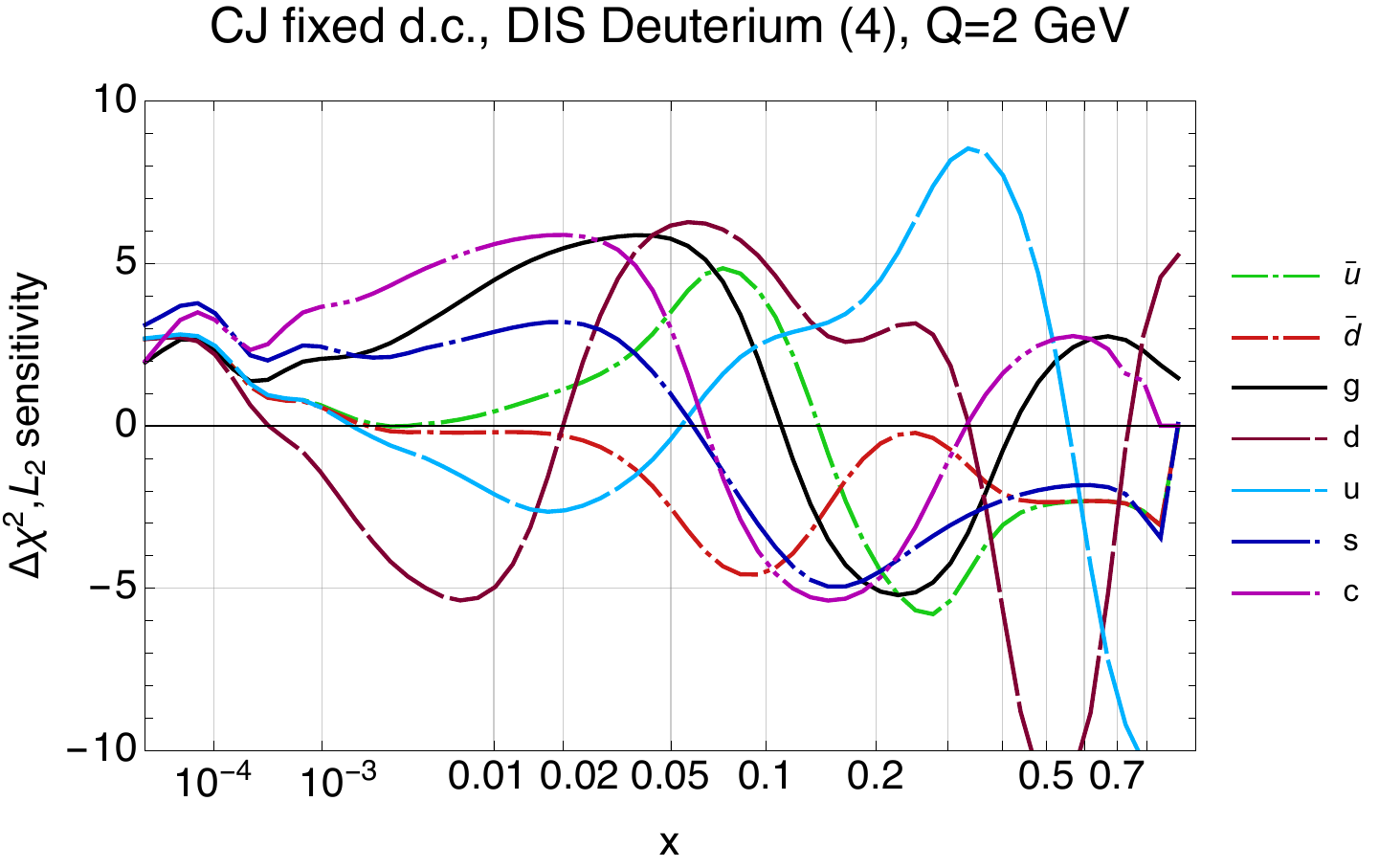} 
\caption{
    A comparison of the PDF pulls of the deuterium DIS data computed according to the $L_2$
    method at 2 GeV for the \texttt{CT fixed d.c.}~(left) and \texttt{CJ fixed d.c.}~(right) 
    fits; both cases are for the scenario with fixed deuteron corrections.
    Here and elsewhere, we place on the vertical axis a self-explanatory label of $\Delta \chi^2$ as we plot
    the $L_2$ sensitivity which approximates this quantity.
    }
\label{fig:deuterium-pulls}
\end{figure*}

The $L_2$ sensitivity provides a fast approximation to the information contained in the
LM scans typified by Fig.~\ref{fig:qval_LM}.  It does so by quantifying the extent to which variations
in the fitted PDFs drive the shifts in the log-likelihood $\chi^2_E$ for each experiment $E$. 
For example, the $L_2$ plots discussed below supply the change in an experiment's value
of $\chi^2$ given a defined increase in the value of a PDF of interest, as a function of $x$ for fixed 
$Q$. That being the case, if two experiments are in relative tension with respect to the
behavior of a given PDF --- say, one favors a larger PDF value in a given $x$ range, the other favors a smaller value --- an increase in the PDF would result in a negative change,
$\Delta \chi^2_{E_1}\! <\! 0$, in the $\chi^2$ for the first experiment, and a positive shift,
$\Delta \chi^2_{E_2}\! >\! 0$, for the second experiment. As shall be seen below, these competing pulls are
visualized by the $L_2$ method as opposing deviations from zero in the negative/positive directions.
In this fashion, the $L_2$ sensitivity can encapsulate, on a flavor-by-flavor basis, the patterns of pulls exerted on the PDFs (often called ``PDF pulls" below for short) by various
experiments or groups of experiments as a function of $x$ and $Q$.
In fact, the method is not limited to the influence of data sets on the PDFs, but can be extended to any observable that can be calculated starting from those.

More formally, the $L_2$ sensitivity yields an approximation of the
shift $\Delta \chi^2_E$ in the value of the log-likelihood for experiment $E$ caused by an upward 1$\sigma$ variation of the chosen PDF or PDF-dependent theoretical prediction. We evaluate the $L_2$ sensitivity in the Hessian approximation as 
\begin{align}
	S_{f,L2}(E)\, &\equiv\, \vec{\nabla} \chi^2_E \cdot \frac{\vec{\nabla} f}{|\vec{\nabla} f|} \nonumber \\
	&
	= \Delta \chi^2_E \, \cos{\varphi(f,\chi^2_E)}\ ,
\label{eq:L2}
\end{align}
where $\cos{\varphi(f,\chi^2_E)}$ represents the cosine of the correlation angle between a PDF of flavor $f$ (or, indeed, 
any PDF derived quantity) and the experimental $\chi^2_E$, evaluated over the
$2N$ Hessian eigenvector sets.
In the Hessian formalism adopted 
both by the CT and CJ frameworks to estimate PDF uncertainties,
this correlation cosine can be computed as indicated in Ref.~\cite{Wang:2018heo}:
\begin{align}
&\cos{\varphi(f,\chi^2_E)} \nonumber \\
& =\frac{1}{4\Delta f \Delta \chi^2_E}\sum_{j=1}^{N}(f_{j}^{+}-f_{j}^{-})([\chi^2_E]_{j}^{+}-[\chi^2_E]_{j}^{-})\ , 
\label{eq:corr-def}
\end{align}
where
\begin{equation}
\Delta f=\left\vert \vec{\nabla}f\right\vert =\frac{1}{2}\sqrt{\sum_{j=1}^{N}\left(f_{j}^{+}-f_{j}^{-}\right)^{2}}\ ,
\label{eq:unc-def}
\end{equation}
and ``$\pm$'' denote the PDF variations along the positive and negative direction of the $j$-th Hessian eigenvector.

\begin{table*}[th]
\center\scriptsize
\begin{tabular}{c|l||c|c|c||c|c|c}
Group \#     & Experiment               & CJ no d.c. & CJ fixed d.c. & CJ no-W\_slac  & CT no d.c. & CT fixed d.c. & CT no nu-A  \\ 
\hline\hline
{\bf 1}      & $\gamma$+jet              & 62/56  & 61/56 & 60/56 & -- & -- & -- \\
\hline
{\bf 2}      & Jets Tevatron             & 37/182 & 36/182  & 36/182  & 225/182 & 225/182 & 229/182\\
\hline
{\bf 3}      & DIS proton                & 3007/2548 & 2973/2548 & 2330/1848 & 1818/1523 & 1812/1523 & 1806/1523\\
\hline
{\bf 4}      & DIS deuteron              & 1363/1389 & 1214/1389 & 704/671 & 401/373 & 387/373 & 381/373 \\
             & \bf SLAC deuteron (only) &\bf 507/582 & \bf 376/582 & -- & -- & -- & -- \\
\hline
{\bf 5}      & WZ Tevatron               & 193/117 & 133/117  & 99/90  & 113/101 & 129/101 & 137/101\\
             & \bf CDF $W$ asym. (only)  & \bf14/13 & \bf 17/13 & \bf --  & \bf --  & \bf --  & \bf -- \\
             & \bf D\O\ $W$ asym. (only) & \bf82/14 & \bf 12/14 & \bf --  & \bf --  & \bf --  & \bf -- \\
\hline
{\bf 6}      & WZ LHC                   & -- & -- & --  & 267/185 & 347/185 & 315/185\\
\hline
{\bf 7}      & Drell-Yan                 & 302/250  & 284/250  & 272/250  & 454/318 & 348/318 & 344/318 \\
\hline
{\bf 8}      & \bf $\nu$-A incl.~DIS    &\bf -- &\bf -- &\bf --  & \bf 269/336 & \bf279/336 & \bf  -- \\
\hline
{\bf 9}      & $t\bar t$ production     & -- & -- & --  & 44/31 & 45/31 & 46/31\\
\hline
{\bf 10}     & $\nu$-A dimuon SIDIS     & -- & -- & --  & 103/149 & 104/149 & 103/149\\
\hline
{\bf 11}     & Jets LHC                 & -- & -- & --  & 594/483 & 595/483  & 598/483\\
\hline\hline
             & \bf TOTAL                 & \bf 4963/4542 & \bf4699/4542 & \bf3501/3097 & \bf 4289/3681  & \bf 4271/3681 & \bf3959/3345 \\
\hline\hline
\end{tabular}
\caption{For each fit, we report the total $\chi^2$ per point, $\chi^2/N_\mathit{pt}$, as well as its breakdown according to the experiment categories listed in Table~\ref{tab:exps_and_categories}. The data sets removed in the \texttt{CJ no-W\_slac} fit are singled out and {\bf emboldened} in their respective categories.
The $\nu$-A inclusive data excluded in the \texttt{CT no nu-A} fit are also {\bf emboldened}.
We note that the $\chi^2/N_\mathit{pt}$ values for \texttt{CJ free d.c.}~are the same as those reported above for
\texttt{CJ fixed d.c.}~given that the central fit of \texttt{CJ fixed d.c.} corresponds to the best fit obtained
when freely varying all parameters in \texttt{CJ free d.c.}
 }
\label{tab:chi2_summary_AA}
\end{table*}

When $L_2$ sensitivities are summed over all experiments, the resulting sum should be close to zero by construction, assuming deviations from a symmetric Gaussian probability distribution are negligible (see~\ref{app:directions} for further discussion). 
As an example, Fig.~\ref{fig:deuterium-pulls} shows the combined $L_2$ sensitivities, $S_{f,L2}(E)$, of the experiments in the DIS deuteron group (\#\textbf{4} in Table~\ref{tab:exps_and_categories}) to each parton flavor separately, calculated according to Eq.~(\ref{eq:L2}) for the CT and CJ \texttt{fixed d.c.}~fits, where the deuteron corrections were fixed to the central value determined in the CJ15 analysis. These figures can be interpreted as the statistical pulls at fixed $Q=2$ GeV from this group of experiments on each PDF flavor, $f(x,Q)$, at the $x$ values specified on the horizontal axis.
One can observe quite large deviations from zero, with  $S_{f,L2}$ values nearly reaching $\pm 10$ units in some regions of $x$. This non-negligible pull by the deuteron DIS experiments is ultimately offset by contributions from other groups of experiments to obtain a zero result (within about one unit of $\chi^2$) when summing over all of these.
It is therefore interesting to investigate which experimental groups pull significantly against the DIS deuteron data sets, as we do below in Sec.~\ref{sec:d/u_pulls} and \ref{sec:gluon_pulls}.
Here and elsewhere, we compute $L_2$ sensitivities at a default scale of $Q=2$ GeV as this is close to the initial scale, 1.3 GeV, for DGLAP evolution, as well as to the $Q$ values accessed in DIS experiments and typical scales used to present predictions for PDFs in nonperturbative QCD models and lattice QCD \cite{Hobbs:2019gob}. We typically do not observe pronounced differences in the $L_2$ sensitivity plots for distinct scales, $Q_{1,2}$, provided $|Q_2 - Q_1| \lesssim \mathcal{O}(100\,\mathrm{GeV})$. A set of companion plots generated at a higher scale, $Q=100$ GeV, may be viewed at Ref.~\cite{CT18website}.

Fig.~\ref{fig:deuterium-pulls} contains a substantial amount of information. For example, looking at the left panel for \texttt{CT fixed d.c.}, the {\bf negative} $S_{f,L2}$ for the $d$ quark at $x \approx 0.25$ indicates that the deuteron DIS data prefer a {\bf higher} $d$ quark at $x\approx 0.25$ than 
the nominal $d$-quark PDF in the full fit. Similarly, the {\bf positive} $S_{f,L2}$ for the $u$ quark at the same $x$ indicates a preference for a relatively lower $u$-quark PDF. In totality, the deuteron DIS data prefer a higher $d/u$ at $x=0.1\!-\!0.4$ than that obtained in the full CT fit. (This preference for an enhanced $d/u$ in this $x$ interval is further confirmed in Fig.~\ref{fig:du_SfL2} of Sec.~\ref{sec:d/u_pulls}.) 

From the right panel of Fig.~\ref{fig:deuterium-pulls}, we also read that the deuteron DIS data in the \texttt{CJ fixed d.c.}~fit prefer an enhanced $d/u$ over a slightly higher interval $x=0.3\!-\!0.7$.
Regarding other flavors and $x$ ranges, in the left panel of \texttt{CT fixed d.c.}~we observe a significant preference of the deuteron DIS group for lower $u$- and $d$-quark PDFs at $x=0.01\!-\!0.1$, in the region relevant for LHC $W$-boson production.  One also notices a preference for a larger gluon PDF in the interval, $x=0.02\!-\!0.1$, relevant for Higgs-boson production at the LHC, and, at slightly lower $x$, for a larger perturbative charm-quark PDF, which is radiatively generated from the gluon. 

Finally, we remark that the Hessian sensitivity is most effective in identifying the top 5-10 experiments or groups of experiments sensitive to variations in the chosen PDF, as has been verified by comparing the rankings obtained from Hessian sensitivities and LM scans~\cite{Wang:2018heo,Hobbs:2019gob}. However, especially for subleading experiments, detailed rankings depend on the chosen definition of the sensitivity indicator and deviations from the simple linear approximation that we assumed when using symmetric finite derivatives in $S_{f,L2}(E)$ as in Eqs.~(\ref{eq:corr-def}) and (\ref{eq:unc-def}).

\begin{table*}[th]
\center\scriptsize
\begin{tabular}{l||c||c||c|c|c||c|c|c}
Experiment               &  $N_\mathit{pt}$   & no d.c.~NLO    & no d.c.~NNLO       &  no d.c.~NNLO-X        & fixed d.c.~NLO       & $\delta^\mathrm{NNLO}$  &  $\delta^\mathrm{NNLO-X}$    & $\delta^\mathrm{fixed\, d.c.}$    \\  
\hline\hline                                                                                                                                                                                
$\gamma$+jet             &  0                 & --             & --                 &  --                   & --                   &  --                     &  --                         &  --                               \\
\hline                                                                                                                                                                                      
Jets Tevatron            &  182               & 225            & 236                &  233                  & 225                  &  11                     &  8                          &  0                                \\
\hline                                                                                                                                                                                      
DIS proton               &  1523              & 1818           & 1865               &  1839                 & 1812                 &  47                     &  21                         &  -6                               \\
\hline                                                                                                                                                                                      
DIS deuteron             &  373               & 401            & 406                &  405                  & 387                  &  5                      &  4                          &  -14                              \\
\hline                                                                                                                                                                                      
WZ Tevatron              &  101               & 113            & 120                &  110                  & 129                  &  7                      &  -3                         &  16                               \\
\hline                                                                                                                                                                                      
WZ LHC                   &  185               & 267            & 259                &  185                  & 347                  &  -8                     &  -82                        &  80                               \\
\hline                                                                                                                                                                                      
Drell-Yan                &  318               & 454            & 364                &  432                  & 348                  &  -90                    &  -22                        &  -106                             \\
\hline                                                                                                                                                                                      
$\nu$-A incl.~DIS        &  336               & 269            & 285                &  260                  & 279                  &  16                     &  -9                         &  10                               \\
\hline                                                                                                                                                                                      
$t\bar{t}$ production    &  31                & 44             & 28                 &  29                   & 45                   &  -16                    &  -15                        &  1                                \\
\hline                                                                                                                                                                                      
$\nu$-A dimuon SIDIS     &  149               & 103            & 107                &  111                  & 104                  &  4                      &  8                          &  1                                \\
\hline                                                                                                                                                                                      
Jets LHC                 &  483               & 594            & 608                &  603                  & 595                  &  14                     &  9                          &  1                                \\
\hline\hline                                                                                                                                                                                
\bf TOTAL                &  \bf 3681          & \bf 4289       &\bf 4277            &  \bf 4208             & \bf 4271             &  \bf -12                &  \bf-81                     &  \bf -18                          \\
\hline\hline
\end{tabular}
\caption{Complementary to Table~\ref{tab:chi2_summary_AA}, we compare the values of $\chi^2$ obtained by CT in the \texttt{no d.c.}~fit at NLO as well as at NNLO
with both default CT18 settings and those of the alternative, CT18X global fit (first
three columns, respectively) with CT \texttt{fixed d.c.}~at NLO. The corresponding shifts away from the $\chi^2$ obtained under CT \texttt{no d.c.}~NLO for each
experimental group, $\delta^\mathrm{NNLO}$, $\delta^\mathrm{NNLO-X}$, and $\delta^\mathrm{fixed\, d.c.}$, are given in the lattermost columns.
}
\label{tab:chi2_summary_CT_small}
\end{table*}


\section{Comparison of deuteron data impact in the CJ15 and CT18 fits}
\label{sec:results}

In accordance with the preceding discussion in Sec.~\ref{sec:low} and \ref{sec:methods}, our analysis will be based on a series of fits named according to the following convention:
\begin{enumerate}
    \item Fits {\it without} deuteron corrections: \texttt{CT no d.c.}, \texttt{CJ no d.c.};
    \item Fits with the {\it fixed} CJ15 correction: \texttt{CT fixed d.c.},  \texttt{CJ fixed d.c.};
    \item A CJ fit in which the off-shellness correction is freely varied: \texttt{CJ free d.c.}; 
    \item Fits with the fixed CJ15 deuteron correction and variations in the fitted data sets: \texttt{CT no nu-A} (removing inclusive $\nu$-A DIS experiments from CT),
          and \texttt{CJ no-W\_slac} (removing the CDF [131] and DØ [132] $W$ boson asymmetry data and SLAC DIS [proton and deuteron] sets from CJ).  
\end{enumerate}
In each enumerated case, the CT and CJ fits are methodologically comparable to each other. The differences between the fits in the first two categories will highlight the nontrivial effects of deuteron corrections. We consider several variations of the \texttt{fixed d.c.}~fits. Firstly, the comparison of the CJ sensitivity plots in categories 2 and 3 demonstrates that freeing the d.c.~parameters in the CJ fit tangibly improves the agreement among all categories of experiments. Secondly, we try to make the CT and CJ fits comparable not only methodologically but also (partially) with respect to data selections. We do this in category 4 by removing the indicated data sets from each fit.

\subsection{Overall agreement of theory and data \label{sec:OverallAgreement}}
We first review the overall quality of these fits by referring to Table~\ref{tab:chi2_summary_AA}, which lists values for the total $\chi^2$ per point ($\chi^2/N_\mathit{pt}$) according to the experimental groups listed in Table~\ref{tab:exps_and_categories}.
The $\chi^2$ values for SLAC DIS, Tevatron $W$-boson asymmetry, and $\nu$-A DIS data sets are shown in Table~\ref{tab:chi2_summary_AA} in separate lines. 

The $\chi^2$ values in Table~\ref{tab:chi2_summary_AA} indicate that the deuteron data agree globally with the published fits of both groups. Deuteron corrections are essential to the CJ analysis, which includes deuteron DIS data at the largest values of $x$ from SLAC as well as the very sensitive D\O\ data on the reconstructed $W$-charge asymmetry \cite{Accardi:2016qay}. Even if the total $\chi^2/N_\mathit{pt}$ may seem only marginally improved, the highlighted data sets require deuteron corrections to reconcile the SLAC DIS deuteron data with the deuteron-independent D\O\ $W$-asymmetry data.

In CT, the inclusion of deuteron corrections also improves the description of the DIS deuteron data (especially the BCDMS $F^d_2$ measurements), producing a
$14$-unit reduction in $\chi^2$ for the $N_\mathit{pt}\!=\!373$ points fitted in Group {\bf \#4}, with an additional $6$-unit reduction once
the inclusive $\nu$-A data are removed. Mainly due to the absence of the SLAC DIS data in CT, this is a smaller relative reduction than that observed for CJ in the left columns of Table~\ref{tab:chi2_summary_AA}.

It is also interesting to note far more substantial shifts in $\chi^2$ within Table~\ref{tab:chi2_summary_AA} among the other CT experimental groups: the introduction of the fixed deuteron correction in CT improves the $\chi^2$ of the DY data (\#\textbf{7}) by more than $100$ units, while 
at the same time, \textit{increasing} the $\chi^2$ of the LHC weak-boson production (\#\textbf{6}) by an opposing $\Delta \chi^2$ change of 80 units. The $\chi^2$ for group 5, ``WZ Tevatron", also increases by 16 units. The inclusive $\nu$-A DIS data set (\#{\textbf{8}}) is fitted well globally, with $\chi^2/N_{pt}=0.80\ (0.83)$ in the \texttt{CT no d.c.} (\texttt{CT fixed d.c.}) fit. In sum, the fixed deuteron correction improves the CT total $\chi^2$ by 18 units for 3670 data points. Since the $\nu$-A DIS data are well-described, removing them from the CT fit actually increases the total $\chi^2/N_{pt}$, but also seems to release some tension with the WZ LHC data, whose $\chi^2$ value decreases by 32 units. While these look like modest changes, overall, we will see next that they do influence some PDF flavors and the local compatibility among select groups of experiments. 
It is instructive to evaluate the shifts in $\chi^2$ discussed above in the context of corresponding variations associated with NNLO effects. For this purpose, we compare in Table~\ref{tab:chi2_summary_CT_small} values of $\chi^2$ obtained with CT
for the default NLO fits explored in this paper, \texttt{CT no d.c.}~and \texttt{CT fixed d.c.}, against two NNLO fits. These latter fits
are the NNLO counterpart of \texttt{CT no d.c.}, which we call \texttt{no d.c.~NNLO}, and an alternative NNLO fit adopting
the modified DIS scale choice of the CT18X NNLO fit discussed in Ref.~\cite{Hou:2019efy}, which we denote
\texttt{no d.c.~NNLO-X}. These additional fits allow us to quantify the impact on $\chi^2$ of the NNLO correction itself, as well
as the effect of perturbative scale variations at NNLO, respectively. In addition, in the trailing columns of Table~\ref{tab:chi2_summary_CT_small} we also show $\chi^2$ differences of each of these fits with respect to our base
\texttt{CT no d.c.}~fit at NLO. For example,
\begin{equation}
    \delta^\mathrm{NNLO} \equiv \chi^2(\mathrm{no\,d.c.\,NNLO}) - \chi^2(\mathrm{no\,d.c.\,NLO})\ .
\end{equation}
These comparisons illustrate that inclusion of the fixed deuteron correction at NLO may impact the theoretical description
of select experimental groups at a level comparable to NNLO corrections and scale variations. This is especially evident for the
DIS deuteron data, for which the 14-unit reduction in $\chi^2$ discussed above for \texttt{CT fixed d.c.}~can be compared to
small 5 and 4-unit increases with \texttt{no d.c.~NNLO} and \texttt{no d.c.~NNLO-X}, respectively. The above-mentioned
$\chi^2$ shift for the Drell-Yan data with \texttt{CT fixed d.c.}~similarly surpasses the corresponding shifts obtained
with the NNLO fits. In contrast, the $t\bar{t}$ data, for instance, are largely insensitive to the deuteron correction at NLO, but
are significantly better-fitted with NNLO corrections.
Ultimately, as future generations of QCD fits pursue higher accuracy at NNLO and beyond, it may therefore also be appropriate
to consider deuteron corrections.

\subsection{Impact of deuteron corrections on the $d/u$ PDF ratio}
\label{sec:d/u_PDF}

As discussed in Sec.~\ref{sec:intro}, experimental information involving deuterium targets, especially measurements of the DIS structure function of the deuteron (Group \#{\bf 4}, ``DIS Deuterium,'' in Table~\ref{tab:exps_and_categories}), has been pivotal for separating the $d$-quark content of the proton from other parton flavors. The leading impact of deuteron corrections thus primarily influences the extracted $d$-PDF at high $x$, where the deuteron most prominently differs from a superposition of a free proton and a free neutron. 
In contrast, the effect of the deuteron corrections on the $u$-type PDFs is comparatively mild, as these are most directly constrained by measurements of the proton's structure function
(Group \#{\bf 3}, ``DIS Proton''). 

To gauge the leading impact of deuteron corrections, in this subsection we now examine the $x$ dependence of the $d/u$ ratio within the CT/CJ frameworks, before proceeding
in Sec.~\ref{sec:weakmix} to an examination of the indirect effects on the lower-$x$ $d_\mathit{val}$ PDF relevant for $\sin^2 \theta_W$ and on the
gluon PDF in Sec.~\ref{sec:gluon_PDF} (the PDF pulls will be considered in Secs.~\ref{sec:d/u_pulls} and \ref{sec:gluon_pulls}).

\begin{figure*}[btp]
	\includegraphics[width=0.47\textwidth]{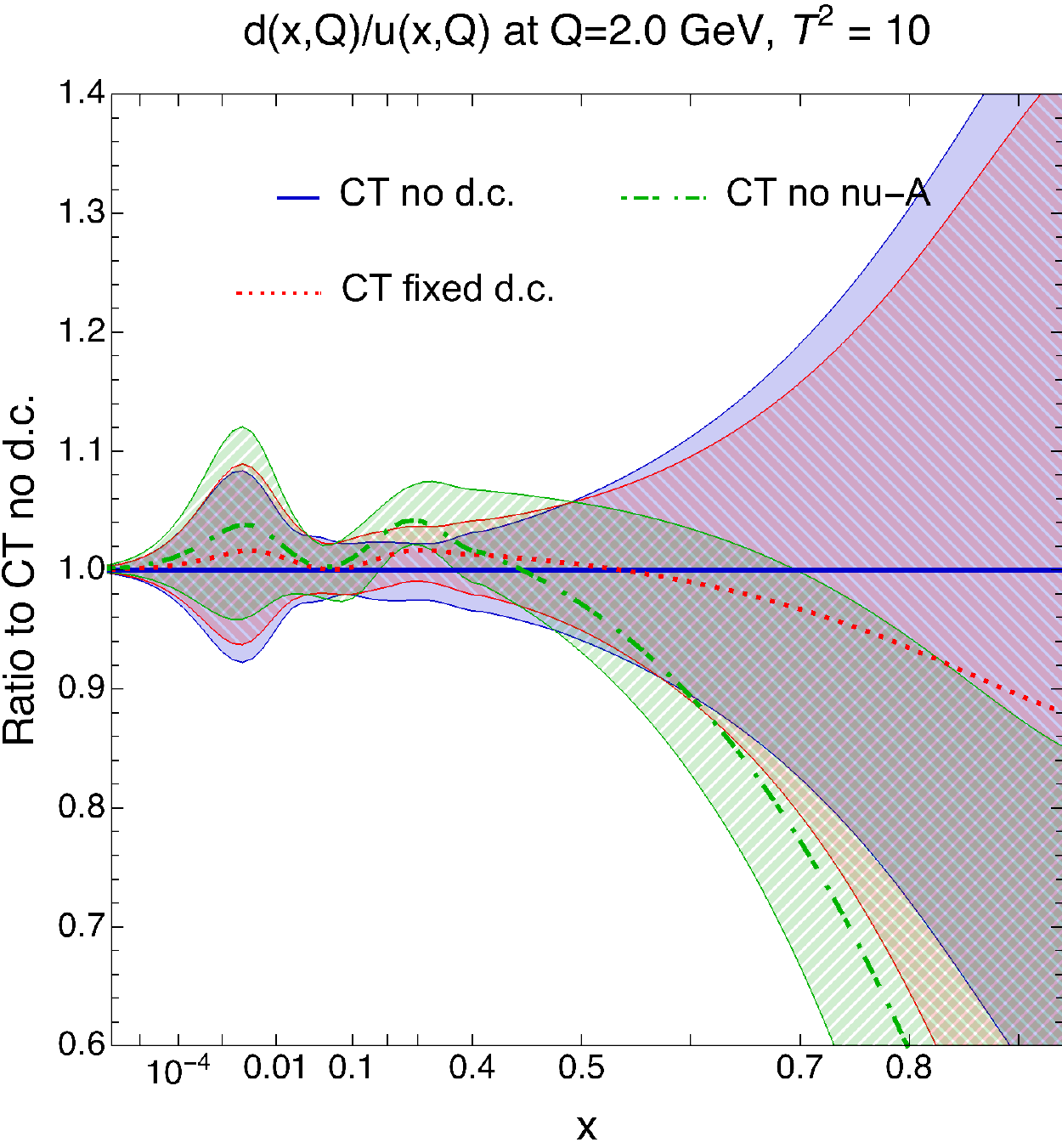} 
	\hfill
	\includegraphics[width=0.47\textwidth]{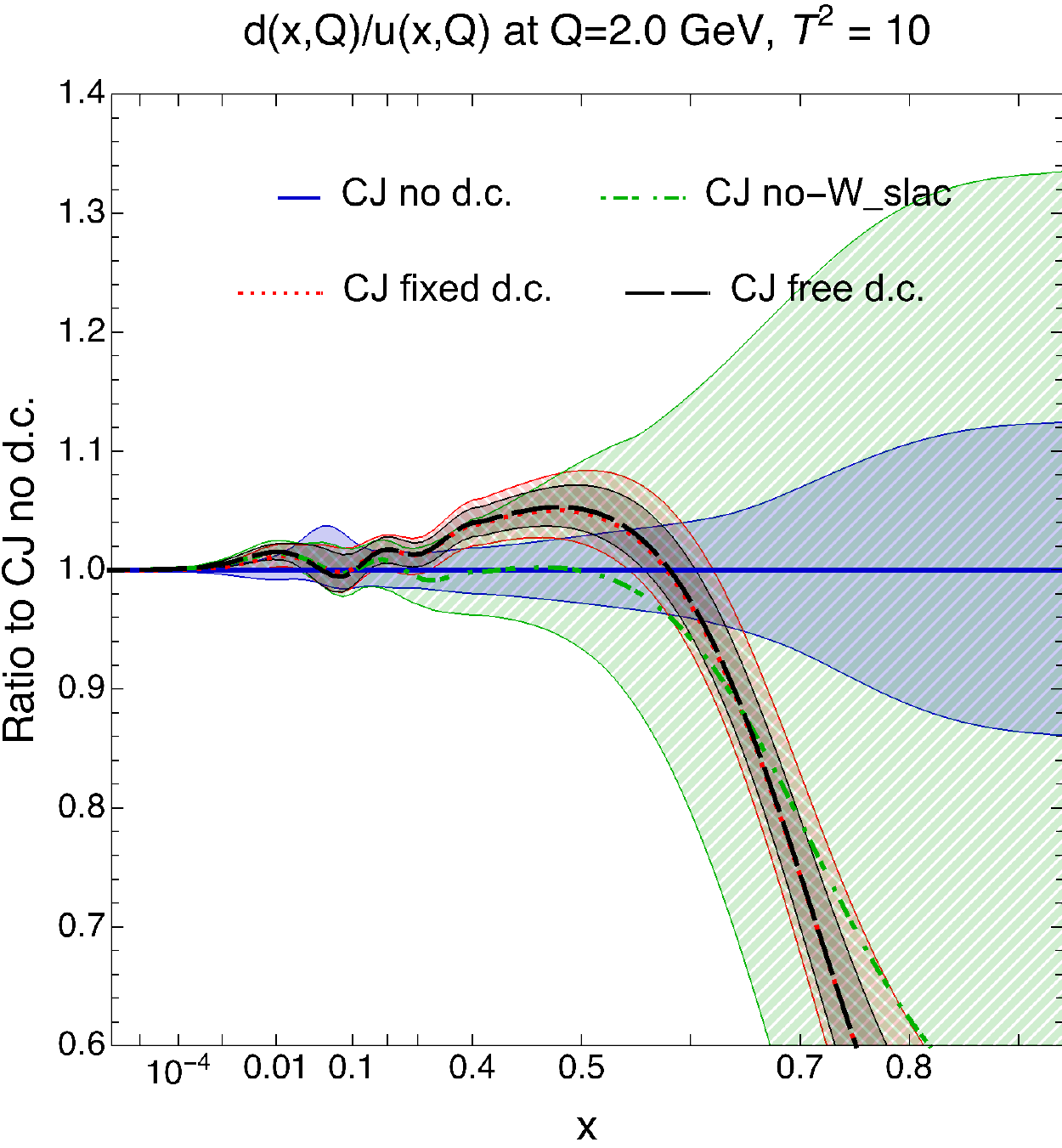} \\
	\includegraphics[width=0.47\textwidth]{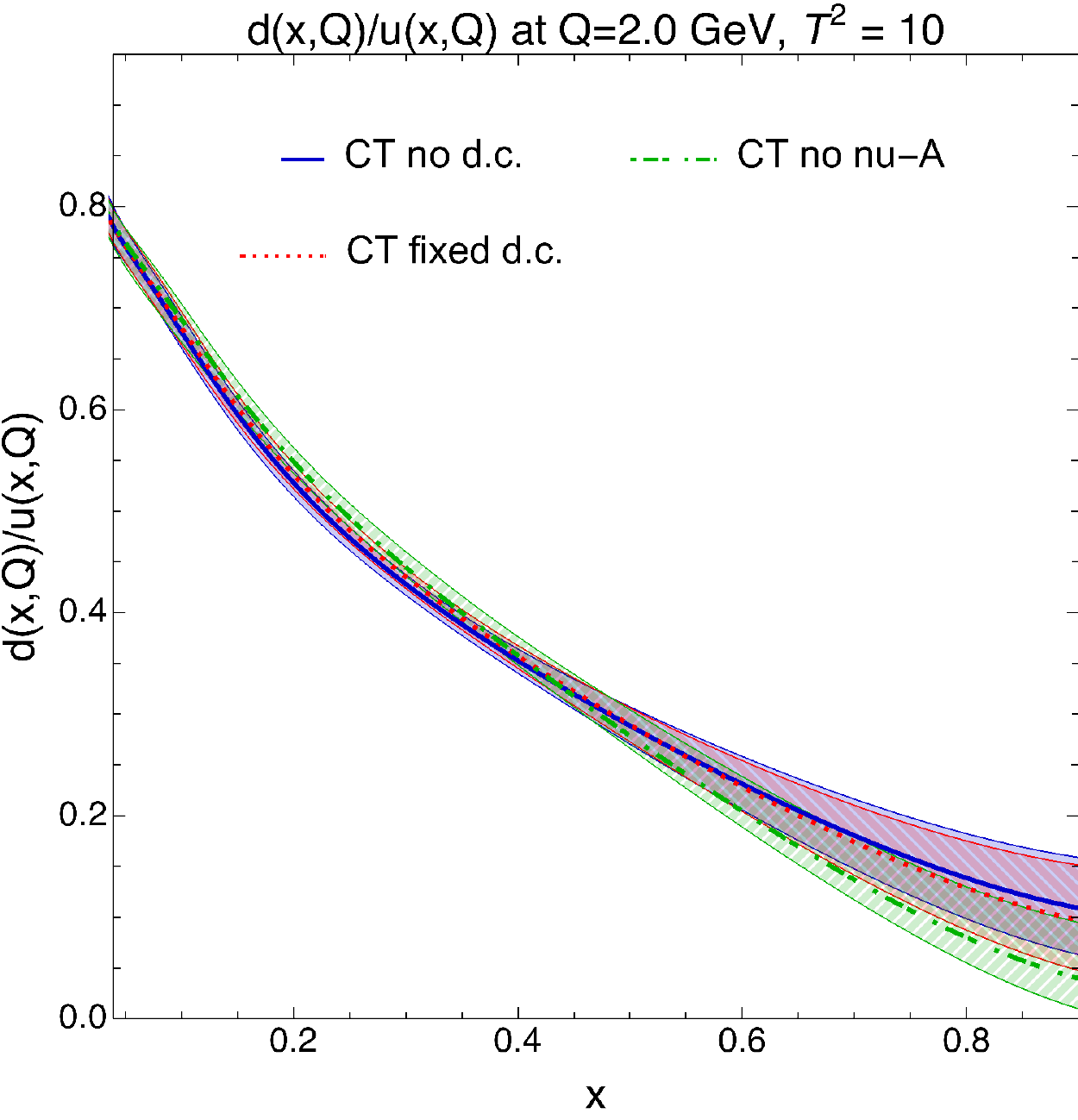} 
	\hfill
	\includegraphics[width=0.47\textwidth]{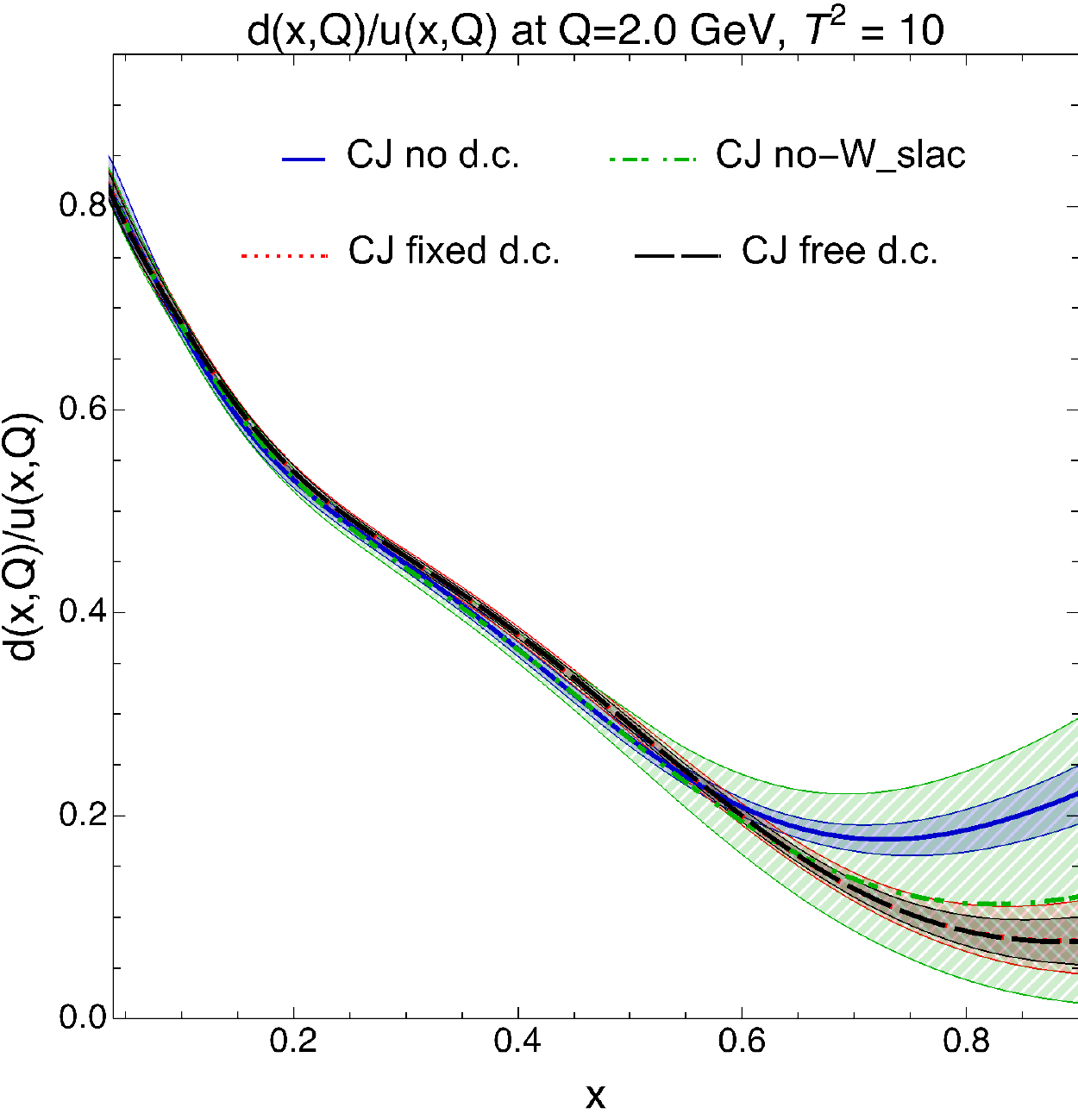} 
\caption{
    Upper row: The PDF ratios $d/u$ and their asymmetric error bands for $T^2=10$ at scale $Q=2$ GeV within the CT (left) and CJ (right) fitting frameworks.
    We normalize all $d/u$ error bands to the ratio from the central \texttt{no d.c.}~fit (without any assumed deuteron correction). The left panel shows the \texttt{CT no d.c.}, \texttt{CT fixed d.c.}, and \texttt{CT no nu-A} error bands. The right
    panel shows the analogous \texttt{CJ no d.c.}, \texttt{CJ fixed d.c.},
    \texttt{CJ no-W\_slac}, and the \texttt{CJ free d.c.}~fits. The abscissas are scaled to
    highlight the impact of the deuteron corrections at large $x$, where the impact is most pronounced, as well as the modest enhancement in the $d/u$ ratio for $x \lesssim 0.01$ in CT at left.
    Lower row: now showing the absolute $d/u$ ratios on a linear $x$-axis scale to highlight the behavior at high $x$.
}
\label{fig:du_comp}
\end{figure*}

Fig.~\ref{fig:du_comp} illustrates the impact of the $F^d_2$ corrections, as introduced in Sec.~\ref{sec:deut}, on $d/u$. The upper row shows the $d/u$ ratio and its error band obtained in the discussed series of fits, normalized to the central value obtained in the fits with no deuteron corrections. 
The lower row shows the unnormalized $d/u$ ratios themselves, using a linear horizontal scale to better visualize the $x>0.1$ interval, and in particular the $x \to 1$ behavior of this quantity. In both rows, the left and right
panels give results for CT and CJ fits, respectively.
We see that the deuteron corrections have a qualitatively similar impact on $d/u$ in both CT and CJ, especially at $x\! \gtrsim\! 0.1$, with evidence of a mild, few-percent enhancement of the fitted $d/u$ ratio over the \texttt{no d.c.}~fits for $0.1\! \lesssim x\! \lesssim 0.5$ once the fixed deuteron correction is included. 
This enhancement turns into a suppression at still higher $x\! \gtrsim\! 0.5$, beyond which $d/u$ is strongly affected by the $2$-body nucleon-nucleon corrections included in the $F^d_2$ calculation. For CJ, this suppression is larger than in the CT case, but compatible with the latter within the respective uncertainties of each fit.

The qualitative $x$ dependence of the deuteron-corrected fit of $d/u$ in the top rows of Fig.~\ref{fig:du_comp} closely follows the $F^N_2/F^d_2$ ratio plotted in Fig.~\ref{fig:deut-SF}, in which $F^N_2$ and $F^d_2$ represent the deuteron structure function computed using the isoscalar and full nuclear predictions,
respectively. Indeed, in the \texttt{no d.c.}~fits, $F^d_2$ is effectively fitted as an isoscalar target, but in the \texttt{CT fixed d.c.}, for example, the $F^d_2$ data for the physical deuteron are corrected to $F^N_2$, which leads to a relative suppression of the fitted $d/u$ PDF ratio for $x\! \gtrsim\! 0.5$. 

In the CJ and CT \texttt{fixed d.c.}~fits (red curves), the d.c.~parameters are held constant at their values obtained from the central CJ15 fit.  If, on the other hand, the d.c.~parameters in Eq.~(\ref{eq:offshell}) are actively fitted with the PDF shape parameters, we obtain the same central PDFs but a narrower uncertainty band on $d/u$, as shown by the \texttt{CJ free d.c.}~error band in the right panels of Fig.~\ref{fig:du_comp}. This reduction in the PDF uncertainty of $d/u$, which at first sight is paradoxical because we have increased the number of fit parameters, is actually a consequence of the correlation
between the treatment of $F^d_2$ structure function data and extracted $d$-PDF. More specifically, when the nucleon off-shellness parameters are freed, the variations in the $d$-quark parameters that were necessary to encompass the $F^d_2$ data in the \text{CJ fixed d.c.}~fit are partly
absorbed by the parameters of Eq.~(\ref{eq:offshell}).
In other words, releasing the off-shell parameters reduces tensions in parameter space, and ultimately also diminishes the
overall experiment-by-experiment $\chi^2_E$ variations in the PDF analysis, as we shall note again below in Sec.~\ref{sec:d/u_pulls} and~\ref{sec:gluon_pulls}.
We have verified that, over the same $x$ range, the relative uncertainty in the determination of other flavors, such as, \textit{e.g.}, the gluon, does correspondingly increase. This is an instance of the fact that, typically, the constraining power of a fit can be enhanced by a greater number of free parameters only in a limited sector of parameter space.

The different choices of data sets in the CT and CJ fits also affect the fitted $d/u$ PDF ratio. 
For example, unlike CJ, the CT fit includes DIS data from inclusive $\nu$-A collisions, multiplied by a phenomenological parametrization of the heavy-nuclear structure function relative to the physical deuteron. 
The green bands in Fig.~\ref{fig:du_comp} (left) are for the \texttt{CT no nu-A} variant of the CT fit. The removal of these data augments the shifts in the CT $d/u$ ratio induced by including the fixed deuteron correction, which now is comparable to the CJ result. The reason for this may simply lie in the lack of further deuteron-to-isoscalar proton plus neutron corrections, or may also be related to possible discrepancies between neutral- and charge-current DIS data or interactions \cite{Kopeliovich:2012kw,Kovarik:2010uv,Paukkunen:2013grz}.

Conversely, the CT fits do not include the low-$W$ SLAC data and the reconstructed $W$-boson asymmetry from the D\O\ experiment that are influential upon the large-$x$ $d$-quark fit in CJ \cite{Accardi:2016qay}. When these are removed from the CJ fit as well --- see the green \texttt{CJ no-W\_slac} bands in the right column of Fig.~\ref{fig:du_comp} --- we obtain an enlarged uncertainty that includes both the deuteron corrected and uncorrected bands.
The uncertainty on $d/u$ at $x\to 1$ in the \texttt{CJ no-W\_slac} fit is wider than in the \texttt{CT no nu-A} fit in part in reflection of different parametrization forms and selection of experiments between the CJ and CT analyses.

\begin{figure*}[th]
	\includegraphics[width=0.47\textwidth]{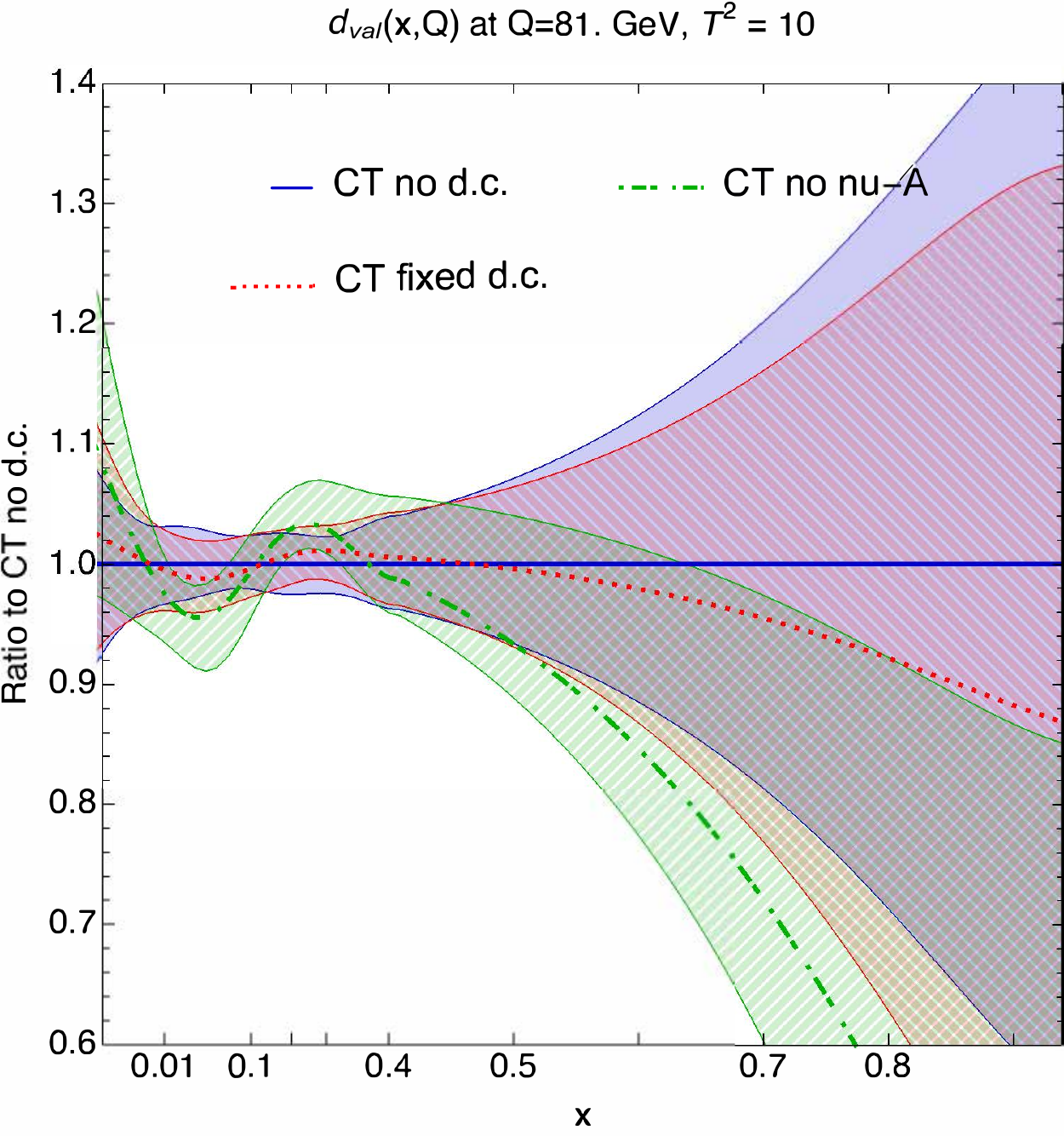} 
	\hfill
	\includegraphics[width=0.47\textwidth]{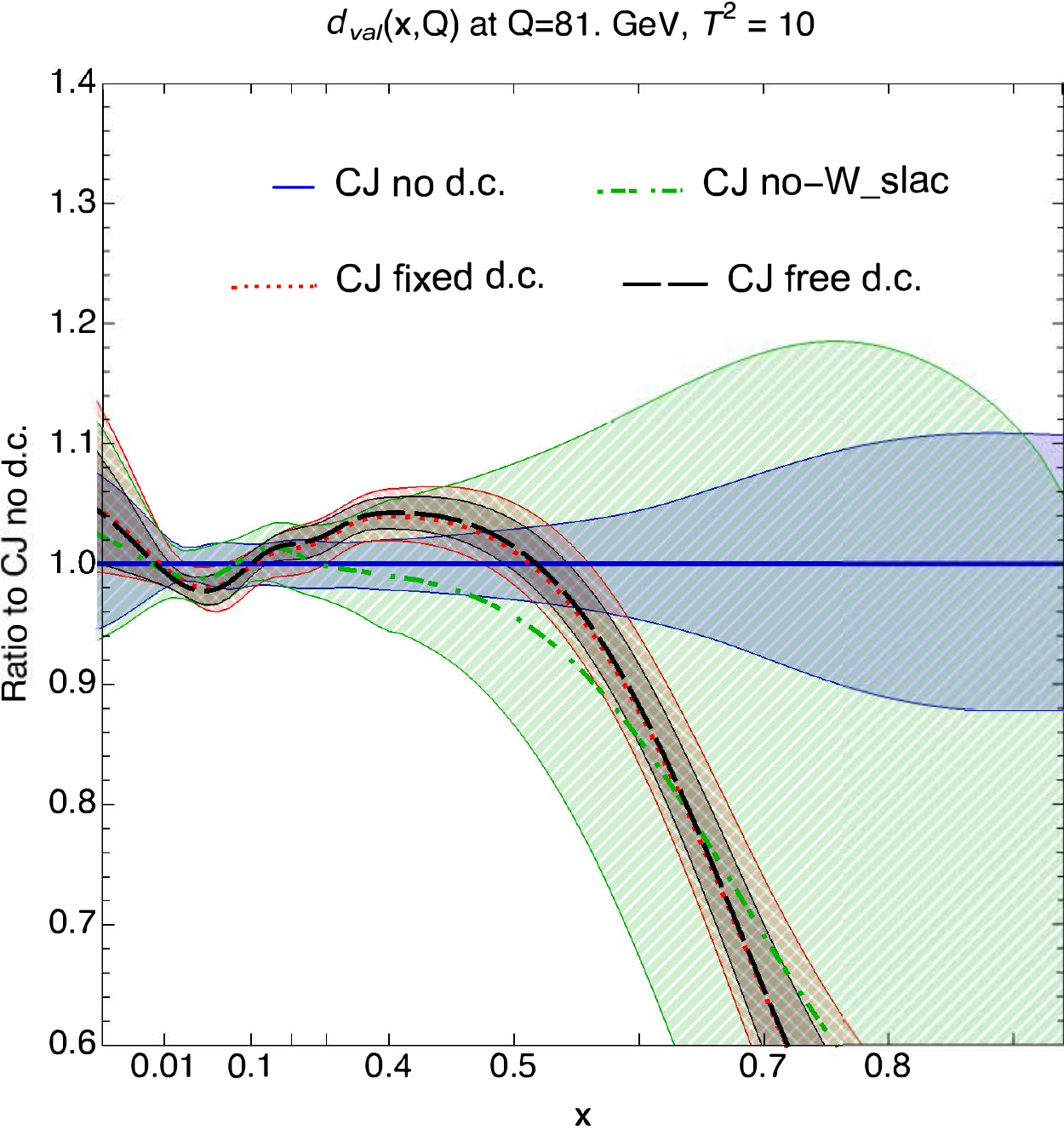} \\
\caption{
    The valence $d$-quark PDFs and their uncertainties, normalized to the central values obtained in the \texttt{no d.c.}~fits. The fits and conventions are the same as for Fig.~\ref{fig:du_comp}. 
}
\label{fig:dv_comp}
\end{figure*}

\subsection{Impact on the valence PDFs in the LHC EW precision region}
\label{sec:weakmix}
The effects of including deuteron structure corrections to $F^d_2$, while most pronounced at $x>0.1$, have some consequences in the low-$x$ region as well. In the CT result shown in Fig.~\ref{fig:du_comp}(left),
modest enhancements in $d/u$ at about $x\!\sim\!10^{-3}$ can be seen for the \texttt{CT fixed d.c.}~fit. While the sea-quark PDFs are relatively unaffected in this kinematic region, the valence component of the $d$-quark and, to a lesser extent, the $u$-quark PDFs in this small-$x$ region are sensitive to the inclusion and theoretical treatment of both neutrino-nucleus and deuterium data at large $x$. This sensitivity is mainly a consequence of corresponding valence sum rules. 

Fig.~\ref{fig:dv_comp} illustrates this feature. In the left panel, the \texttt{fixed d.c.}~CT best-fit valence PDF (red dotted line), shown in this case at $Q=81$~GeV~$\approx M_W$,  prefers a slightly lower $d_{val}$ at $x\approx 0.008-0.13$ than in the \texttt{no d.c.}~fit, and a slightly higher one at $x\approx 0.13-0.45$. This deviation becomes still more pronounced if the neutrino-nucleus DIS data are removed (leading to the green dot-dashed line). In the right panel, we observe for CJ a qualitatively similar trend and associated $x$ dependence over slightly shifted $x$ regions of $0.01-0.1$ and $0.1-0.53$.

In Sec.~\ref{sec:intro}, we illustrated in Figs.~\ref{fig:s2w_corr} and \ref{fig:qval_LM} that the weak-mixing angle measurements at the LHC using the forward-backward asymmetry, $A_\mathrm{FB}$, are sensitive to the valence $d$- and $u$-PDFs in an $x$-region about $\sim\!0.03$, {\it i.e.}, where Fig.~\ref{fig:dv_comp} indicates a dependence of these PDFs upon the treatment of the deuteron/heavy-nucleus data at higher-$x$ values.

\begin{figure*}[th]
	\includegraphics[width=0.47\textwidth]{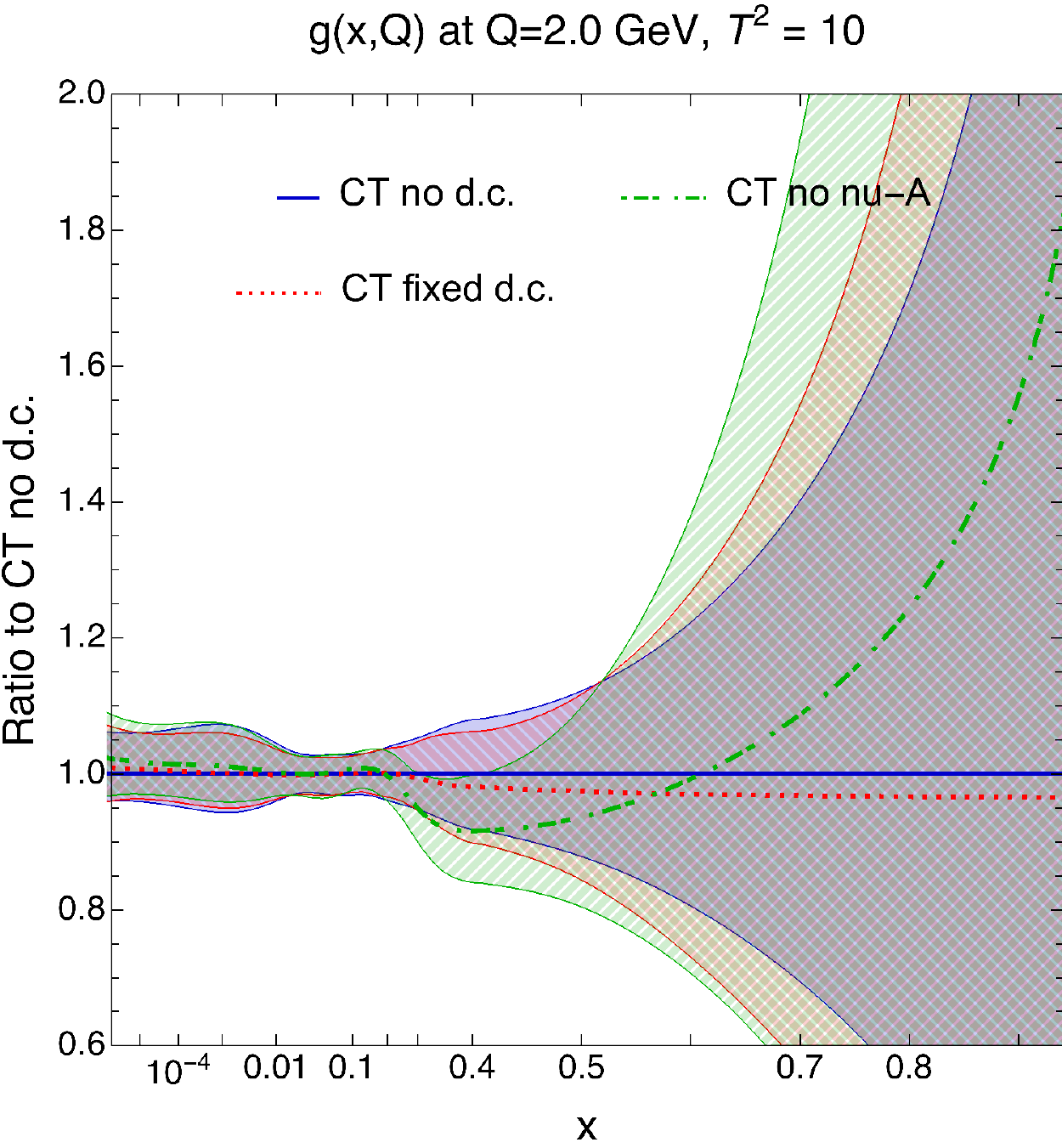} 
	\hfill
	\includegraphics[width=0.47\textwidth]{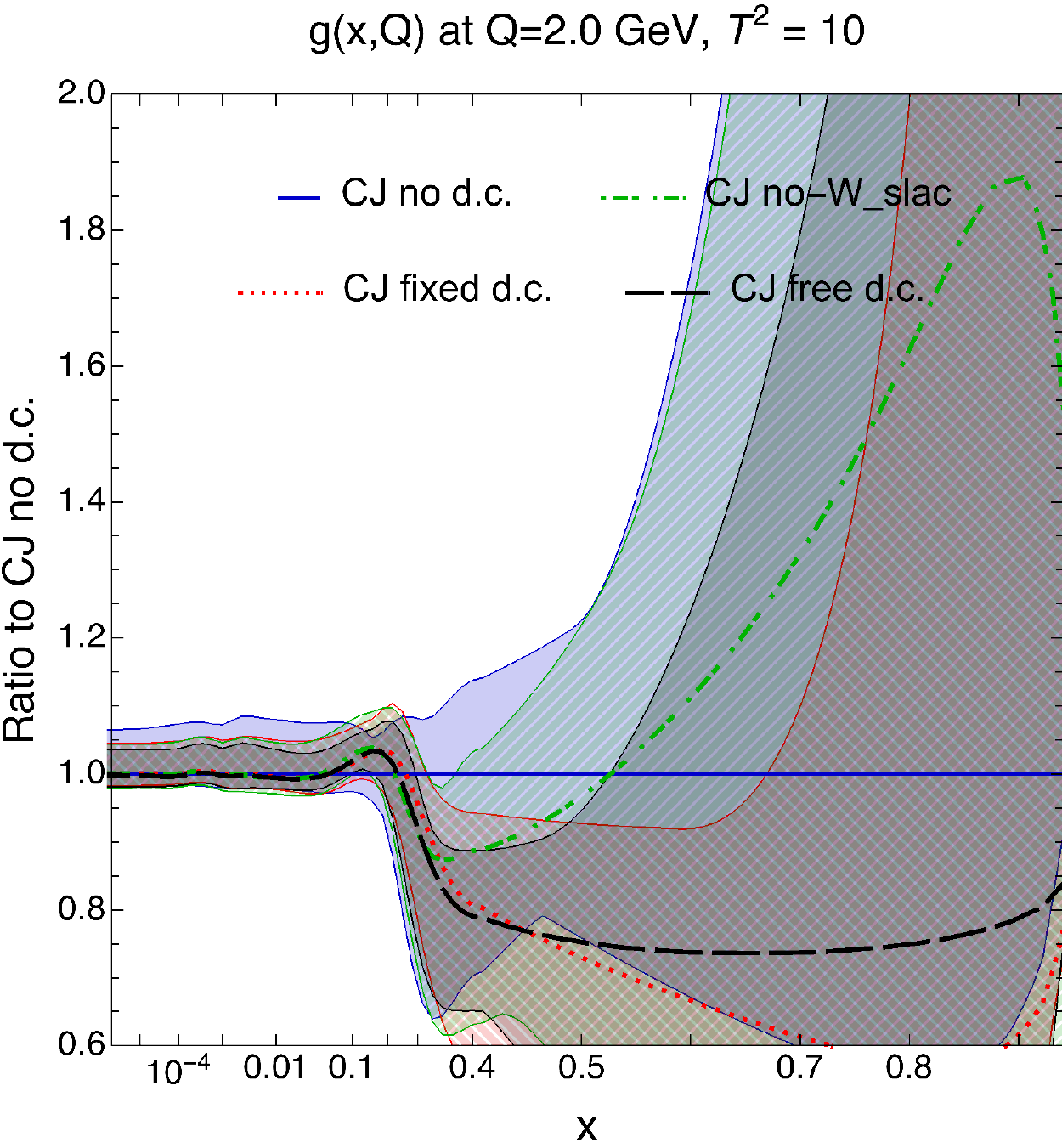} \\
\caption{
    Same as Fig.~\ref{fig:dv_comp}, but now for the gluon PDF.
}
\label{fig:g_comp}
\end{figure*}

\subsection{Impact of deuteron corrections on the gluon PDF} 
\label{sec:gluon_PDF}

The deuteron data sets can also impact the gluon density through $Q^2$-scaling violations;
this is particularly true when these measurements cover a large range of the four-momentum squared, $Q^2$, of the exchanged boson. 
Similarly to the case for $d_\mathit{val}$ in Sec.~\ref{sec:intro}, the Lagrange Multiplier scans \cite{Hou:2019efy} and PDF sensitivity techniques \cite{Wang:2018heo,Hobbs:2019gob} in the CT18 analysis collectively demonstrate that the gluon at $x>0.1$ receives significant constraints from the DIS
deuterium data. Constraints from the extensive fixed-target DIS data are in fact competitive with the HERA DIS data, which probe the lower-$x$ region, and also with LHC and Tevatron inclusive jet production, which cover a wide $x$ range but involve complex arrays of systematic effects which remain under active study. 

To address this point, in Fig.~\ref{fig:g_comp} we plot the error bands for the gluon PDF at $Q\!=\!2$~GeV as determined in the series of fits discussed at the beginning of this Section, again normalized to the central value obtained in the \texttt{no d.c.}~fits.  
As seen in the left panel of Fig.~\ref{fig:g_comp}, the gluon PDF in the CT fits exhibits a modest sensitivity to the chosen deuteron correction treatment, with a dependence
that is somewhat moderated by the adopted $W^2 > 12.25$ GeV$^2$ cut. Still, as with $d/u$, there is a qualitative tendency for the fixed deuteron correction to reduce the high-$x$
gluon PDF, with this modification being enhanced by the exclusion of the inclusive $\nu$-A DIS data. Like CT, the CJ fits seen in Fig.~\ref{fig:g_comp} (right), which include the SLAC DIS data, similarly display a relative suppression of the gluon for $x\! \gtrsim\! 0.3$ once deuteron corrections are taken into account. While this effect is of moderate size, it is nonetheless statistically significant in the context of the $T^2=10$ tolerance used to determine the uncertainty bands. For CJ, it will be interesting to confirm this effect by fitting the full JLab 6 GeV inclusive data set \cite{Jlab6-inprep}, and, even more so, the JLab 12 GeV data which will augment the precision of the available DIS measurements over a wide $Q^2$ range at large $x$, once available. 

\subsection{Valence-sector PDF pulls: the $d/u$ ratio}
\label{sec:d/u_pulls}

\begin{figure*}[p]
\centering
	\includegraphics[height=0.32\textheight]{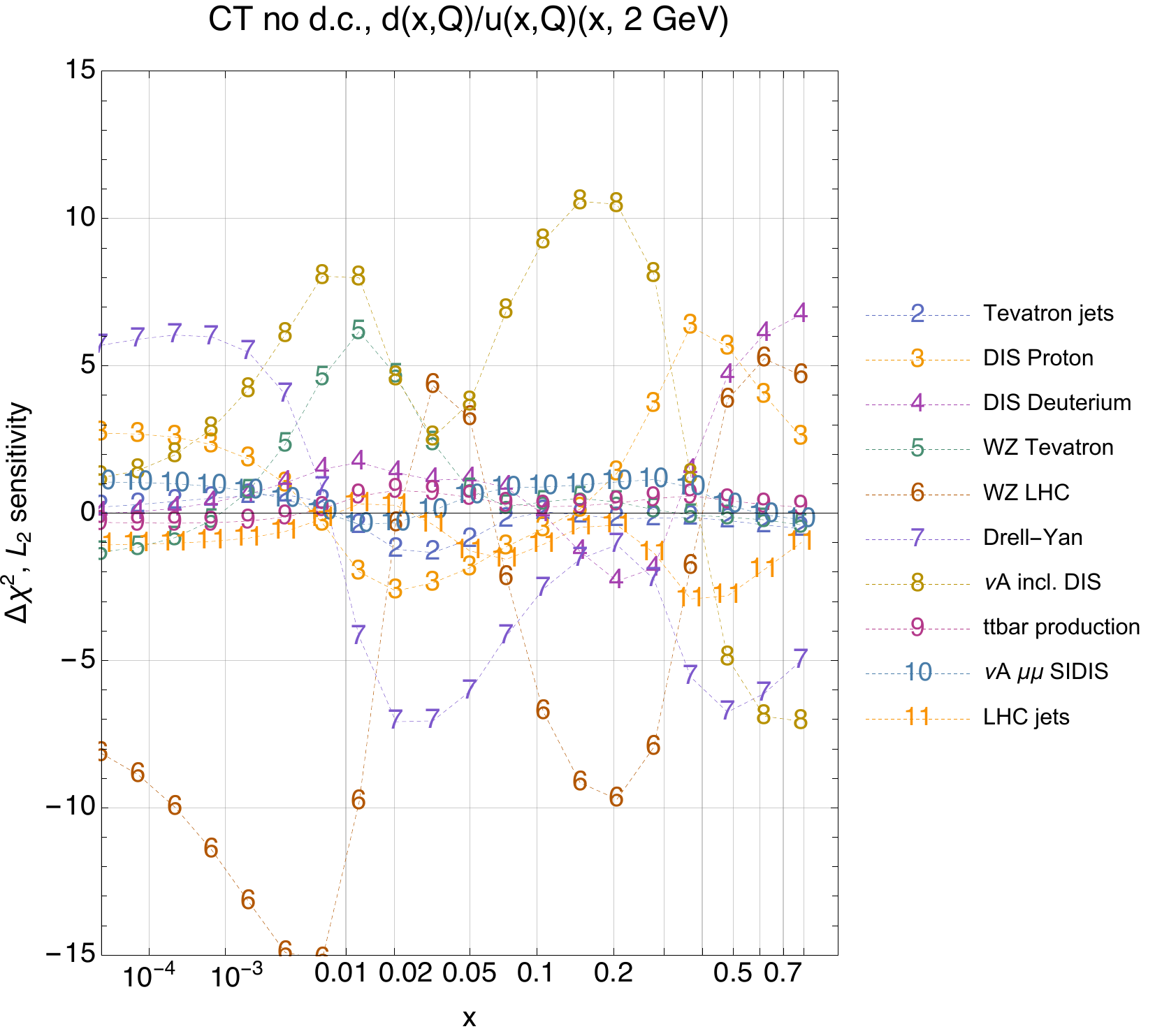}
	\includegraphics[height=0.32\textheight]{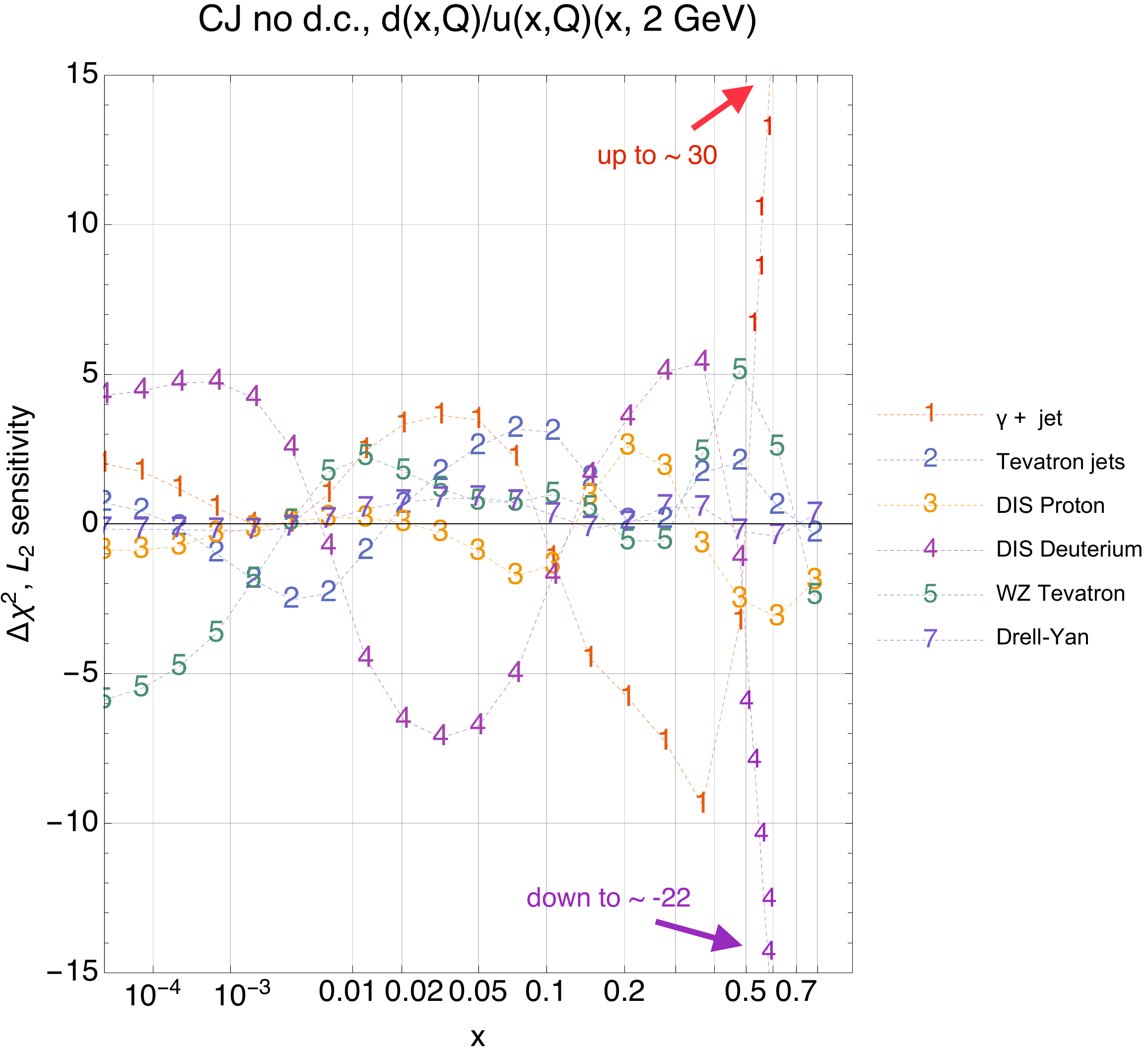} \\
	\includegraphics[height=0.32\textheight]{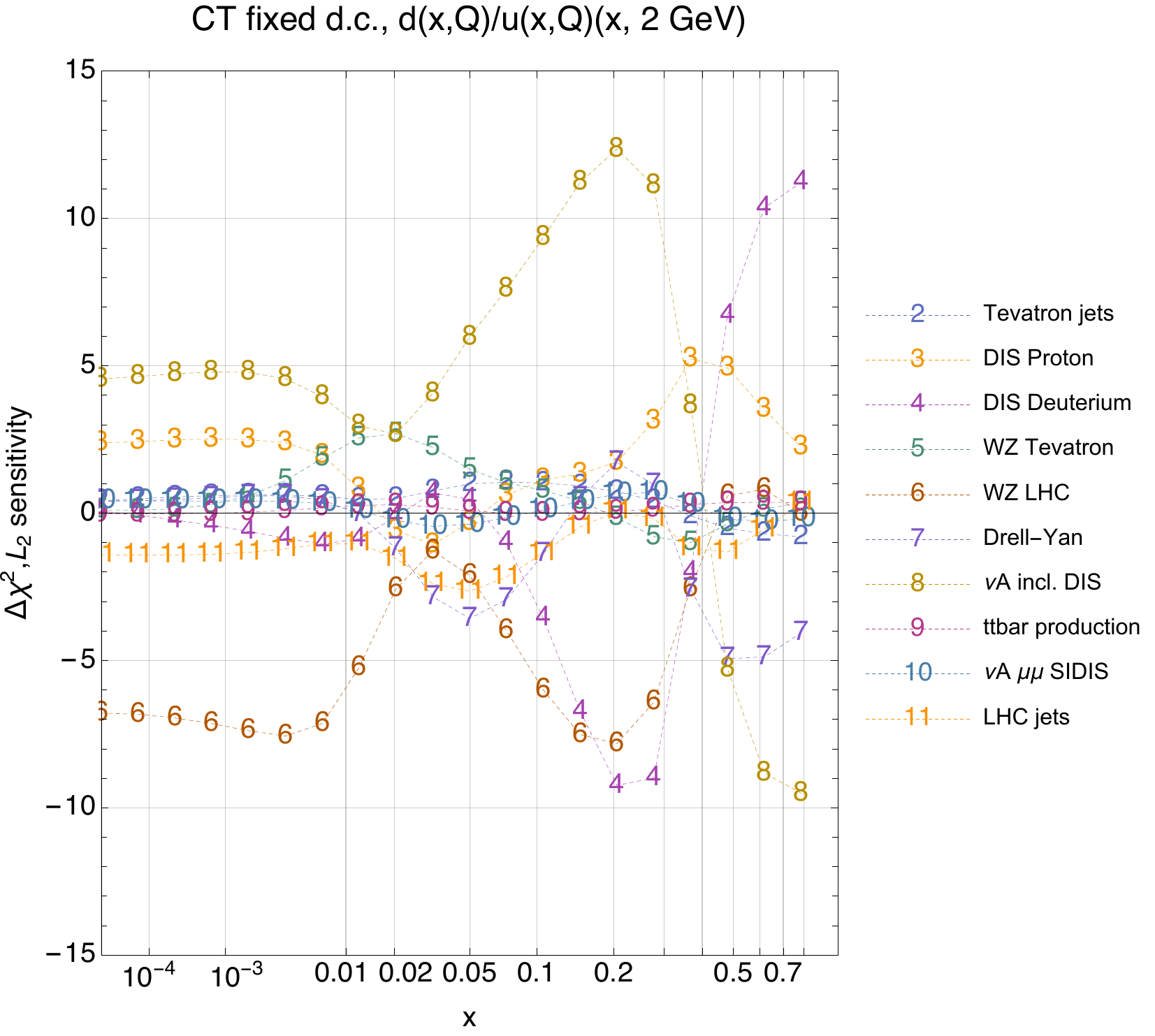} \includegraphics[height=0.32\textheight]{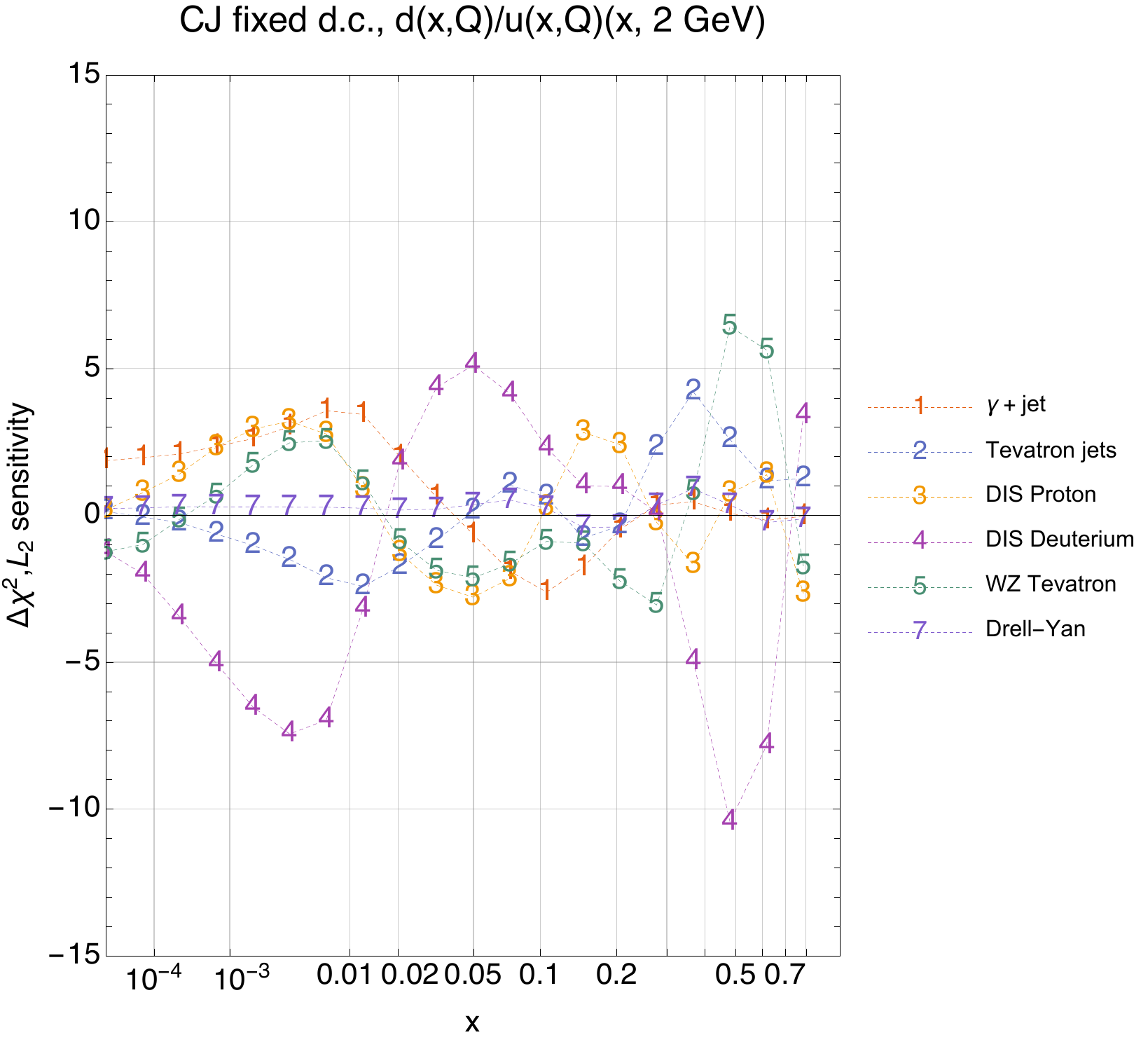} \\
	\includegraphics[height=0.32\textheight]{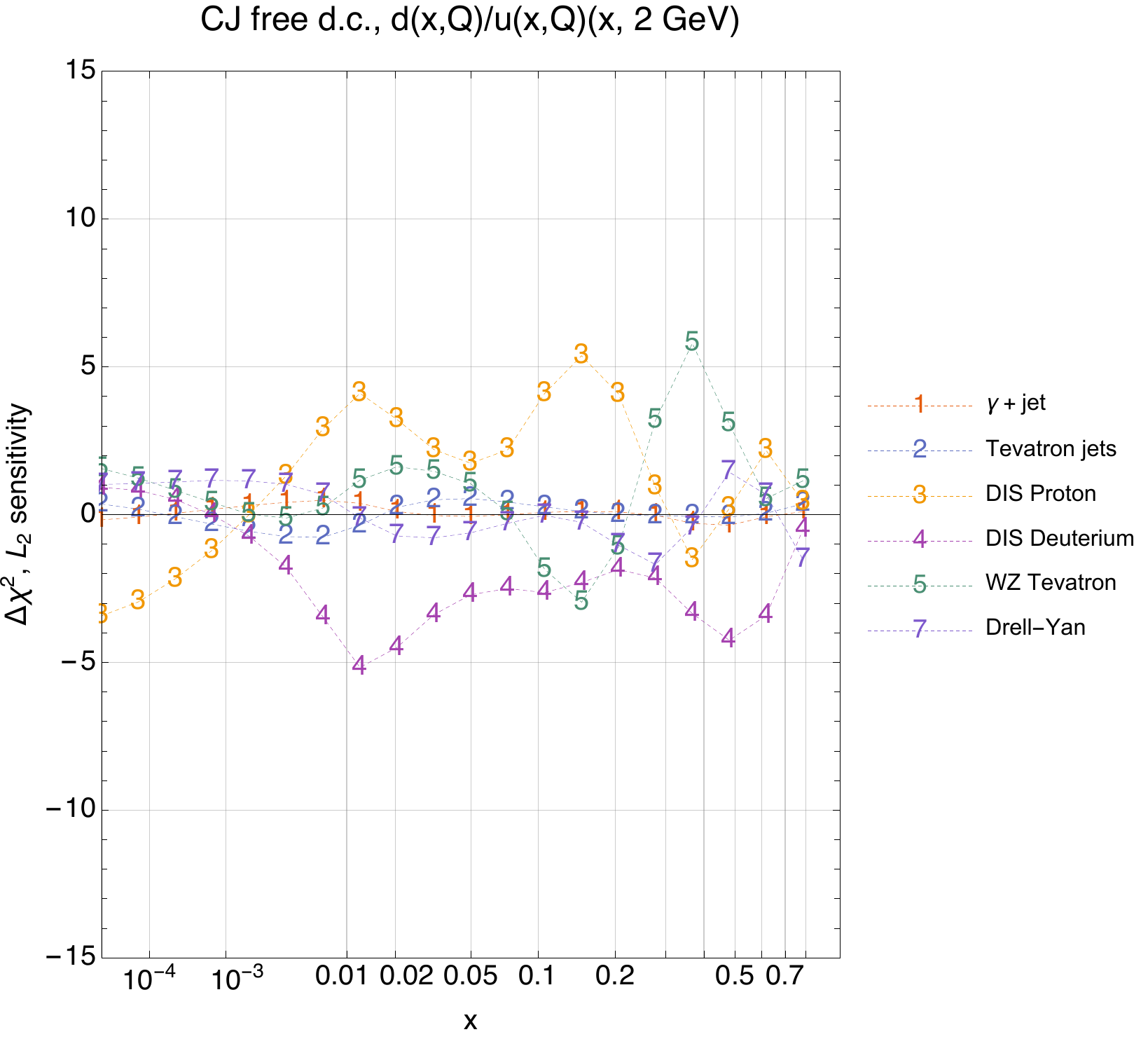} 
\caption{
The $L_2$ sensitivities computed according to Eq.~(\ref{eq:L2}) for $Q=2$ GeV, giving the pulls on the
    $d/u$ PDF ratio of the process-dependent data sets fitted by CT (left) and CJ (right). Upper, middle, lower rows:  results for the \texttt{no d.c.}, \texttt{fixed d.c.}, and \texttt{free d.c.}~fits discussed at the beginning of Sec.~\ref{sec:results}.
}
\label{fig:du_SfL2}
\end{figure*}

\begin{figure*}[th]
	\centering
	\includegraphics[width=0.48\textwidth]{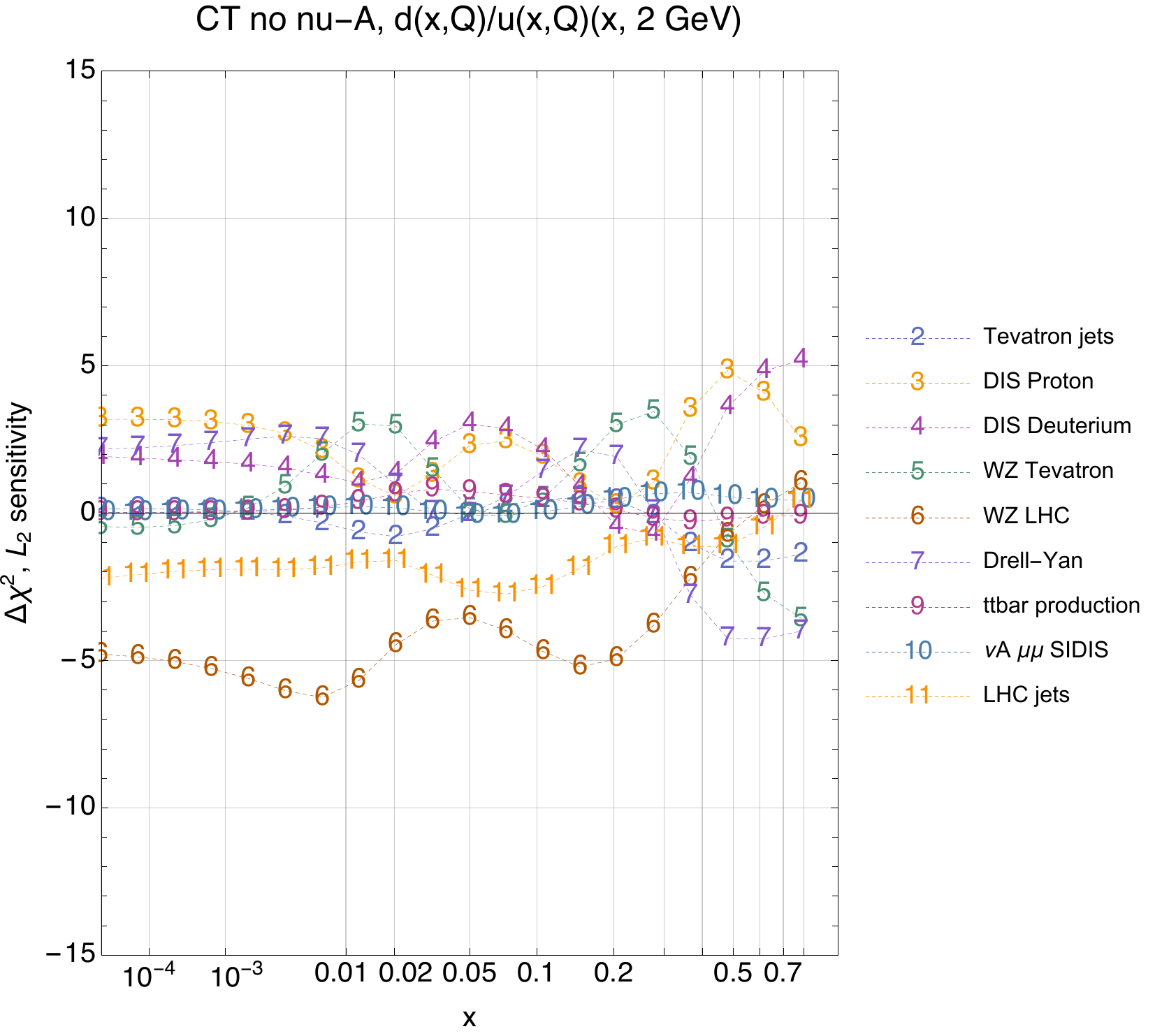} 
	\includegraphics[width=0.48\textwidth]{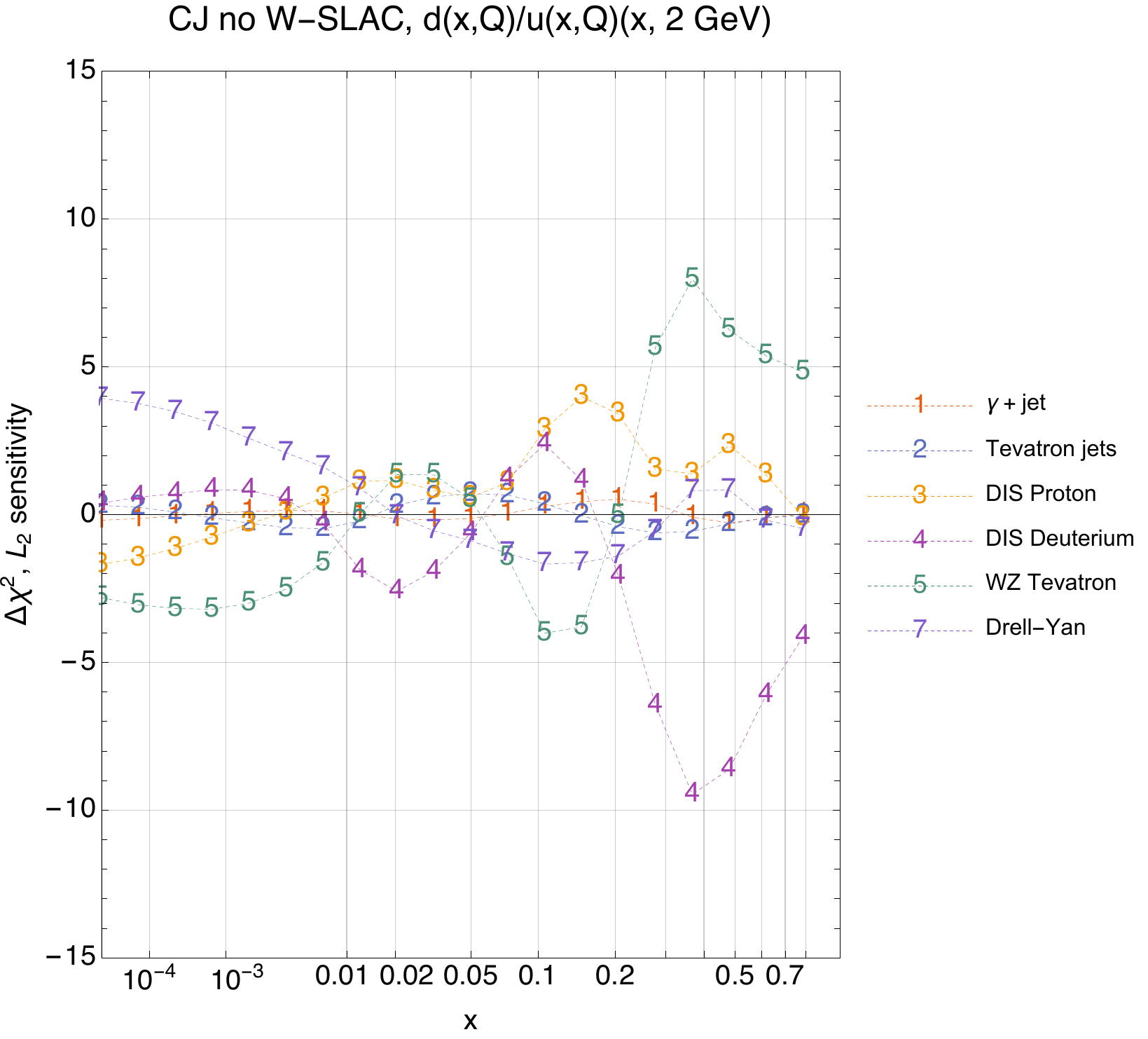} 
\caption{
    As in Fig.~\ref{fig:du_SfL2}, we plot the PDF pulls on $d/u$ at 2 GeV with fixed deuteron corrections present, but, in this case, removing select
    experiments which have shown significant competing pulls with respect to the DIS deuterium
    sets. For CT (left panel), we remove the inclusive $\nu$-A data (Group \#{\bf 8}), while for CJ, we remove the SLAC DIS experiments (part of Group \#{\bf 4})
    and $W$-asymmetry information from the Tevatron (part of Group \#{\bf 5}).
    }
\label{fig:du_SfL2_exps}
\end{figure*}

In Fig.~\ref{fig:du_SfL2}, we plot the $L_2$ sensitivities of the groups of experiments to the $d/u$ PDF ratio for varied implementations of the deuteron corrections. $L_2$ sensitivities for fits exploring data-set variations are shown in Fig.~\ref{fig:du_SfL2_exps}, which we also discuss below.

Starting with the \texttt{no d.c.}~fits that either do not include (CT) or remove (CJ) deuteron corrections, we notice that, in both cases, the landscape of PDF pulls tends to be dominated by a few competing groups of experiments, which differ between the two fitting frameworks.
For CT in the left panel, these are the LHC $W/Z$ production (Group \#{\bf 6}) and inclusive
nuclear DIS (Group \#{\bf 8}), which possess the sharpest opposing pulls at $x\!\sim\! 0.2$, for example, in the direction
of either favoring or disfavoring a larger $d/u$ ratio, respectively. At slightly higher $x\! >\! 0.3$ values, which are of particular interest from the perspective of QCD-informed models of the $d/u$ behavior as $x \to 1$, these are joined by the DIS-deuterium (\#{\bf 4}) and Drell-Yan (\#{\bf 7}) groups of experiments. 
Turning to the CJ case, displayed in the top-right panel, it is the DIS deuterium (\#{\bf 4}) and gamma-jet (\#{\bf 1}) groups that dominate the landscape of PDF pulls --- and quite strikingly at large $x$ --- with lesser but also significant pulls from the Tevatron $W/Z$ production data (\#{\bf 5}). The large-$x$ pulls are expected, since the SLAC data are quite sensitive to nuclear dynamics in the deuteron target, as already noted.

Once fixed deuteron-structure corrections are introduced into the respective \texttt{fixed d.c.}~fits, the relative
patterns of PDF pulls experience an intriguing series of shifts, which we display in the middle row of Fig.~\ref{fig:du_SfL2}.
For CJ, the deuteron corrections largely resolve the huge tensions between the photon+jet and DIS deuteron data, because dynamical nuclear effects in the latter are now included in the theoretical calculation of the deuteron DIS cross section without forcing the $d$-quark to deform to compensate for the missing nuclear effect. A residual tension between the DIS deuteron (\#\textbf{4}) and $W/Z$ Tevatron data (\#\textbf{5}) data is still visible at $x\!\approx\! 0.5$.
While the large-$x$ tensions are reduced, the small-$x$ pulls visibly change in shape for several flavors. 

For CT, the introduction of the fixed deuteron correction detailed in Sec.~\ref{sec:deut} instead preserves the qualitative $x$ dependence of the $L_2$ pulls ({\it i.e.}, the shapes) but softens their magnitude in a few cases, for example, for the Drell-Yan (\#\textbf{7}) and the small-$x$ LHC $W/Z$ production (\#\textbf{6}) data.
Notably, the DIS-deuterium sensitivity shifts to closely resemble that of the LHC $W/Z$ data (\#{\bf 6}) in favoring a larger $d/u$ for $x \sim 0.2$. At the same time, the size of the competing pulls at high $x\! \gtrsim\! 0.3$ between the DIS-deuterium and inclusive $\nu$-A data (\#{\bf 8}) is enhanced, while the opposing pulls of other experiments are modestly reduced over the same range in $x$. This is especially clear for the CT Drell-Yan data (\#{\bf 7}), which in the \texttt{CT no d.c.}~fit had preferred a softer $d/u$ ratio at low $x\! <\! 0.01$  and an enlarged value of $d/u$ over $0.01\! \lesssim\! x\! \lesssim\! 0.2$. In \texttt{CT fixed d.c.}, these preferences mostly vanish.
At the same time, outside this interval of very high $x$, the opposing pulls of the inclusive $\nu$-A data (Group~\#\textbf{8}) and both the DIS deuterium (\#\textbf{4}) and LHC $W/Z$ (\#{\bf 6}) experiments increase and sharpen for $x\gtrsim 0.01$. In fact, this is the same collection of experiments for which in Table~\ref{tab:chi2_summary_AA} we observed  increases (in the case of Groups \#{\bf 6} and~\#\textbf{8}) in their respective values of $\chi^2/N_\mathit{pt}$ upon introducing deuteron corrections for $F^d_2$.
Both the $\chi^2$ values in Table~\ref{tab:chi2_summary_AA} and  the $L_2$ analysis for CT therefore indicate a noticeable rearrangement of the pulls of the inclusive neutrino-nucleus DIS data and select other experiments introduced by the correction to the deuteron DIS data. This rearrangement, being presently of similar order with respect to other contributing effects, will require attention in the future.

In the last row of Fig.~\ref{fig:du_SfL2} we present the $L_2$ pulls of 
the \texttt{CJ free d.c.}~fit in which the deuteron off-shellness degrees-of-freedom in Eq.~(\ref{eq:offshell}) are released.
Comparing the vertical extents of the peaks with those of the \texttt{CJ fixed d.c.}~fit in the middle row, we see that freeing the offshell parameters moderates the PDF pulls over the whole $x$ range, especially those at $x\! >\! 0.3$ between the DIS deuteron (\#{\bf 4}) and the WZ Tevatron CJ Drell-Yan (\#{\bf 5}) data.
This behavior can be generically understood as a consequence of increasing the number of free parameters, but is not guaranteed, for example, in the presence of incompatible data sets.
The \texttt{CJ free d.c.}~plot is thus an indication of a good level of consistency between the considered data sets, when the PDFs and deuteron corrections are fitted together. 

\begin{figure*}[p]
	\centering
	\includegraphics[height=0.32\textheight]{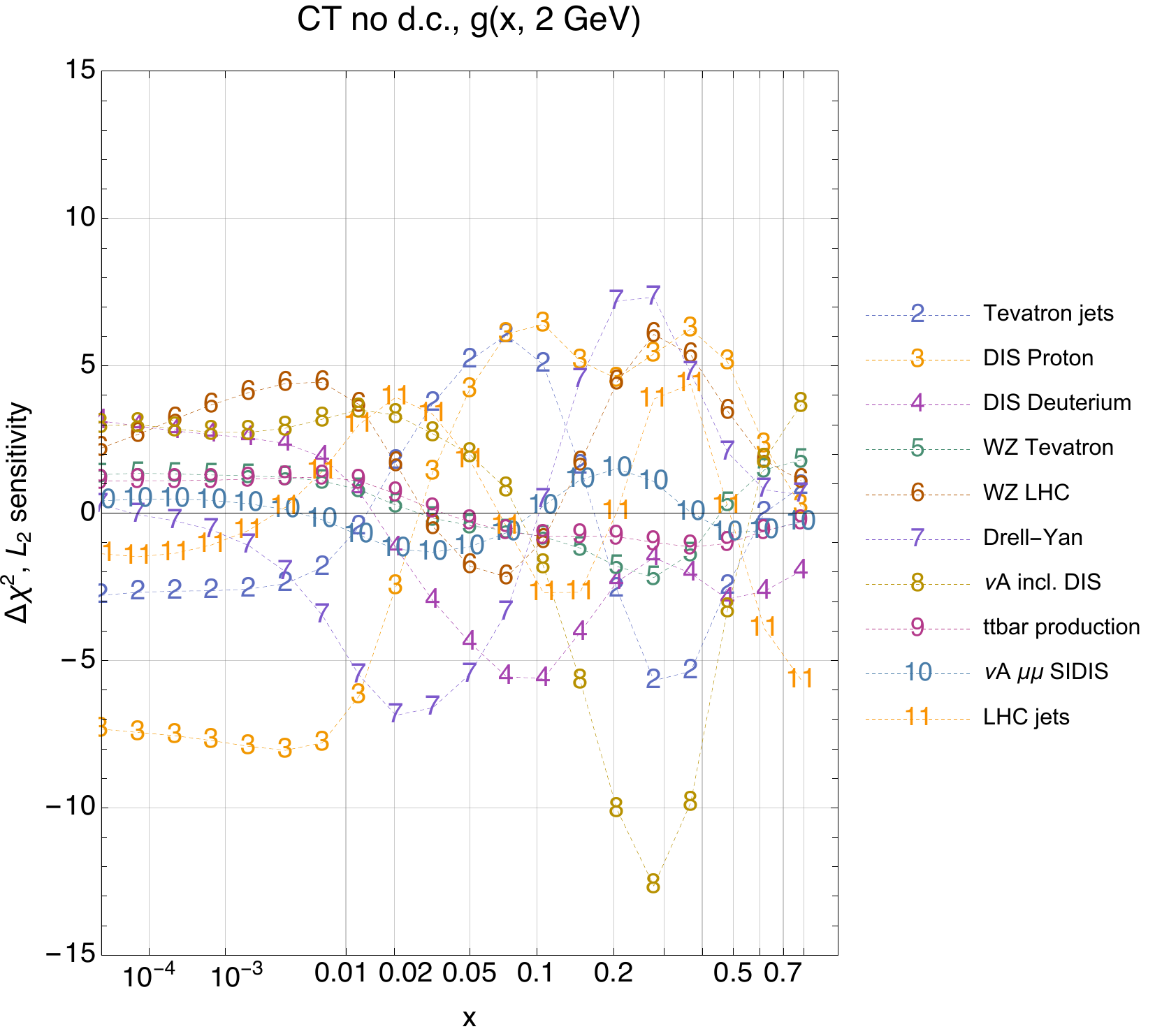}
	\includegraphics[height=0.32\textheight]{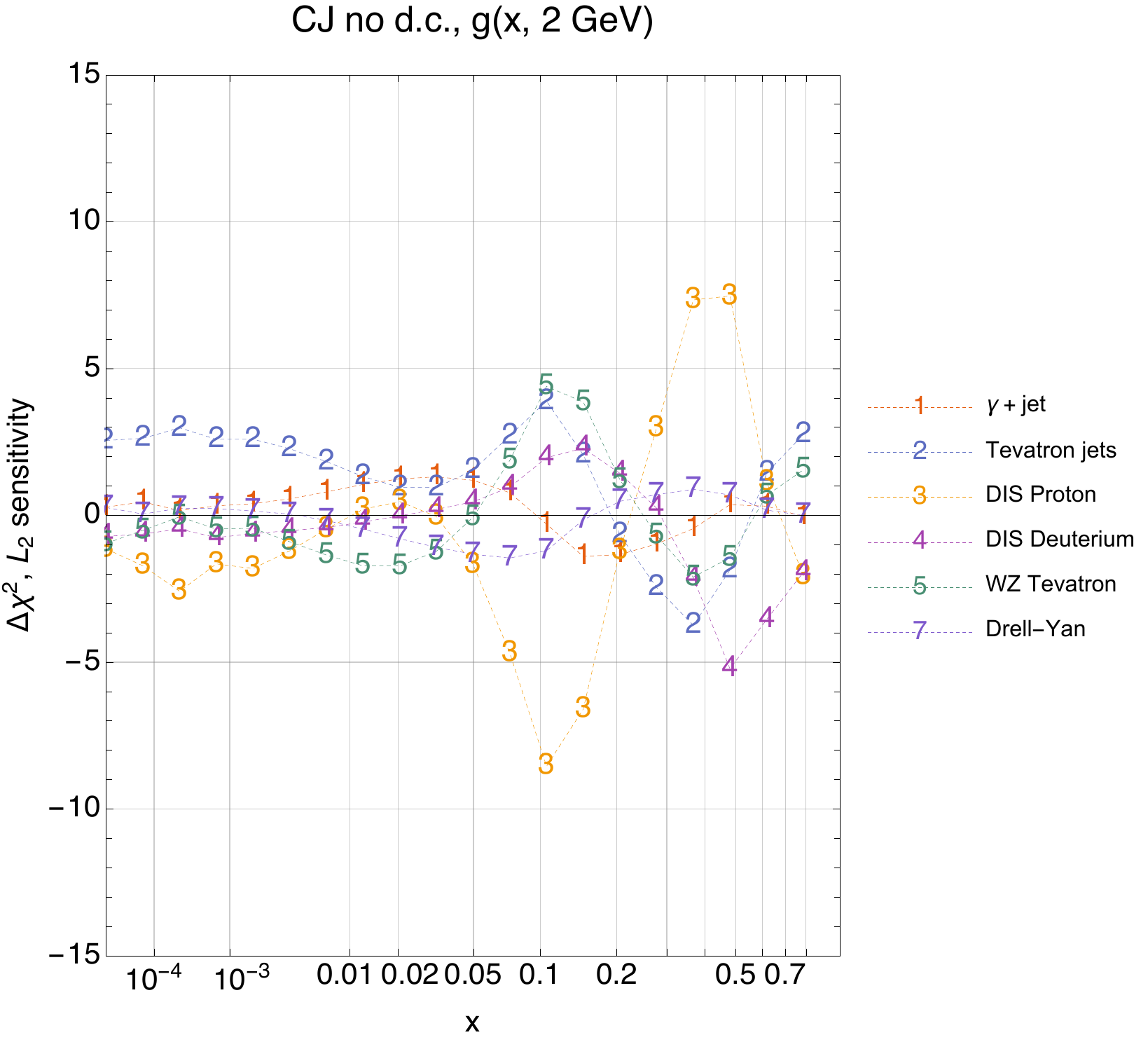}\\
	\includegraphics[height=0.32\textheight]{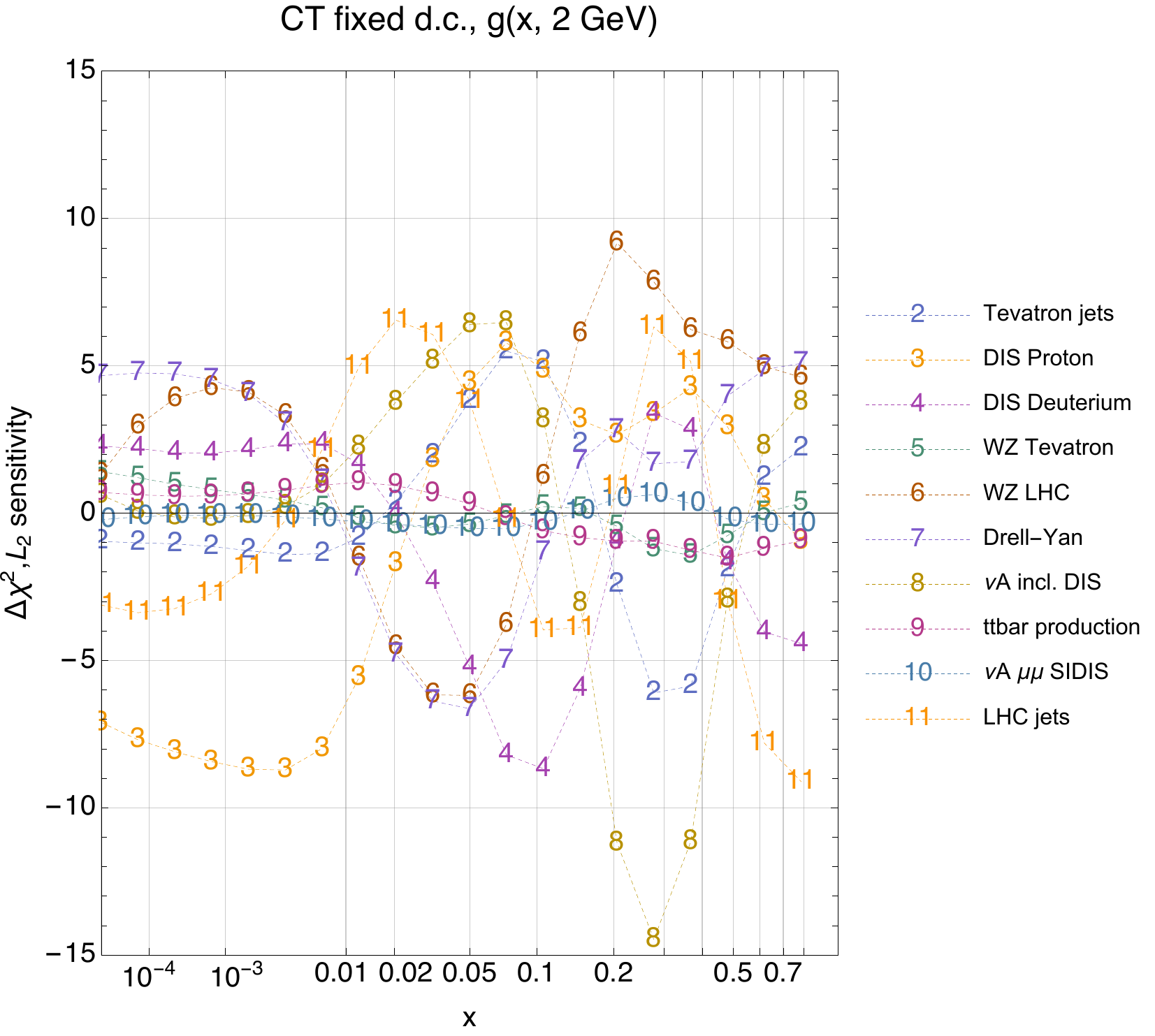} 
	\includegraphics[height=0.32\textheight]{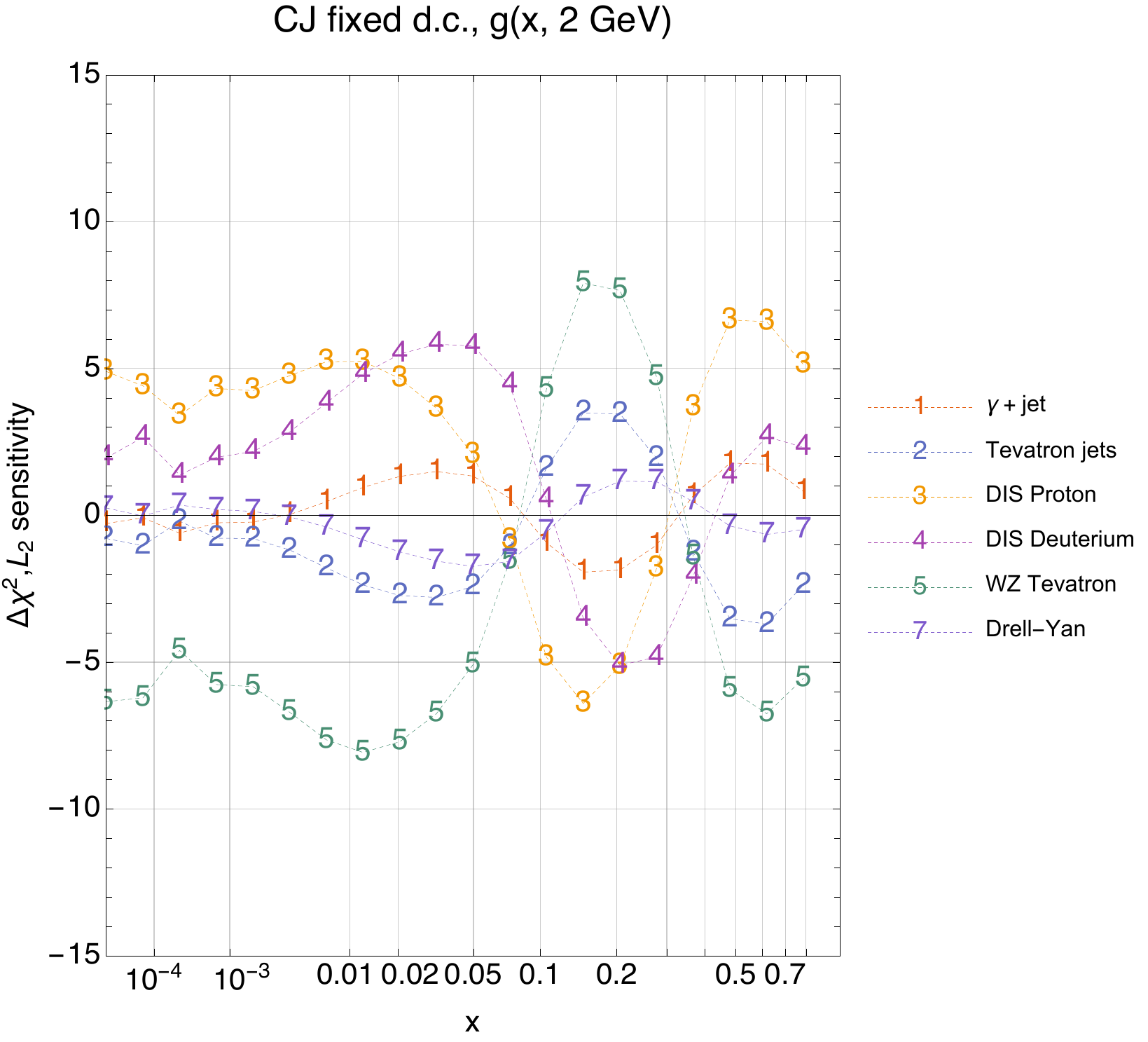} \\
	 	\includegraphics[height=0.32\textheight]{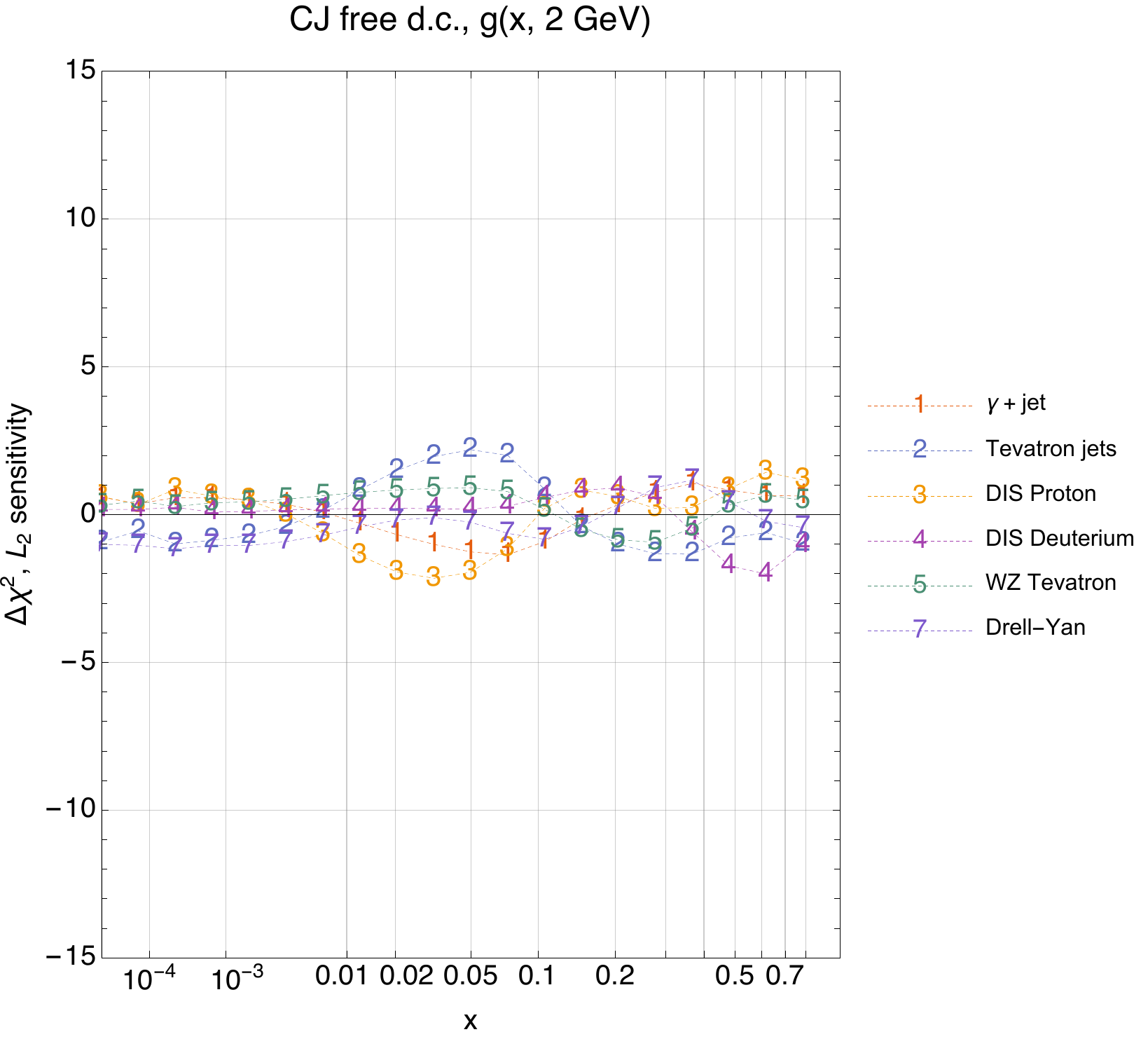} 
\caption{
    Analogous to Fig.~\ref{fig:du_SfL2}, for the PDF pulls on $g(x,Q)$ at $Q=2$ GeV.
    }
\label{fig:glu_SfL2}
\end{figure*}

The results discussed so far suggest a nontrivial relationship between the treatment of the DIS deuterium data and the description of
other data sets in each fitting framework:
{\bf the 
impact of deuteron-structure corrections in a global fit like CT and CJ cannot generally be
expected to apply to deuterium data alone, but have secondary effects on the patterns of pulls of other data sets.}

It is therefore interesting to study variations in the choices of experimental data sets in both fits, in particular, removing from each analysis those data sets that showed especially strong sensitivity to deuteron corrections or otherwise played a major role in the foregoing discussion.

For CT, we remove the entire collection of inclusive $\nu$-A measurements (Group \#{\bf 8}), and refit with the fixed deuteron corrections of Sec.~\ref{sec:deut} in place; the resulting $L_2$ sensitivity plot is displayed in the left panel of Fig.~\ref{fig:du_SfL2_exps}. Overall, the magnitude of the PDF pulls is slightly reduced, with the biggest change occurring for the DIS Deuterium Group (\#\textbf{4}), which is now more closely aligned with the pulls exerted by the DIS proton data (\#{\bf 3}) throughout the plotted domain in $x$.  
When considered in parallel with Fig.~\ref{fig:du_SfL2}, and in the light of the previous discussion of Fig.~\ref{fig:du_SfL2}, the left panel of Fig.~\ref{fig:du_SfL2_exps} suggests a connection between the pulls of the DIS deuteron and $\nu$-A data in fits with and without
deuteron corrections. For both groups of experiments, the interplay between the theoretical description of deuterium and heavy-nuclear data is relevant. To that end, \textbf{investigating the systematic treatment of nuclear effects for light and heavy nuclei is a critical subject for future global analyses that aim to use such data for constraining the nucleon PDFs to higher accuracy. }

A similar consideration arises for CJ. As we have  discussed, the combination of $W$-boson charge asymmetries and SLAC DIS data is strongly constraining on the $d/u$ ratio at large $x$, and the d.c.~treatment influences also the PDF pulls at smaller $x$ values (as seen in the right panels of Fig.~\ref{fig:du_SfL2}). We therefore remove these data sets from the fit to obtain the \texttt{CJ no-W\_slac}~fit shown in the right panel of Fig.~\ref{fig:du_SfL2_exps}. In this instance,
the removal of the combined $W$ and SLAC DIS data relieves tensions seen in \texttt{CJ fixed d.c.},
for $x\! <\! 0.1$. However, the large-$x$ tension between DIS deuteron and WZ Tevatron data (that now only include the $W \to \ell$ decay lepton asymmetries) remains largely intact and can in fact also be seen in the CT panel on the left of Fig.~\ref{fig:du_SfL2_exps}. It remains to be seen whether this is of experimental origin, or due to an as yet incomplete treatment of nuclear corrections in the deuteron target.

\begin{figure*}[t]
	\centering
	\includegraphics[width=0.48\textwidth]{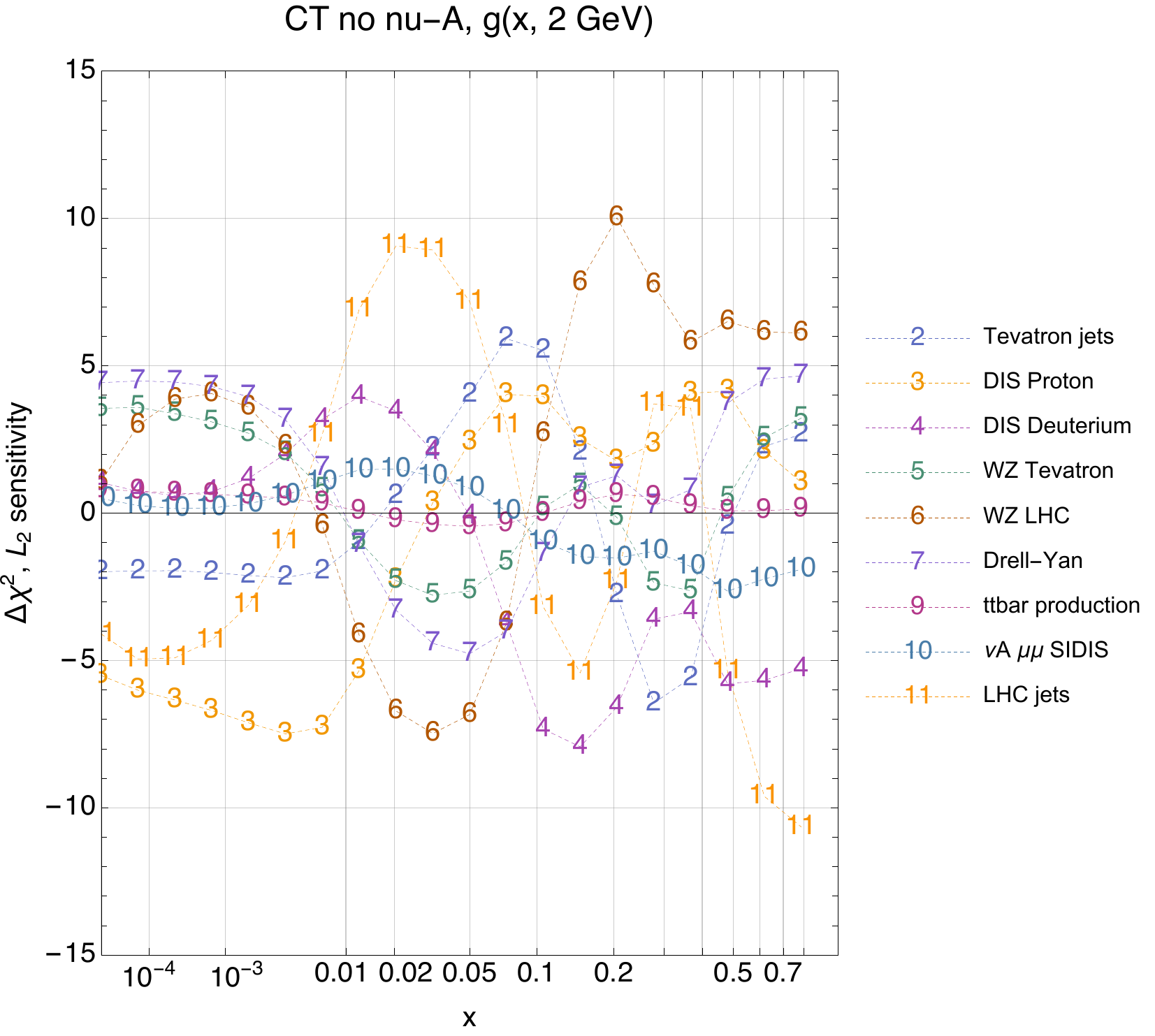} 
	\includegraphics[width=0.48\textwidth]{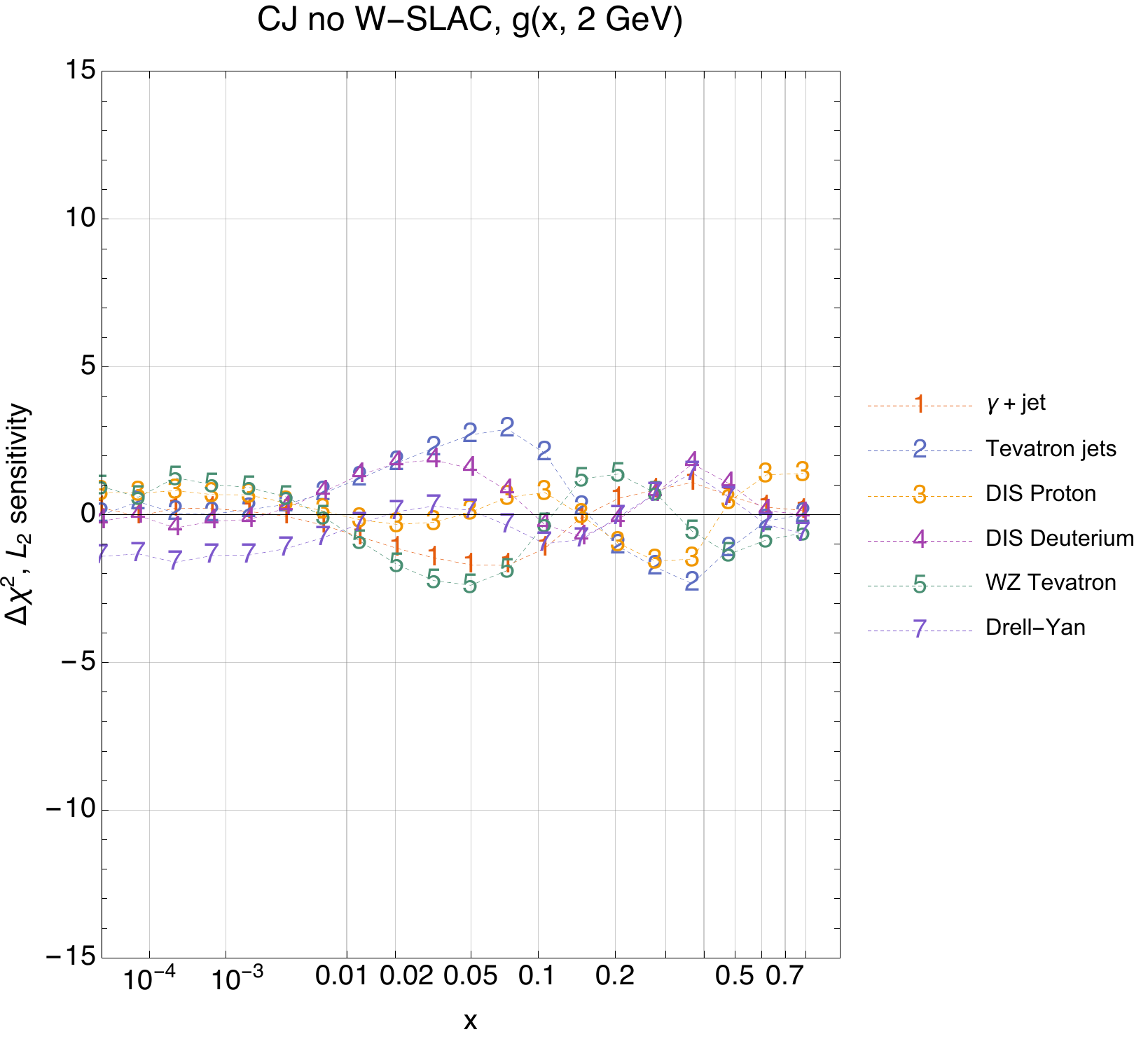} 
\caption{
    PDF pulls on $g$ at 2 GeV  after removing the inclusive $\nu$-A experiments for CT (left) and $W$ charge asymmetry and SLAC DIS data from CJ (right), with deuteron corrections fixed.
    }
\label{fig:glu_SfL2_exps}
\end{figure*}

\subsection{PDF pulls in the gluon and light-quark sea sectors}
\label{sec:gluon_pulls}

At first sight, it might seem reasonable to suppose that deuteron-structure corrections, being most sizable at high $x$ and more immediately connected to extractions
of the $d$-quark, would be relatively inconsequential for determinations of the gluon PDFs. 
In actuality, constraining the gluon PDF through DIS data requires an adequate prediction of the scale dependence of both proton and deuteron DIS data sets, with the latter simultaneously sensitive to the ($Q^2$ dependent) deuteron corrections
discussed in Sec.~\ref{sec:deut}. Moreover, the momentum sum rule requires that the changes in the total momentum fraction from the large-$x$ and small-$x$ quark and gluon PDFs compensate one another. The practical implementation
of the deuteron correction can therefore impact $g(x,Q)$ over a still broader range beyond high $x$. 

We therefore turn to an examination of the pulls on the gluon PDF in fits with and without deuteron corrections, presented in Fig.~\ref{fig:glu_SfL2}, before examining CT and CJ fits with the modified data sets in Fig.~\ref{fig:glu_SfL2_exps}.
Comparing the \texttt{CJ no d.c.}~fit in the upper-right panel of Fig.~\ref{fig:glu_SfL2} to the \texttt{CJ fixed d.c.}~fit in the middle-right panel, one sees that adding a fixed deuteron correction clearly aligns the pulls of the DIS proton group (\#{\bf 3}) and DIS deuteron group (\#{\bf 4}) on the gluon. The $x$ dependence of these pulls was effectively uncorrelated without the deuteron correction. In the presence of the fixed correction, however, they are aligned and pronounced over the whole $x$ range and are opposed mostly by the strong pull of the $WZ$ Tevatron data (Group \#{\bf 5}). 
Furthermore, after the off-shell parameters in the CJ deuteron correction are freed (lowermost panel), the tension between Groups \#{\bf 3}, \#{\bf 4}, and \#{\bf 5} is relieved, resulting in a more consistent data set, with weak pulls ($\lesssim 3$ units) everywhere. It is also interesting to note that a similar effect arose in the $d/u$ sector discussed in the previous subsection, and therefore seems to be a robust feature of fitting the deuteron corrections simultaneously with the PDF parameters.

In the two CT fits in the upper left and middle left panels of Fig.~\ref{fig:glu_SfL2}, a somewhat different pattern emerges.
Inclusion of the deuteron correction in \texttt{CT fixed d.c.} does have the effect of partially aligning the pulls of the DIS Proton
(Group \#{\bf 3}) and DIS deuteron (Group \#{\bf 4}) on the gluon,
but this effect is restricted to a narrower interval, $x\in [0.2, 0.5]$. The pulls from the other groups are relatively unaffected by the deuteron correction, although we see some realignment of the pulls from LHC $WZ$ and fixed-target Drell-Yan production experiments (Groups \#{\bf 6} and \#{\bf 7}). 
The weaker dependence of the gluon pulls in the CT fit on the deuteron correction compared to the CJ case is most likely due to the absence of the SLAC DIS data from the former fit. One can therefore investigate the effect of the removal of the SLAC data on the CJ fit. Intriguingly, as shown in the right panel of Fig.~\ref{fig:glu_SfL2_exps}, simultaneously
removing the SLAC DIS data and the Tevatron $W$-boson asymmetry data largely alleviates the competing gluon pulls, which are now smaller than those observed in the CT fit, especially from the LHC sets not included in CJ.

Clearly, the gluon pulls in the CT fit are due to the data other than the large-$x$ SLAC and $W$ production data. In particular,
in the absence of the large-$x$ SLAC and $W$ production data in the CT fit,
we notice a strong preference for a harder gluon at $x\approx 0.3$ from the $\nu$-A DIS experiments (Group \#{\bf 8}) both with and without the nuclear correction. In fact, the preference of the CDHSW $\nu$-A DIS deuteron data for a harder gluon at large $x$ had already been identified in the CT18 analysis \cite{Hou:2019efy}, although the net effect of including this data set in the CT18 fit does not exceed the PDF uncertainty. However, as the left panel of Fig.~\ref{fig:glu_SfL2_exps} shows, removing these data from the fit does not substantially alter the pulls of the remaining experiments shown in the CT plots of Fig.~\ref{fig:glu_SfL2}, which are led by the jet and $W/Z$ measurements from the LHC.
%

\section{Conclusion}
\label{sec:conc}
In this analysis, we have for the first time undertaken a
comparative analysis of two global fitting frameworks, CTEQ-JLab (CJ) and CTEQ-TEA (CT), using the
$L_2$ sensitivity statistical metric developed in Refs.~\cite{Wang:2018heo,Hobbs:2019gob}. This metric allowed our study to take advantage of the complementary strengths of the two frameworks: the extended experimental coverage and various theoretical developments implemented in the two approaches, as well as the flexible PDF parametrizations available in CT and the unique capabilities of CJ in describing low-energy and nuclear dynamics. In doing so, we made a number of technical adjustments to each framework (discussed in detail in the appendix) in order to reconcile the CT and CJ treatment of PDF uncertainties and thereby render them sufficiently similar to be meaningfully juxtaposed.

We have, in particular, concentrated on evaluating the impact on PDF determinations of nuclear corrections which take into account the two-baryon structure of the deuteron. In fact, as discussed in Sec.~\ref{sec:intro}, DIS and Drell-Yan measurements on deuterium are very informative in providing flavor separation of $d$-type quarks from other parton flavors (under an assumption of nucleon charge symmetry). At the same time, the introduction of deuterium data into proton PDF fits brings along its own uncertainties associated with nuclear and power-suppressed effects. Global analyses take diverse approaches in handling the deuteron and heavy-nuclear effects, from selection of the least affected experimental data \cite{Hou:2019efy,Ball:2017nwa,Bailey:2020ooq}, to including some fixed~\cite{Hou:2019efy} or free \cite{Accardi:2016qay,Bailey:2020ooq} nuclear corrections and performing Bayesian marginalization \cite{Ball:2020xqw,Ball:2018twp} with respect to the nuclear parameters. It is therefore important to understand the role of the deuteron corrections in a controlled setting, by isolating them from other factors that affect the existing PDF ensembles at comparable levels.\footnote{See examples in \cite{Hou:2019efy} for comparable variations in $S_{f,L2}(E)$ caused by various assumptions.} 

By examining the fitted PDFs and resulting PDF pulls of experimental data under several
theoretical scenarios for the treatment of deuteron corrections,  in Sec.~\ref{sec:results} we have gathered a substantial number of results that clarify these questions.
We reiterate here our overriding conclusions based on this investigation:
\begin{itemize}
    \item 
    While the compilation of $\chi^2$ values in Sec.~\ref{sec:OverallAgreement} indicates good {\it global} agreement of CJ and CT NLO theoretical predictions with deuteron data sets, the $L_2$ sensitivity additionally provides insights about {\it local} compatibility of fitted experiments in an $x$-dependent fashion. In the case of CJ, the model of deuteron dynamic effects is crucial for the description of the informative low-$Q$ DIS data from SLAC. The dependence on the deuteron correction is reduced in the CT analysis with its more conservative cut on $W^2$. Still, including the CJ deuteron correction reduces $\chi^2$ for the aggregated DIS-deuteron experiments by about $\sim\!\!14$ units and also reduces the cumulative $\chi^2$ for vector-boson production data sets by several tens of units, with the modifications potentially comparable to the NNLO scale dependence in an analysis like CT. Another effect of the deuteron correction is to alleviate the competing pulls of deuteron and some other experiments in the large-$x$ region.
    \item The impact of a fixed deuteron correction is particularly evident in the high-$x$
    distribution for the $d$-quark, or the associated $d/u$ PDF ratio. A number of commonalities exist between the CT and CJ analyses in the qualitative effect of this
    correction on the extracted high-$x$ PDFs. The fixed deuteron correction of Sec.~\ref{sec:deut}
    generally leads to the suppression of the $d/u$ ratio at $x\! >\! 0.5$ relative to the scenario without the deuteron correction. 
    \item Due to the influence of sum rules and nontrivial correlations among fitted PDFs of
    different parton flavors, deuteron corrections to DIS structure functions at large $x$ can
    have important secondary effects on, {\it e.g.}, the gluon or sea-quark PDFs over a range
    of $x$, as well as the $d_\mathit{val}$ distribution at lower $x\! \sim\! 0.03$ of relevance
    to precision studies in the electroweak sector.
    \item In both fitting frameworks, the modifications caused by the deuteron-structure corrections are moderated by the inclusion of some non-deuteron data sets; for CT, these are inclusive neutrino DIS data on heavy nuclear targets, while for CJ, a combination of high-$x$ SLAC DIS data and reconstructed boson-level Tevatron charge asymmetry requires special attention. Disentangling the interplay among these fitted experiments will require a further study at NNLO accuracy, including additional investigation of the implementation of theoretical corrections for nuclear data sets (including both deuteron and heavier targets) and the treatment of $W/Z$ data.
\end{itemize}

As the drive to realize next-generation accuracy in PDF analyses gains speed with preparations for the High-Luminosity LHC \cite{ApollinariG.:2017ojx}, Electron-Ion Collider \cite{Accardi:2012qut}, and Long-Baseline Neutrino Facility \cite{Acciarri:2015uup}, we recommend consideration of deuteron corrections and broader nuclear effects in PDF analyses, as well as continued phenomenological and model-based studies \cite{Melnitchouk:1994rv,Bickerstaff:1985ax,Melnitchouk:1992eu,Miller:2013hla} of deuteron structure in parallel. Deuteron effects will become particularly unavoidable with increasing
PDF precision and in PDF-benchmarking studies, most obviously for the $d$-PDF at $x\! \gtrsim\! 0.2$ and beyond, but, ultimately, for consistency of the extracted gluon density and over a widening
range of $x$.  Consideration of the parton-level violation of the charge symmetry in the deuteron \cite{Hobbs:2011vy} may become relevant as precision goals advance still further. As emphasized in Sec.~\ref{sec:intro} and \ref{sec:weakmix},
the achievement of ultimate precision in tests of the SM in the electroweak sector will partly depend upon the successful treatment of the issues described in this work.

\begin{acknowledgements}
We appreciate insightful discussions with Aurore Courtoy, Lucian Harland-Lang, Robert Thorne, and our colleagues within the CTEQ-JLab and CTEQ-TEA collaborations, with special thanks to Jun Gao, Shujie Li, Wally Melnitchouk, and C.-P. Yuan. Funding agency and grant acknowledgements appear
on the first page of this manuscript.
\end{acknowledgements}


\section{Appendix: Comparison procedure and technical details}
\label{app:directions}

To meaningfully compare two distinct global fits on a common footing as done in this article, it has been necessary to  conciliate their methodologies, cf.~Sec.~\ref{sec:mods}. Part of this consisted in making a few technical adjustments to ensure the mutual compatibility of the CJ and CT computations of the $L_2$ sensitivity introduced in Sec.~\ref{sec:L2} and employed in Sec.~\ref{sec:results}. In this Appendix, we provide more detail about these adjustments.

\begin{figure*}[tb]
\hspace*{-10pt}	\includegraphics[height=0.2325\textheight]{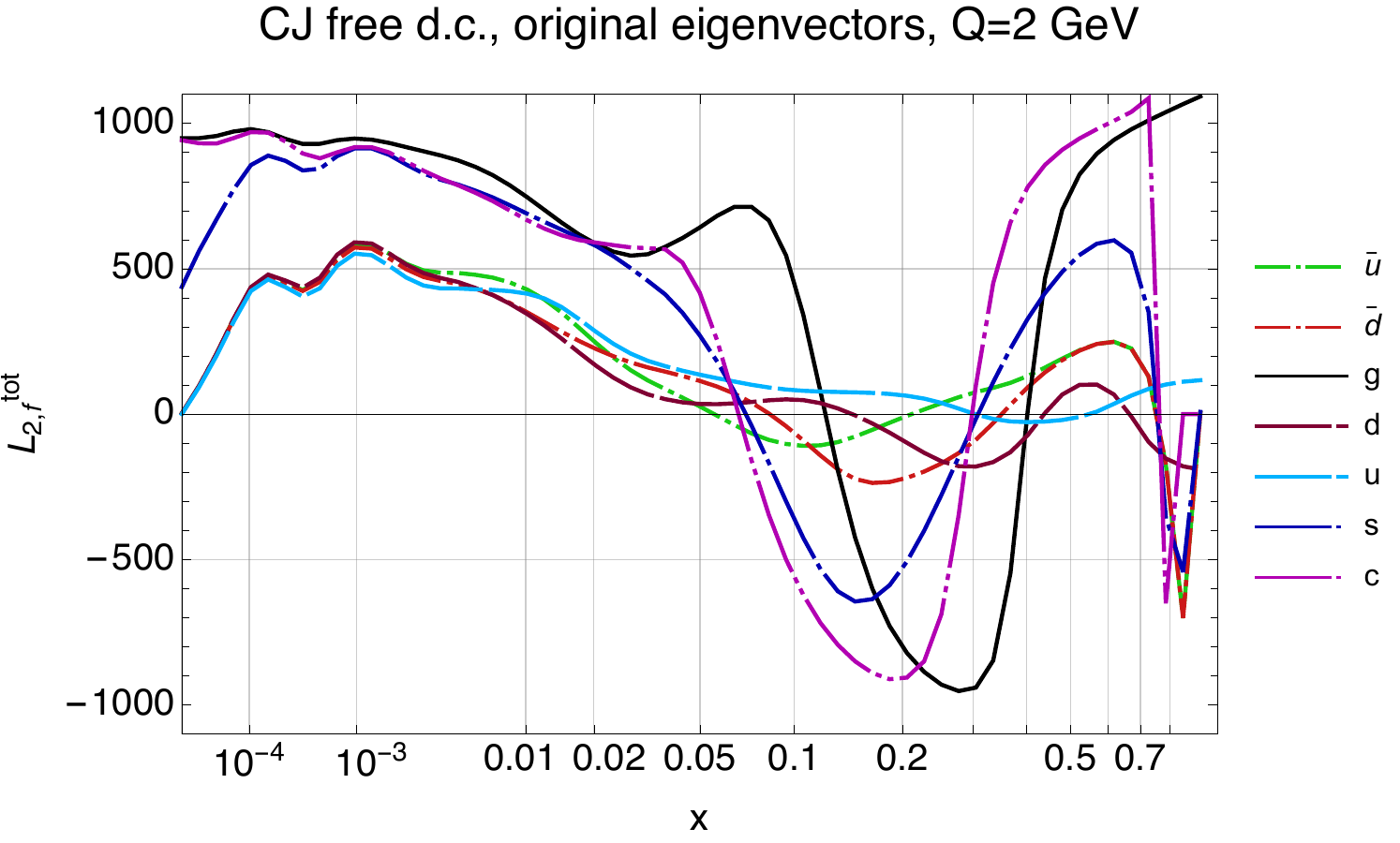} \quad
	\includegraphics[height=0.23\textheight]{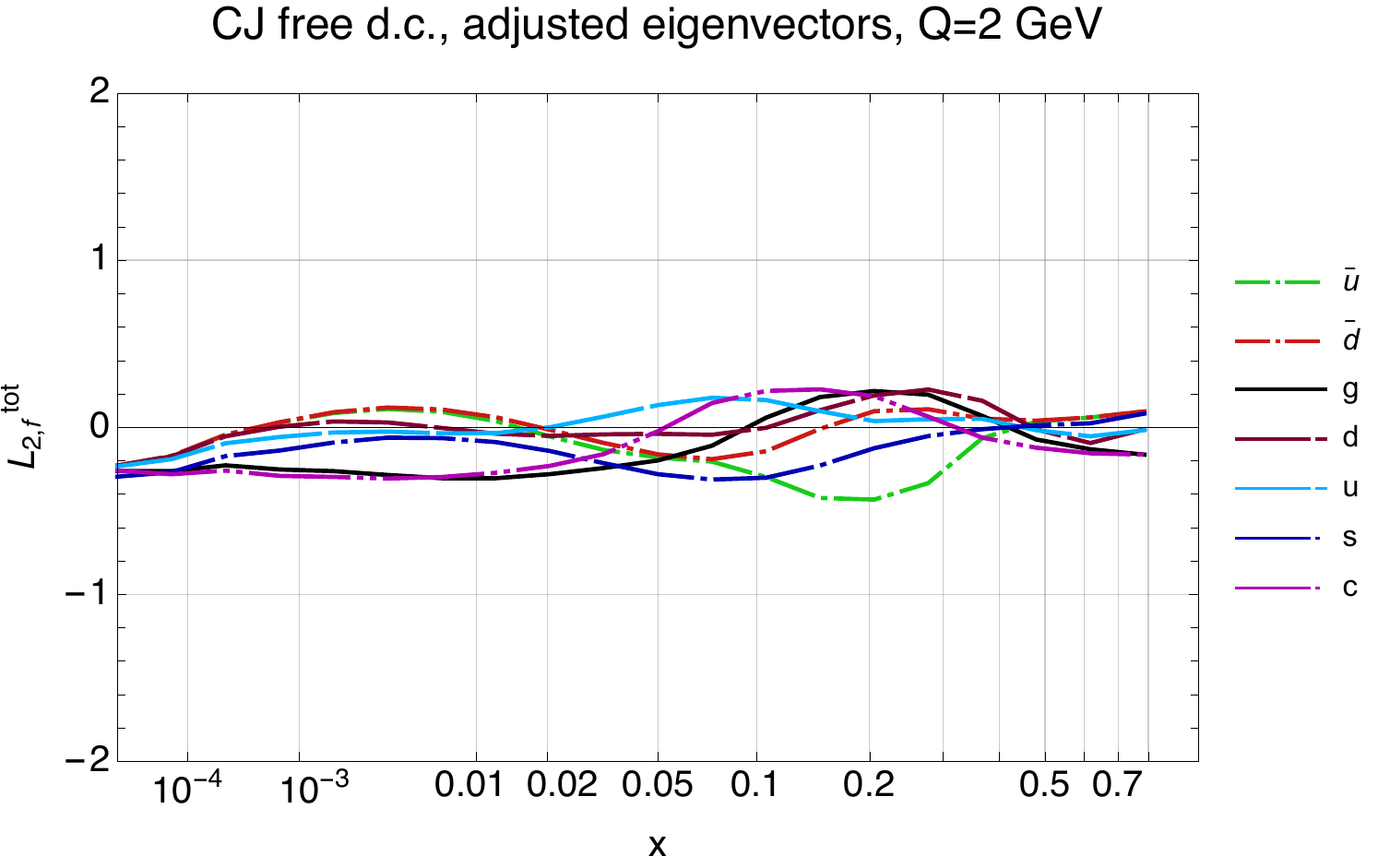}
	\includegraphics[height=0.23\textheight]{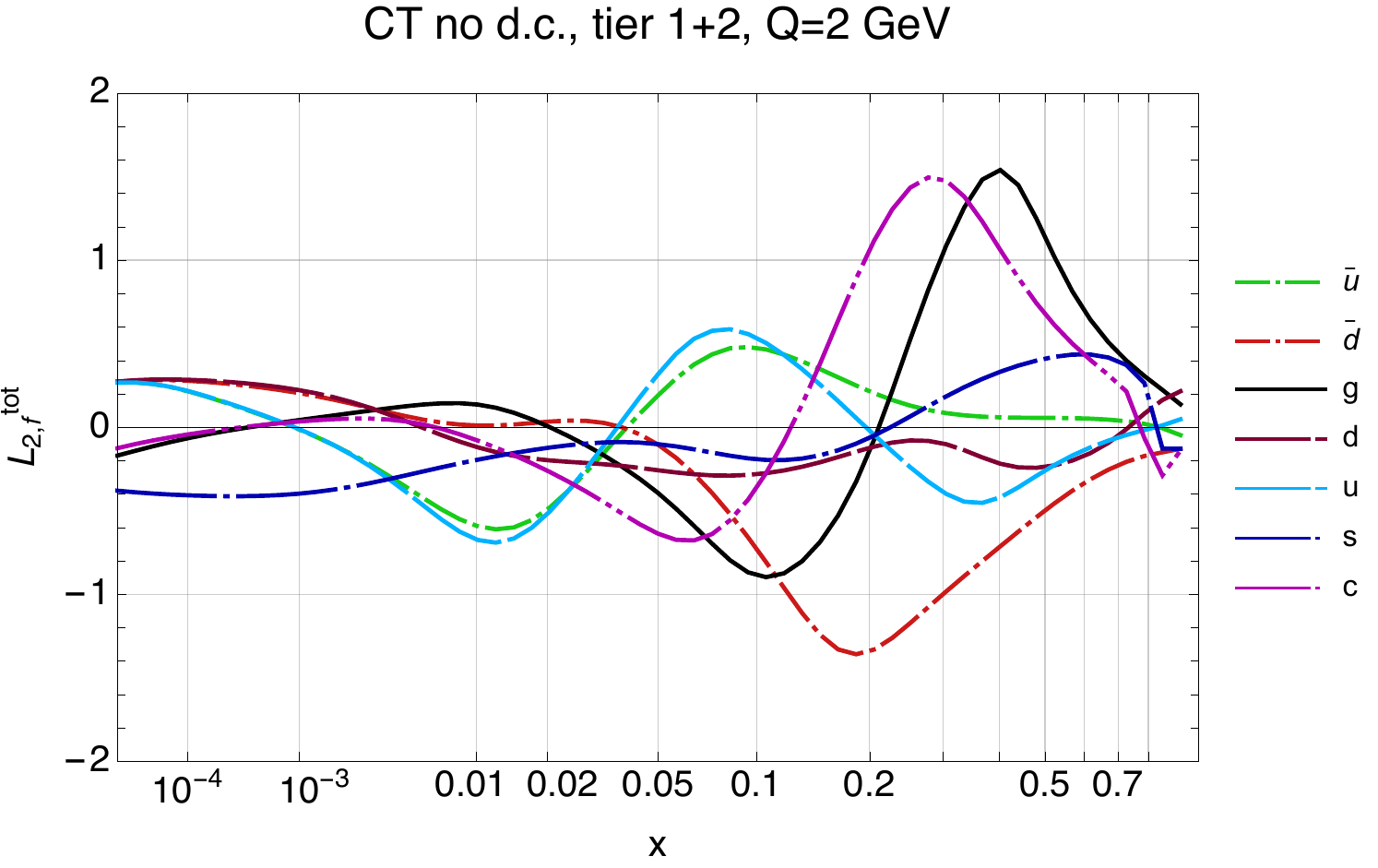} \quad
	\includegraphics[height=0.23\textheight]{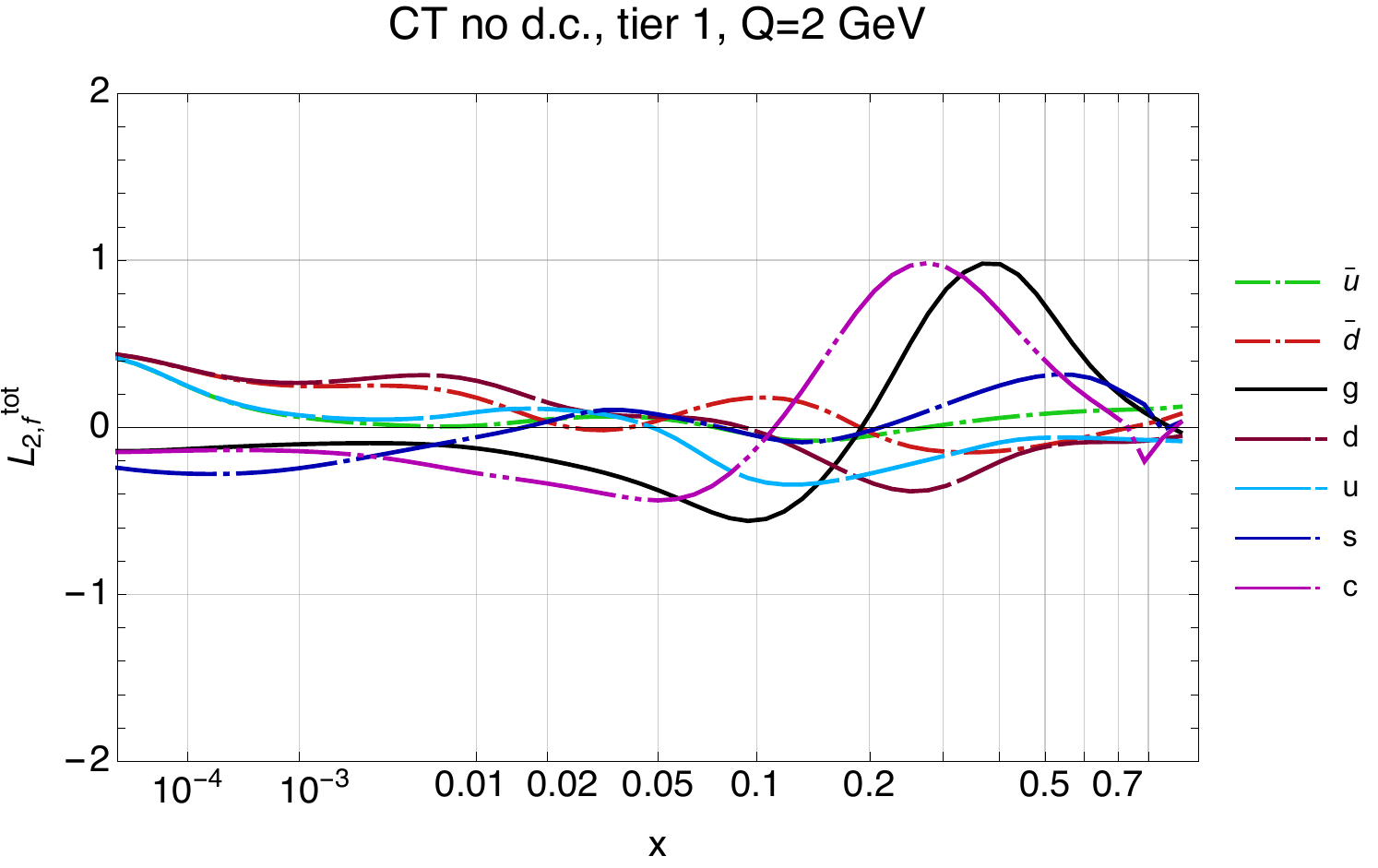} 
\caption{
Total sensitivities, $L^\mathrm{tot}_2$, summed over all experiments as in Eq.~\ref{L2tot}. 
The top left panel shows $L^\mathrm{tot}_2$ for the \texttt{CJ free d.c.}~PDF ensemble with the standard CJ15 computation of the Hessian eigenvectors. 
The top right panel illustrates the same result after adjusting each \texttt{CJ free d.c.}~eigenvector to have $\Delta\chi^2=T^2=10$ in both the positive and negative direction, thereby correcting the departures from Gaussianity observed for poorly constrained eigenvectors in the standard CJ Hessian analysis.
In the bottom row, the left panel shows $L^\mathrm{tot}_2$ 
for the \texttt{CT no d.c.}~PDF ensemble computed with the CT18 two-tier uncertainty constraints with $T^2=10$. On the right, the same but including only the Tier-1 uncertainty constraint that do not introduce \textit{a priori} deviations from the Gaussianity condition.
}
\label{fig:L2tot}
\end{figure*}

The $L_2$ sensitivity can be interpreted as a fast approximation, based on the Hessian error formalism \cite{Stump:2001gu}, of the $\Delta \chi^2_E$ shifts contained in the LM scans shown in the panels of Fig.~\ref{fig:qval_LM}. For that reason, we expect the sum of $L_2$ sensitivities 
over all fitted experiments $E$ to vanish for each parton flavor $f$; {\it i.e.},
\begin{equation}
   L^\mathrm{tot}_{2,f}\, \equiv\, \sum_E S_{f,L2}(E)\, \approx\, 0\ \quad \forall f \ 
   .
\label{L2tot}
\end{equation}
This desired result --- that the ``total'' $L_2^
\mathrm{tot}$ sensitivities approximately sum to zero for all flavors within a given fit --- represents an ideal scenario in which all data sets demonstrate full mutual compatibility,
and uncertainties are small enough to validate a quadratic approximation for the $\chi^2$ function  
around the central PDF parameters, $\vec{a}_0$. In practice, however, neither condition was perfectly
realized when generating the Hessian eigenvector sets, causing $L_2^\mathrm{tot}$ to deviate from zero for both the CJ15 and CT18 eigenvector ensembles. We illustrate the graphs of $L_2^\mathrm{tot}$ obtained with the standard eigenvector computations, for tolerance $T^2=10$, in the left panels of Fig.~\ref{fig:L2tot}. In the upper figure, the total sensitivities in the standard \texttt{CJ free d.c.}~show very pronounced deviations from zero. In the lower figure, we see milder but not entirely negligible deviations from zero in the counterpart default \texttt{CT no d.c.}~fit. 

\begin{table*}[tbh]
\center
\begin{tabular}{c|c|c|l}
      &  $\delta \tau^E$  &  Expt ID  & Experiment \\ 
    \hline
    1  & 37.2555             &  160      & HERA run I+II  \\ 
    2  &  32.6485            &  204      & E866 proton proton  Drell-Yan process   \\
    3  &  28.1273            &  545      & CMS 8 TeV 19.7 $\mathrm{fb}^{-1}$, single incl. jet cross sec., R = 0.7  \\
    4  &  24.7211            &  250      & LHCb 8 TeV 2.0 $\mathrm{fb}^{-1}$ W/Z cross sec.   \\
    5  &  24.2715            &  101      & BCDMS $F_2^p$   \\
    6  &  20.5027            &  542      & CMS 7 TeV 5 $\mathrm{fb}^{-1}$, single incl. jet cross sec., R = 0.7 \\
    7  &  19.1007            & 245      &  LHCb 7 TeV 1.0 $\mathrm{fb}^{-1}$ W/Z forward rapidity cross sec.    \\
    8  & 18.9541            & 102       & BCDMS $F_2^d$ \\
    9  & 17.935             & 203       & E866f Drell-Yan process \\
    10 & 17.5018            & 544       &ATLAS 7 TeV 4.5 $\mathrm{fb}^{-1}$, single incl. jet cross sec., R = 0.6 \\
    ...&  ...                &  ...      & ...       \\
    39 & 0.446136            &  145      & H1 $\sigma_r^b$  \\
   \hline
\end{tabular}
\caption{Impact of a given experiment on the difference between the Tier-1+2 and Tier-1 CT total sensitivities, quantified by the $\delta\tau_E$ distance defined in Eq.~\eqref{eq:CT18-L1distance}. The first 3 experiments are well separated from the remaining ones, which are more closely spaced.}
\label{tab:exp_cat_CT}
\end{table*}

We have identified the sources of this behavior and remedied this situation as follows.
\begin{itemize}
\item
\textbf{CJ}:
The standard CJ PDF error sets are obtained by the Hessian analysis around the best-fit PDF parameters, $\vec{a}_0$, with each eigenvector scaled by a given tolerance factor $T$ to nominally produce an increase of $T^2$ above the minimum in the $\chi^2$ function ($T=1.646$ in the CJ15 analysis \cite{Accardi:2016qay}, $T=\sqrt{10}$ in this paper).
For the computation of the sensitivities, the $\chi^2$ function is instead scanned along {\it every} eigenvector starting from the best-fit parameters, $\vec{a}_0$, until parameters $\vec a_i$ are found in the plus- and minus-directions such that 
\begin{align}
    \Delta\chi^2(\vec a_{2i+1}) &= \Delta\chi^2(\vec a_{2i}) = T^2 \nonumber \\
    &\hspace*{2cm} \forall\ i = 1, \ldots, N_{par} \ .
\end{align}
This way, the error PDFs correspond exactly to a given likelihood $\mathcal{L} \propto e^{-T^2/2}$, and, by construction, $L_2^\mathrm{tot}=0$ except for numerical uncertainties, see Eq.~\eqref{eq:corr-def}.
The total sensitivities after this adjustment of the eigenvector excursions are shown in the upper right panel of Fig.~\ref{fig:L2tot}. All $L_2^\mathrm{tot}$ values are now well within $\pm 1$ unit from zero.
\item 
\textbf{CT}: 
The published general-purpose CT fits \cite{Hou:2019efy,Dulat:2015mca,Lai:2010vv} apply a two-tier procedure to determine the published error bands. Excursions along each eigenvector are constrained both by the Tier-1 penalty imposed by the increase of the global $\chi^2$ above $T^2$ units, and by the dynamical Tier-2 penalty based on effective Gaussian variables that ensure that no single-experiment $\chi^2_E$ value increases above its best-fit value by more than its uncertainty at the requested confidence level. The Tier-2 procedure may lead to unequal excursions in the plus- and minus-directions along a given eigenvector, therefore introducing a departure from the condition for $L_2^\mathrm{tot}=0$. For this reason, we revert to the Tier-1 PDF error determination, which nominally satisfies the desired condition on $L_2^\mathrm{tot}$. The resulting total sensitivities are compared to the standard CT18 calculation in the lower row of Fig.~\ref{fig:L2tot}.\footnote{As a technical observation, both Tier-1 and Tier-1+2 predictions in this case are computed assuming $T^2=10$, without modifying the excursions of the effective Gaussian parameters \cite{Lai:2010vv} in the Tier-2 penalty compared to the default CT18 setup.} All total sensitivities with the Tier-1 error sets are within $\pm 1$ unit from zero, with a somewhat enhanced deviation observed for the gluon PDF and for the charm PDF that follows it. 
\end{itemize}

The very bad CJ $L_2^\mathrm{tot}$ obtained before the adjustment in the upper left panel of Fig.~\ref{fig:L2tot}
was traced primarily to a substantial deviation from Gaussianity observed along a small number of eigenvector directions. This can happen in a fit where certain linear combinations of parameters are poorly constrained by the data, especially when exploiting an extended parametrization such as in the CJ case, where additional parameters are included to allow for a non-vanishing $d/u$ quark at large $x$, to describe off-shell deformation of bound nucleons in a deuteron target, and to fit higher-twist corrections to the standard twist-2 calculation of electron-nucleon DIS. 
In fact, we have verified that the very large values of the total sensitivity are primarily driven by the combined HERA data set, which accounts for most of the calculated sensitivities across all parton flavors, with a secondary large contribution provided by the D\O~$W$-asymmetry measurements. 

In the CT analysis, both the Tier-1+2 and Tier-1 total sensitivities in the lower row of Fig.~\ref{fig:L2tot} are of a natural order, being generally $\approx\! 0$, especially in the
Tier-1 calculation at lower-right. This is aside from the high-$x$ $L_2^\mathrm{tot}$ values for the $g$- and $c$-PDFs, which are somewhat larger than in the CJ15 adjusted analysis,
but still $\lesssim\! 1$ in this case.
The Tier-1 calculation is less affected by non-Gaussianities and produces a smaller total sensitivity than the default Tier-1+2 analysis. 
As with the CJ analysis, it is interesting to identify the key experiments that affect the differences between the CT Tier-1+2 and Tier-1 total sensitivities. 
Since these differences are comparatively small, we use a different procedure than in the CJ case, where it was sufficient to analyze the size of the experiment-by-experiment
sensitivities before parameter adjustment.
We notice that the departure from $L_2^\mathrm{tot}\approx 0$ is determined by the $\chi^2$ {\it imbalance},
\begin{equation}
    \tau^E_i \equiv [\chi^2 _{2i}]^{E} - [\chi^2 _{2i-1}]^{E},\ \ \ i = 1, \ldots, N_\mathit{par}\ .
\end{equation}
between the PDF error sets in each positive and negative eigenvector direction, see Eq.~\eqref{eq:corr-def}. We can therefore quantify the impact of a given experiment on the observed differences between the total sensitivities for the Tier-1+2 and Tier-1 CT analysis by calculating
\begin{equation}
    \delta \tau^{E} = |\overrightarrow{\tau}^{E}_\mathrm{Tier-1+2} - \overrightarrow{\tau}^{E}_\mathrm{Tier-1}|\ .
\label{eq:CT18-L1distance}
\end{equation}
Note that we could also have considered other metrics, such as the quadratic distance between the respective $\chi^2$ imbalances. However, our goal here is to identify the experiments with the largest impact on the reduction of the total sensitivity, rather than performing a detailed quantitative analysis and ranking of each experiment. 
 
We collect the largest values of the $\delta\tau^E$ metric in Table~\ref{tab:exp_cat_CT}, where experiments have been ordered from the highest to lowest impact. Intriguingly, the most affected experiment in the CT case is also the HERA I+II combined data set. We believe that the underlying reason(s) for this commonality with the CJ analysis are the high precision and large kinematic coverage of the HERA data, which are therefore sensitive to small variations in the PDFs and subject to secondary constraints, such as those imposed on the gluon distribution by PDF sum rules. 


\bibliographystyle{spphys}
\bibliography{myreferences}

\begin{thebibliography}{100}
\providecommand{\url}[1]{{#1}}
\providecommand{\urlprefix}{URL }
\expandafter\ifx\csname urlstyle\endcsname\relax
  \providecommand{\doi}[1]{DOI \discretionary{}{}{}#1}\else
  \providecommand{\doi}{DOI \discretionary{}{}{}\begingroup
  \urlstyle{rm}\Url}\fi

\bibitem{Stump:2001gu}
D.~Stump, J.~Pumplin, R.~Brock, D.~Casey, J.~Huston, J.~Kalk, H.L. Lai, W.K.
  Tung, Phys. Rev. \textbf{D65}, 014012 (2001).
\newblock \doi{10.1103/PhysRevD.65.014012}

\bibitem{Nadolsky:2001yg}
P.M. Nadolsky, Z.~Sullivan, eConf \textbf{C010630}, P510 (2001)

\bibitem{Nadolsky:2008zw}
P.M. Nadolsky, H.L. Lai, Q.H. Cao, J.~Huston, J.~Pumplin, D.~Stump, W.K. Tung,
  C.P. Yuan, Phys. Rev. \textbf{D78}, 013004 (2008).
\newblock \doi{10.1103/PhysRevD.78.013004}

\bibitem{Armbruster:2018}
A.~Armbruster,  (2018).
\newblock Private communication

\bibitem{Jimenez-Delgado:2013sma}
P.~Jimenez-Delgado, W.~Melnitchouk, J.F. Owens, J. Phys. G \textbf{40}, 093102
  (2013).
\newblock \doi{10.1088/0954-3899/40/9/093102}

\bibitem{Gao:2017yyd}
J.~Gao, L.~Harland-Lang, J.~Rojo, Phys. Rept. \textbf{742}, 1 (2018).
\newblock \doi{10.1016/j.physrep.2018.03.002}

\bibitem{Kovarik:2019xvh}
K.~Kova\v{r}\'\i{}k, P.M. Nadolsky, D.E. Soper, Rev. Mod. Phys. \textbf{92}(4),
  045003 (2020).
\newblock \doi{10.1103/RevModPhys.92.045003}

\bibitem{Ethier:2020way}
J.J. Ethier, E.R. Nocera, Ann. Rev. Nucl. Part. Sci. \textbf{70}, 43 (2020).
\newblock \doi{10.1146/annurev-nucl-011720-042725}

\bibitem{Accardi:2016qay}
A.~Accardi, L.T. Brady, W.~Melnitchouk, J.F. Owens, N.~Sato, Phys. Rev.
  \textbf{D93}(11), 114017 (2016).
\newblock \doi{10.1103/PhysRevD.93.114017}

\bibitem{Alekhin:2017kpj}
S.~Alekhin, J.~Blümlein, S.~Moch, R.~Placakyte, Phys. Rev. \textbf{D96}(1),
  014011 (2017).
\newblock \doi{10.1103/PhysRevD.96.014011}

\bibitem{Hou:2019efy}
T.J. Hou, et~al., Phys. Rev. D \textbf{103}(1), 014013 (2021).
\newblock \doi{10.1103/PhysRevD.103.014013}

\bibitem{Ball:2017nwa}
R.D. Ball, et~al., Eur. Phys. J. \textbf{C77}(10), 663 (2017).
\newblock \doi{10.1140/epjc/s10052-017-5199-5}

\bibitem{Bailey:2020ooq}
S.~Bailey, T.~Cridge, L.A. Harland-Lang, A.D. Martin, R.S. Thorne, Eur. Phys.
  J. C \textbf{81}(4), 341 (2021).
\newblock \doi{10.1140/epjc/s10052-021-09057-0}

\bibitem{Accardi:2016muk}
A.~Accardi, PoS \textbf{DIS2015}, 001 (2015).
\newblock \doi{10.22323/1.247.0001}

\bibitem{Wang:2018heo}
B.T. Wang, T.J. Hobbs, S.~Doyle, J.~Gao, T.J. Hou, P.M. Nadolsky, F.I. Olness,
  Phys. Rev. \textbf{D98}(9), 094030 (2018).
\newblock \doi{10.1103/PhysRevD.98.094030}

\bibitem{Hobbs:2019gob}
T.~Hobbs, B.T. Wang, P.M. Nadolsky, F.I. Olness, Phys. Rev. D \textbf{100}(9),
  094040 (2019).
\newblock \doi{10.1103/PhysRevD.100.094040}

\bibitem{Dulat:2015mca}
S.~Dulat, T.J. Hou, J.~Gao, M.~Guzzi, J.~Huston, P.~Nadolsky, J.~Pumplin,
  C.~Schmidt, D.~Stump, C.P. Yuan, Phys. Rev. \textbf{D93}(3), 033006 (2016).
\newblock \doi{10.1103/PhysRevD.93.033006}

\bibitem{Fu:2020mxl}
Y.~Fu, S.~Yang, M.~Liu, L.~Han, T.J. Hou, C.~Schmidt, C.~Wang, C.P. Yuan, Chin.
  Phys. C \textbf{45}(5), 053001 (2021).
\newblock \doi{10.1088/1674-1137/abe36d}

\bibitem{Khalek:2018mdn}
R.~Abdul~Khalek, S.~Bailey, J.~Gao, L.~Harland-Lang, J.~Rojo, Eur. Phys. J. C
  \textbf{78}(11), 962 (2018).
\newblock \doi{10.1140/epjc/s10052-018-6448-y}

\bibitem{Hou:2019gfw}
T.J. Hou, Z.~Yu, S.~Dulat, C.~Schmidt, C.P. Yuan, Phys. Rev. D
  \textbf{100}(11), 114024 (2019).
\newblock \doi{10.1103/PhysRevD.100.114024}

\bibitem{Pumplin:2001ct}
J.~Pumplin, D.~Stump, R.~Brock, D.~Casey, J.~Huston, J.~Kalk, H.L. Lai, W.K.
  Tung, Phys. Rev. \textbf{D65}, 014013 (2001).
\newblock \doi{10.1103/PhysRevD.65.014013}

\bibitem{Arneodo:1996qe}
M.~Arneodo, et~al., Nucl. Phys. \textbf{B483}, 3 (1997).
\newblock \doi{10.1016/S0550-3213(96)00538-X}

\bibitem{Abramowicz:2015mha}
H.~Abramowicz, et~al., Eur. Phys. J. C \textbf{75}(12), 580 (2015).
\newblock \doi{10.1140/epjc/s10052-015-3710-4}

\bibitem{Virchaux:1991jc}
M.~Virchaux, A.~Milsztajn, Phys. Lett. B \textbf{274}, 221 (1992).
\newblock \doi{10.1016/0370-2693(92)90527-B}

\bibitem{virchaux1992measurement}
M.~Virchaux, A.~Milsztajn, Physics Letters B \textbf{274}(2), 221 (1992)

\bibitem{Benvenuti:1989rh}
A.C. Benvenuti, et~al., Phys. Lett. \textbf{B223}, 485 (1989).
\newblock \doi{10.1016/0370-2693(89)91637-7}

\bibitem{Benvenuti:1989fm}
A.C. Benvenuti, et~al., Phys. Lett. \textbf{B237}, 592 (1990).
\newblock \doi{10.1016/0370-2693(90)91231-Y}

\bibitem{Seligman:1997mc}
W.~Seligman, et~al., Phys. Rev. Lett. \textbf{79}, 1213 (1997).
\newblock \doi{10.1103/PhysRevLett.79.1213}

\bibitem{Holt:2010vj}
R.J. Holt, C.D. Roberts, Rev. Mod. Phys. \textbf{82}, 2991 (2010).
\newblock \doi{10.1103/RevModPhys.82.2991}

\bibitem{Roberts:2013mja}
C.D. Roberts, R.J. Holt, S.M. Schmidt, Phys. Lett. B \textbf{727}, 249 (2013).
\newblock \doi{10.1016/j.physletb.2013.09.038}

\bibitem{Lin:2017snn}
H.W. Lin, et~al., Prog. Part. Nucl. Phys. \textbf{100}, 107 (2018).
\newblock \doi{10.1016/j.ppnp.2018.01.007}

\bibitem{Lin:2020rut}
M.~Constantinou, et~al.
\newblock {Parton distributions and lattice QCD calculations: toward 3D
  structure} (2020)

\bibitem{Courtoy:2020fex}
A.~Courtoy, P.M. Nadolsky, Phys. Rev. D \textbf{103}(5), 054029 (2021).
\newblock \doi{10.1103/PhysRevD.103.054029}

\bibitem{Brodsky:1973kr}
S.J. Brodsky, G.R. Farrar, Phys. Rev. Lett. \textbf{31}, 1153 (1973).
\newblock \doi{10.1103/PhysRevLett.31.1153}

\bibitem{Brodsky:1974vy}
S.J. Brodsky, G.R. Farrar, Phys. Rev. D \textbf{11}, 1309 (1975).
\newblock \doi{10.1103/PhysRevD.11.1309}

\bibitem{Farrar:1975yb}
G.R. Farrar, D.R. Jackson, Phys. Rev. Lett. \textbf{35}, 1416 (1975).
\newblock \doi{10.1103/PhysRevLett.35.1416}

\bibitem{Ball:2020xqw}
R.D. Ball, E.R. Nocera, R.L. Pearson, Eur. Phys. J. C \textbf{81}(1), 37
  (2021).
\newblock \doi{10.1140/epjc/s10052-020-08826-7}

\bibitem{Martin:2009iq}
A.D. Martin, W.J. Stirling, R.S. Thorne, G.~Watt, Eur. Phys. J. \textbf{C63},
  189 (2009).
\newblock \doi{10.1140/epjc/s10052-009-1072-5}

\bibitem{Eskola:2016oht}
K.J. Eskola, P.~Paakkinen, H.~Paukkunen, C.A. Salgado, Eur. Phys. J.
  \textbf{C77}(3), 163 (2017).
\newblock \doi{10.1140/epjc/s10052-017-4725-9}

\bibitem{Kovarik:2015cma}
K.~Kova\v{r}\'{\i}k, et~al., Phys. Rev. \textbf{D93}(8), 085037 (2016).
\newblock \doi{10.1103/PhysRevD.93.085037}

\bibitem{Deflorian:2011fp}
D.~de~Florian, R.~Sassot, P.~Zurita, M.~Stratmann, Phys. Rev. \textbf{D85},
  074028 (2012).
\newblock \doi{10.1103/PhysRevD.85.074028}

\bibitem{Hirai:2007sx}
M.~Hirai, S.~Kumano, T.H. Nagai, Phys. Rev. \textbf{C76}, 065207 (2007).
\newblock \doi{10.1103/PhysRevC.76.065207}

\bibitem{Khanpour:2016pph}
H.~Khanpour, S.~Atashbar~Tehrani, Phys. Rev. \textbf{D93}(1), 014026 (2016).
\newblock \doi{10.1103/PhysRevD.93.014026}

\bibitem{AbdulKhalek:2019mzd}
R.~Abdul~Khalek, J.J. Ethier, J.~Rojo, Eur. Phys. J. \textbf{C79}(6), 471
  (2019).
\newblock \doi{10.1140/epjc/s10052-019-6983-1}

\bibitem{Alekhin:2017fpf}
S.I. Alekhin, S.A. Kulagin, R.~Petti, Phys. Rev. D \textbf{96}(5), 054005
  (2017).
\newblock \doi{10.1103/PhysRevD.96.054005}

\bibitem{Kulagin:2004ie}
S.A. Kulagin, R.~Petti, Nucl. Phys. A \textbf{765}, 126 (2006).
\newblock \doi{10.1016/j.nuclphysa.2005.10.011}

\bibitem{osti_879078}
D.A. Mason,   (2006).
\newblock \doi{10.2172/879078}.
\newblock \urlprefix\url{https://www.osti.gov/biblio/879078}

\bibitem{Goncharov:2001qe}
M.~Goncharov, et~al., Phys. Rev. D \textbf{64}, 112006 (2001).
\newblock \doi{10.1103/PhysRevD.64.112006}

\bibitem{Kayis_Topaksu_2008}
A.~Kayis-Topaksu, G.~Önengüt, R.~van Dantzig, M.~de~Jong, R.~Oldeman,
  M.~Güler, S.~Kama, U.~Köse, M.~Serin-Zeyrek, P.~Tolun, et~al., Nuclear
  Physics B \textbf{798}(1-2), 1–16 (2008).
\newblock \doi{10.1016/j.nuclphysb.2008.02.013}.
\newblock \urlprefix\url{http://dx.doi.org/10.1016/j.nuclphysb.2008.02.013}

\bibitem{Frankfurt:1988nt}
L.L. Frankfurt, M.I. Strikman, Phys. Rept. \textbf{160}, 235 (1988).
\newblock \doi{10.1016/0370-1573(88)90179-2}

\bibitem{Arneodo:1992wf}
M.~Arneodo, Phys. Rept. \textbf{240}, 301 (1994).
\newblock \doi{10.1016/0370-1573(94)90048-5}

\bibitem{Kopeliovich:2012kw}
B.Z. Kopeliovich, J.G. Morfin, I.~Schmidt, Prog. Part. Nucl. Phys. \textbf{68},
  314 (2013).
\newblock \doi{10.1016/j.ppnp.2012.09.004}

\bibitem{Malace:2014uea}
S.~Malace, D.~Gaskell, D.W. Higinbotham, I.~Cloet, Int. J. Mod. Phys. E
  \textbf{23}(08), 1430013 (2014).
\newblock \doi{10.1142/S0218301314300136}

\bibitem{Berge:1989hr}
J.P. Berge, et~al., Z. Phys. \textbf{C49}, 187 (1991).
\newblock \doi{10.1007/BF01555493}

\bibitem{Moreno:1990sf}
G.~Moreno, et~al., Phys. Rev. D \textbf{43}, 2815 (1991).
\newblock \doi{10.1103/PhysRevD.43.2815}

\bibitem{Rondio:1993mf}
E.~Rondio, Nucl. Phys. A \textbf{553}, 615C (1993).
\newblock \doi{10.1016/0375-9474(93)90668-N}

\bibitem{Ball:2018twp}
R.D. Ball, E.R. Nocera, R.L. Pearson, Eur. Phys. J. C \textbf{79}(3), 282
  (2019).
\newblock \doi{10.1140/epjc/s10052-019-6793-5}

\bibitem{Kulagin:1989mu}
S.A. Kulagin, Nucl. Phys. A \textbf{500}, 653 (1989).
\newblock \doi{10.1016/0375-9474(89)90233-9}

\bibitem{Owens:2012bv}
J.F. Owens, A.~Accardi, W.~Melnitchouk, Phys. Rev. D \textbf{87}(9), 094012
  (2013).
\newblock \doi{10.1103/PhysRevD.87.094012}

\bibitem{Melnitchouk:1994rv}
W.~Melnitchouk, A.W. Schreiber, A.W. Thomas, Phys. Lett. B \textbf{335}, 11
  (1994).
\newblock \doi{10.1016/0370-2693(94)91550-4}

\bibitem{Ellis:1982cd}
R.K. Ellis, W.~Furmanski, R.~Petronzio, Nucl. Phys. \textbf{B212}, 29 (1983).
\newblock \doi{10.1016/0550-3213(83)90597-7}

\bibitem{Qiu:1990xxa}
J.w. Qiu, G.F. Sterman, Nucl. Phys. \textbf{B353}, 105 (1991).
\newblock \doi{10.1016/0550-3213(91)90503-P}

\bibitem{Georgi:1976ve}
H.~Georgi, H.D. Politzer, Phys. Rev. D \textbf{14}, 1829 (1976).
\newblock \doi{10.1103/PhysRevD.14.1829}

\bibitem{DeRujula:1976baf}
A.~De~Rujula, H.~Georgi, H.D. Politzer, Annals Phys. \textbf{103}, 315 (1977).
\newblock \doi{10.1016/S0003-4916(97)90003-8}

\bibitem{Schienbein:2007gr}
I.~Schienbein, et~al., J. Phys. G \textbf{35}, 053101 (2008).
\newblock \doi{10.1088/0954-3899/35/5/053101}

\bibitem{Brady:2011uy}
L.T. Brady, A.~Accardi, T.J. Hobbs, W.~Melnitchouk, Phys. Rev. D \textbf{84},
  074008 (2011).
\newblock \doi{10.1103/PhysRevD.84.074008}.
\newblock [Erratum: Phys.Rev.D 85, 039902 (2012)]

\bibitem{Abazov:2008er}
V.~Abazov, et~al., Phys. Lett. B \textbf{666}, 435 (2008).
\newblock \doi{10.1016/j.physletb.2008.06.076}

\bibitem{Aaltonen:2008eq}
T.~Aaltonen, et~al., Phys. Rev. D \textbf{78}, 052006 (2008).
\newblock \doi{10.1103/PhysRevD.78.052006}.
\newblock [Erratum: Phys.Rev.D 79, 119902 (2009)]

\bibitem{Abazov:2008ae}
V.~Abazov, et~al., Phys. Rev. Lett. \textbf{101}, 062001 (2008).
\newblock \doi{10.1103/PhysRevLett.101.062001}

\bibitem{Collaboration:2010ry}
F.~Aaron, et~al., Eur. Phys. J. C \textbf{71}, 1579 (2011).
\newblock \doi{10.1140/epjc/s10052-011-1579-4}

\bibitem{Aktas:2004az}
A.~Aktas, et~al., Eur. Phys. J. C \textbf{40}, 349 (2005).
\newblock \doi{10.1140/epjc/s2005-02154-8}

\bibitem{Abramowicz:1900rp}
H.~Abramowicz, et~al., Eur. Phys. J. C \textbf{73}(2), 2311 (2013).
\newblock \doi{10.1140/epjc/s10052-013-2311-3}

\bibitem{Malace:2009kw}
S.~Malace, et~al., Phys. Rev. C \textbf{80}, 035207 (2009).
\newblock \doi{10.1103/PhysRevC.80.035207}

\bibitem{Whitlow:1991uw}
L.~Whitlow, E.~Riordan, S.~Dasu, S.~Rock, A.~Bodek, Phys. Lett. B \textbf{282},
  475 (1992).
\newblock \doi{10.1016/0370-2693(92)90672-Q}

\bibitem{airapetian2011inclusive}
A.~Airapetian, N.~Akopov, Z.~Akopov, E.~Aschenauer, W.~Augustyniak, R.~Avakian,
  A.~Avetissian, E.~Avetisyan, S.~Belostotski, N.~Bianchi, et~al., Journal of
  High Energy Physics \textbf{2011}(5), 126 (2011)

\bibitem{Rondio:1997se}
E.~Rondio, Nucl. Phys. B Proc. Suppl. \textbf{54A}, 139 (1997).
\newblock \doi{10.1016/S0920-5632(97)00029-7}

\bibitem{tkachenko2014measurement}
S.~Tkachenko, N.~Baillie, S.~Kuhn, J.~Zhang, J.~Arrington, P.~Bosted,
  S.~B{\"u}ltmann, M.~Christy, D.~Dutta, R.~Ent, et~al., Physical Review C
  \textbf{89}(4), 045206 (2014)

\bibitem{Abe:1996us}
F.~Abe, et~al., Phys. Rev. Lett. \textbf{77}, 2616 (1996).
\newblock \doi{10.1103/PhysRevLett.77.2616}

\bibitem{Abe:1998rv}
F.~Abe, et~al., Phys. Rev. Lett. \textbf{81}, 5754 (1998).
\newblock \doi{10.1103/PhysRevLett.81.5754}

\bibitem{Acosta:2005ud}
D.~Acosta, et~al., Phys. Rev. D \textbf{71}, 051104 (2005).
\newblock \doi{10.1103/PhysRevD.71.051104}

\bibitem{Aaltonen:2010zza}
T.A. Aaltonen, et~al., Phys. Lett. B \textbf{692}, 232 (2010).
\newblock \doi{10.1016/j.physletb.2010.06.043}

\bibitem{D0:2014kma}
V.M. Abazov, et~al., Phys. Rev. D \textbf{91}(3), 032007 (2015).
\newblock \doi{10.1103/PhysRevD.91.032007}.
\newblock [Erratum: Phys.Rev.D 91, 079901 (2015)]

\bibitem{Abazov:2007pm}
V.~Abazov, et~al., Phys. Rev. D \textbf{77}, 011106 (2008).
\newblock \doi{10.1103/PhysRevD.77.011106}

\bibitem{Abazov:2007jy}
V.M. Abazov, et~al., Phys. Rev. D \textbf{76}, 012003 (2007).
\newblock \doi{10.1103/PhysRevD.76.012003}

\bibitem{Aaltonen:2009ta}
T.~Aaltonen, et~al., Phys. Rev. Lett. \textbf{102}, 181801 (2009).
\newblock \doi{10.1103/PhysRevLett.102.181801}

\bibitem{Abazov:2013dsa}
V.M. Abazov, et~al., Phys. Rev. Lett. \textbf{112}(15), 151803 (2014).
\newblock \doi{10.1103/PhysRevLett.112.151803}.
\newblock [Erratum: Phys.Rev.Lett. 114, 049901 (2015)]

\bibitem{Abazov:2013rja}
V.M. Abazov, et~al., Phys. Rev. D \textbf{88}, 091102 (2013).
\newblock \doi{10.1103/PhysRevD.88.091102}

\bibitem{Aaij:2015gna}
R.~Aaij, et~al., JHEP \textbf{08}, 039 (2015).
\newblock \doi{10.1007/JHEP08(2015)039}

\bibitem{Aaij:2015vua}
R.~Aaij, et~al., JHEP \textbf{05}, 109 (2015).
\newblock \doi{10.1007/JHEP05(2015)109}

\bibitem{Khachatryan:2016pev}
V.~Khachatryan, et~al., Eur. Phys. J. C \textbf{76}(8), 469 (2016).
\newblock \doi{10.1140/epjc/s10052-016-4293-4}

\bibitem{Aaij:2015zlq}
R.~Aaij, et~al., JHEP \textbf{01}, 155 (2016).
\newblock \doi{10.1007/JHEP01(2016)155}

\bibitem{Aad:2015auj}
G.~Aad, et~al., Eur. Phys. J. C \textbf{76}(5), 291 (2016).
\newblock \doi{10.1140/epjc/s10052-016-4070-4}

\bibitem{Chatrchyan:2013mza}
S.~Chatrchyan, et~al., Phys. Rev. D \textbf{90}(3), 032004 (2014).
\newblock \doi{10.1103/PhysRevD.90.032004}

\bibitem{Chatrchyan:2012xt}
S.~Chatrchyan, et~al., Phys. Rev. Lett. \textbf{109}, 111806 (2012).
\newblock \doi{10.1103/PhysRevLett.109.111806}

\bibitem{Aad:2011dm}
G.~Aad, et~al., Phys. Rev. \textbf{D85}, 072004 (2012).
\newblock \doi{10.1103/PhysRevD.85.072004}

\bibitem{Towell:2001nh}
R.~Towell, et~al., Phys. Rev. D \textbf{64}, 052002 (2001).
\newblock \doi{10.1103/PhysRevD.64.052002}

\bibitem{E866lanl}
\url{https://p25ext.lanl.gov/e866}

\bibitem{Webb:2003ps}
J.C. Webb, et~al. (2003)

\bibitem{Yang:2000ju}
U.K. Yang, et~al., Phys. Rev. Lett. \textbf{86}, 2742 (2001).
\newblock \doi{10.1103/PhysRevLett.86.2742}

\bibitem{Sirunyan:2017azo}
A.M. Sirunyan, et~al., Eur. Phys. J. C \textbf{77}(7), 459 (2017).
\newblock \doi{10.1140/epjc/s10052-017-4984-5}

\bibitem{Aad:2015mbv}
G.~Aad, et~al., Eur. Phys. J. C \textbf{76}(10), 538 (2016).
\newblock \doi{10.1140/epjc/s10052-016-4366-4}

\bibitem{Chatrchyan:2014gia}
S.~Chatrchyan, et~al., Phys. Rev. D \textbf{90}(7), 072006 (2014).
\newblock \doi{10.1103/PhysRevD.90.072006}

\bibitem{Aad:2014vwa}
G.~Aad, et~al., JHEP \textbf{02}, 153 (2015).
\newblock \doi{10.1007/JHEP02(2015)153}.
\newblock [Erratum: JHEP 09, 141 (2015)]

\bibitem{Khachatryan:2016mlc}
V.~Khachatryan, et~al., JHEP \textbf{03}, 156 (2017).
\newblock \doi{10.1007/JHEP03(2017)156}

\bibitem{Lai:2010nw}
H.L. Lai, J.~Huston, Z.~Li, P.~Nadolsky, J.~Pumplin, D.~Stump, C.P. Yuan, Phys.
  Rev. \textbf{D82}, 054021 (2010).
\newblock \doi{10.1103/PhysRevD.82.054021}

\bibitem{Pumplin:2002vw}
J.~Pumplin, D.R. Stump, J.~Huston, H.L. Lai, P.M. Nadolsky, W.K. Tung, JHEP
  \textbf{07}, 012 (2002).
\newblock \doi{10.1088/1126-6708/2002/07/012}

\bibitem{Amoroso:2020lgh}
S.~Amoroso, et~al., in \emph{{11th Les Houches Workshop on Physics at TeV
  Colliders}: {PhysTeV Les Houches}} (2020)

\bibitem{CT18website}
CT18PDFs.
\newblock \url{https://hep.pa.msu.edu/cteq/public/ct18.html} and
  \url{https://tinyurl.com/ct18pdfs-1}

\bibitem{Kovarik:2010uv}
K.~Kovarik, I.~Schienbein, F.I. Olness, J.Y. Yu, C.~Keppel, J.G. Morfin, J.F.
  Owens, T.~Stavreva, Phys. Rev. Lett. \textbf{106}, 122301 (2011).
\newblock \doi{10.1103/PhysRevLett.106.122301}

\bibitem{Paukkunen:2013grz}
H.~Paukkunen, C.A. Salgado, Phys. Rev. Lett. \textbf{110}(21), 212301 (2013).
\newblock \doi{10.1103/PhysRevLett.110.212301}

\bibitem{Jlab6-inprep}
CTEQ-JLab, I.~Fernando, et~al., in preparation, 2021

\bibitem{ApollinariG.:2017ojx}
G.~Apollinari, I.~B\'ejar~Alonso, O.~Bruning, P.~Fessia, M.~Lamont, L.~Rossi,
  L.~Tavian, CERN Technical Design Report V. 0.1 \textbf{4/2017} (2017).
\newblock \doi{10.23731/CYRM-2017-004}

\bibitem{Accardi:2012qut}
A.~Accardi, et~al., Eur. Phys. J. \textbf{A52}(9), 268 (2016).
\newblock \doi{10.1140/epja/i2016-16268-9}

\bibitem{Acciarri:2015uup}
R.~Acciarri, et~al.,   (2015).
\newblock \doi{arXiv: 1512.06148 [physics.ins-det]}

\bibitem{Bickerstaff:1985ax}
R.P. Bickerstaff, M.C. Birse, G.A. Miller, Phys. Rev. Lett. \textbf{53}, 2532
  (1984).
\newblock \doi{10.1103/PhysRevLett.53.2532}

\bibitem{Melnitchouk:1992eu}
W.~Melnitchouk, A.W. Thomas, Phys. Rev. D \textbf{47}, 3783 (1993).
\newblock \doi{10.1103/PhysRevD.47.3783}

\bibitem{Miller:2013hla}
G.A. Miller, Phys. Rev. C \textbf{89}(4), 045203 (2014).
\newblock \doi{10.1103/PhysRevC.89.045203}

\bibitem{Hobbs:2011vy}
T.J. Hobbs, J.T. Londergan, D.P. Murdock, A.W. Thomas, Phys. Lett. B
  \textbf{698}, 123 (2011).
\newblock \doi{10.1016/j.physletb.2011.02.040}

\bibitem{Lai:2010vv}
H.L. Lai, M.~Guzzi, J.~Huston, Z.~Li, P.M. Nadolsky, J.~Pumplin, C.P. Yuan,
  Phys. Rev. \textbf{D82}, 074024 (2010).
\newblock \doi{10.1103/PhysRevD.82.074024}

\end{thebibliography}
\end{document}